\documentclass[12pt]{article}
\usepackage{amsmath,amsfonts}
\usepackage{multirow}
\usepackage{amssymb}
\usepackage[colorlinks=true,urlcolor=blue]{hyperref}
\usepackage{graphicx}
\usepackage{undertilde}
\usepackage{ulem}
\usepackage{cancel}
\hypersetup{                
backref = true,             
pagebackref = true,                 
hyperindex = true,          
colorlinks = true,          
breaklinks = true,          
urlcolor = blue,            
linkcolor = blue,           
bookmarks = true,           
bookmarksopen = true,       
citecolor=blue,
}

\usepackage{hyperref}
\usepackage{breakurl}
\hypersetup{breaklinks=true}
\def\bepro{\begin{proposition}}
\def\enpro{\end{proposition}}
\newtheorem{theorem}{Theorem}[section]

\newtheorem{proposition}[theorem]{Proposition}

\newcommand{\qeed}{\hfill\textrm{QED}\break\null}

\advance \textheight by \topskip
\makeatletter

\def\cc#1{\kern .7em\hfill #1 \hfill\kern .7em}

\newcommand{\rr}{\color{magenta}}

\newcommand{\beqa}{\begin{eqnarray}}
\newcommand{\eeqa}{\end{eqnarray}}
\newcommand{\epm}{\begin{pmatrix}}
\newcommand{\bpm}{\end{pmatrix}}
\newcommand{\noi}{\noindent}
\newcommand{\nn}{\nonumber}

\newcommand{\ii}{\text{i}}

\def\d{\mathrm{d}}

\title{An overview of generalised Kac-Moody algebras on compact real manifolds}
\begin{document}
\maketitle
\begin{center}

Rutwig Campoamor-Stursberg$^{1\ast}$, Marc de Montigny$^{2\dagger}$, Michel Rausch de Traubenberg$^{3\ddagger}$
\medskip

$^1$  Instituto de Matem\'atica Interdisciplinar and Dpto. Geometr\'\i a y Topolog\'\i a, UCM,E-28040 Madrid, Spain\\

$^2$ Facult\'e Saint-Jean, University of Alberta, 8406 91 Street, Edmonton, Alberta T6B 0M9, Canada\\

$^3$ Universit\'e de Strasbourg, CNRS, IPHC UMR7178, F-67037 Strasbourg Cedex, France\\  

\end{center}

\noindent $^\ast$ Email: rutwig@ucm.es

\noindent $^\dagger$ Email:  mdemonti@ualberta.ca

\noindent $^\ddagger$ Email:  Michel.Rausch@iphc.cnrs.fr 

\medskip 

\begin{abstract} {
A generalised notion of Kac-Moody algebra is defined using smooth maps from a compact real manifold  $\mathcal{M}$ to a finite-dimensional Lie group, by means of complete orthonormal bases for a Hermitian inner product on the manifold and a Fourier expansion. The Peter--Weyl theorem for the case of manifolds related to compact Lie groups and coset spaces is discussed, and appropriate Hilbert bases for the space $L^{2}(\mathcal{M})$ of square-integrable functions are constructed. It is shown that such bases are characterised by the representation theory of the compact Lie group, from which a complete set of labelling operator is obtained. The existence of central extensions of generalised Kac-Moody algebras is analysed using a duality property of Hermitian operators on the manifold, and the corresponding root systems are constructed. Several applications of physically relevant compact groups and coset spaces are discussed.  }
\end{abstract}

\section{Introduction}
Kac-Moody algebras have been used in theoretical physics from the beginning 1980s onwards in various different contexts, such as  string theory, the study of  critical phenomena in two-dimensional statistical systems, Yang--Mills theory as well as in applications to exact solvable models (see {\it e.g.} \cite{dms} and references
therein).  Besides the axiomatic construction, Kac-Moody algebras (or more precisely, affine Lie algebras) can be obtained from affine extensions of the loop algebra of smooth maps from the unit circle $\mathbb{S}^{1}$ into a simple Lie group
\cite{Kac, Kac2, Moo, Mdo,ps,go}. Another infinite-dimensional Lie algebra widely encountered in two-dimensional Conformal Field Theory,  as well as in string theory, is the Virasoro algebra, the central extension of the Witt algebra, that is, the centrally extended Lie algebra of polynomial vector fields on the circle  $\mathbb{S}^1$ \cite{bpz} (and {\it e.g.}  \cite{Fuks} and references
therein).  Various types of generalisations of Kac-Moody algebras  respectively affine Lie algebras \footnote{In the following
(unless otherwise stated) we will always refer to Kac-Moody algebras instead of   affine  Lie algebras in order to be coherent with the physical literature.}  have been proposed in the literature, such as the so-called quasi-simple Lie algebras in Ref. \cite{KT}, the generalised Kac-Moody algebras in Ref. \cite{Frap}, the Borcherds algebra \cite{Borc}, as well as related structures like the Monster algebra \cite{Griess,Conw} and the Monstrous Moonshine 
\cite{Conw2,Bor2,Gann}. 

As a matter of fact, Kac-Moody and Virasoro algebras are deeply related to  the one-dimensional compact manifold $\mathbb S^1$.
 In this context, it is natural to expect that physical theories with more than two dimensions involve richer structures.
 It is with such possibilities in mind that we discuss hereafter a generalisation of  the notion of Kac-Moody algebras
 associated to compact manifolds ${\cal M}$ of dimension higher than one. In particular, we shall restrict ourselves to certain
 type of manifolds, namely compact Lie groups
${\cal M}=G_c$ or coset spaces ${\cal M} = G_c/H$, where $H \subset G_c$ is a closed subgroup. The reason for these choices lies on the fact that the harmonic functions
on the corresponding manifold ${\cal M}$ can be classified in terms of the representation theory of the Lie group $G_c$.
 The algebras described in this paper do not belong to the general classification of Kac-Moody algebras given by Kac \cite{Kac2}; rather, they represent generalisations of affine Lie algebras which, as we will observe below, admit roots but not simple roots and thus no Cartan matrices (except, as we will prove in Section \ref{Sec4}, when the number of central charges, or `order of centrality', is equal to one). Moreover, unlike the usual Kac-Moody algebras, we can construct all the generators of our generalised algebras.   

The notion of generalised Kac-Moody algebras is motivated by various phenomena in higher-dimensional physics, and possess the salient feature of being fully specified by harmonic expansions on ${\cal M}$. These algebras are potentially of use in  the Kaluza-Klein theory (see {\it e.g.} \cite{ss,bl,dpn}, with the latter reference 
being motivated in the supergravity context), where the space-time
takes the form $K = \mathbb R^{1,3}  \times {\cal M}$. Symmetries in $K$, in particular the N\oe ther theorem, lead naturally
to such generalised Kac-Moody algebras. Similarly, this type of structure emerges naturally through the consideration of  current algebras \cite{ad,gjt}.
 For instance, the authors of reference \cite{dd} analysed the symmetries corresponding to the massive states in the Fourier expansion for a Kaluza-Klein compactification in five dimensions, with these symmetries involving Kac-Moody and Virasoro algebras without central extensions. These authors suggested that similar infinite-dimensional symmetries should also appear in more complicated  higher-dimensional theories with non-Abelian symmetry of the extra dimensions.
 The relevant point is that this type of algebras admits central extensions.  These central extensions can be introduced in two
different but related ways, either by introducing two-cocycles in their Lie brackets or, more physically,  adding 
Schwinger terms \cite{Scw} to the current algebra.

  
The structure of the paper is the following:
In Section \ref{Sec2} we
define generalised Kac-Moody algebras  by means of the set of smooth maps from a compact real manifold to a real or complex finite-dimensional Lie group, in terms of a complete orthonormal basis for the Hermitian scalar product on the manifold with a Fourier expansion.
 In the following, we shall restrict our discussion to manifolds related to compact Lie groups, mainly due to technical reasons. Although there is no doubt that the case of non-compact Lie groups is full of interest, with potential applications to non-Euclidean spaces and general manifolds, their study require techniques somewhat different from  those used in this work. The main difference between compact and non-compact Lie groups is that  unitary representations of the former are finite dimensional, whilst  those of the latter are infinite dimensional. In addition, non-compact Lie groups exhibit irreducible unitary representations occurring outside the space of square integrable functions on the group, implying that a more general positive measure on the space of irreducible unitary representations must be defined, thus leading to more general integral formulae as the Plancherel formula instead of the Peter--Weyl theorem of the compact case \cite{HC}. Another technical difficulty resides in the division into discrete and continuous series, specifically in the context of the normalisation problem for discrete and the continuous spectra.
 For these reasons, in this paper we shall only offer a glimpse of the (rather different) constructions based on non-compact groups, the general analysis of which  would be beyond the scope of our work. In Section \ref{Sec3}, we discuss the Fourier expansion on manifolds taken as Lie groups and coset spaces, and discuss the Peter-Weyl theorem in this context.
  The corresponding Hilbert basis $\cal B$ of $L^2({\cal M})$ is appropriately identified. As elements of ${\cal B}$ are characterised
by the representation theory of $G_c$, we consider 
 the labelling problem and identify  a minimal set of operators, beyond the usual Casimir operators and
Cartan subalgebra of $G_c$, to identify unambiguously all elements of  $\cal B$.
With these considerations, in Section \ref{Sec4}, we construct the generalised Kac-Moody algebras for the  case where the  underlying manifold is a compact Lie group and the coset space  is a factor space  of a compact Lie group by a  closed subgroup.
It is shown that these algebras admit central extensions related by some kind of duality to certain Hermitian operators
of ${\cal M}$. The root system of the centrally extended algebra is identified and some elements of the representation theory are
given, at least for the simplest case, corresponding to the $n-$dimensional tori ${\cal M}= \mathbb T^n$.
In Section \ref{sec:ap}, some applications of the construction are presented in detail. Finally, in Section  \ref{Conclusion}, some conclusions are drawn and potential generalisations of the approach discussed.


\section{Algebras associated to compact manifolds\label{Sec2}}

In the following we shall assume that ${\cal M}$ is a compact real manifold. Let  $L^2({\cal M})$ denote the
space of square integrable functions on ${\cal M}$ and $\d\mu({\cal M})$ the integration measure on ${\cal M   }$. If 
 ${\cal B}=\{\rho_I(m), I\in {\cal I}\}$ is a  complete  orthonormal basis for the Hermitian scalar product on ${\cal M}$, with ${\cal I}$ a countable set, the identity 
\beqa
(\rho_I,\rho_J) =\int_{{\cal M}} \d\mu({\cal M}) \overline{\rho^I(m)} {\rho_J(m)} = \delta^I_J\ , \ \  m \in {\cal M} \ , \nn
\eeqa
is satisfied. As a consequence, a function $\Phi \in L^2({\cal M})$ can be described in terms of the basis ${\cal B}$ as 
\beqa
\Phi(m) &=& \sum_{I \in {\cal I}} \Phi^I \rho_I(m) \equiv \Phi^I\rho_I(m) \ , \nn
\eeqa
where  
\beqa
\Phi^I&=&(\rho_I,\Phi)\  \nn
\eeqa
 correspond to  the expansion coefficients. An important question concerns the problem whether, given two elements $\rho_I, \rho_J \in {\cal B}$, the 
product still belongs to the space $L^2({\cal M})$. In this work, this will be the case, as we assume that all functions are bounded, {\it i.e.}, $|\rho_I|<M_I$  for some $M_I$,
so that $\rho_I\rho_J\in L^2({\cal M})$. This enables us to consider the Fourier expansion of product of  elements of ${\cal B}$
\beqa
\label{eq:rhorho}
\rho_I(m) \rho_J(m) = c_{IJ}{}^K \rho_K(m) \ , 
\eeqa
with $c_{IJ}{}^K\in \mathbb C $. In general, it is difficult to derive precise formulae for the coefficients  $c_{IJ}{}^K$, but we shall discuss some examples where they  can be explicitly computed, at least partially.

 In this paper we will restrict our analysis to manifolds associated to compact Lie groups $G_c$. Therefore, the functions
$\rho_I$  of the orthonormal basis ${\cal B}$ can be organised  using  the representation theory of $G_c$,
{\it i.e.}, each $\rho_I$ belongs to a given representation of $G_c$. As $G_c$ is compact,  the functions $\rho_I$ are automatically bounded. 
In addition,  the product \eqref{eq:rhorho} can be evaluated using  representations
of $G_c$ and the corresponding Clebsch-Gordan coefficients. It is important to observe that for $\rho$ (resp. $\rho'$)
belonging to a representation ${\cal D}$ (resp. ${\cal D}'$), the fact that $\rho$ and $\rho'$ are commuting functions implies that 
the product $\rho \rho'$ belongs to  ${\cal S}({\cal D} \otimes {\cal D}')$, where ${\cal S}$ denotes the symmetric
tensor product of ${\cal D}$ and ${\cal D}'$.

Consider now a  simple complex or real finite-dimensional Lie group $G$ with Lie algebra $\mathfrak g$. We denote its basis elements by $T_a$ with $a=1,\cdots, \dim \mathfrak g$. The Lie bracket is given by
\beqa
[T_a,T_b]=\ii f_{ab}{}^c T_c \ . \nn
\eeqa
It is well known that a Kac-Moody algebra can be associated  to the Lie algebra $\mathfrak g$ {\it via} the set of smooth maps from the circle $\mathbb S^1$ to $G$ \cite{Kac, Kac2, Moo,ps,go}. Similarly,  the notion of generalised Kac-Moody algebra associated to the manifold ${\cal M}$, denoted by $\mathfrak g({\cal M})$, can be defined by using the set of smooth maps from ${\cal M}$ to  $G$ as described hereafter \cite{KT,Frap,Bars}. Let  $G({\cal M})$ denote the group of smooth maps from ${\cal M}$ to $G$ and let  $g\in G$, so that
\beqa
\label{eq:g}
g=e^{\ii \;\theta^a T_a} 
\eeqa
holds if $G$ is compact. In these conditions, any element of $G$ can be represented by the exponential of an appropriate element of $\mathfrak g$, while for the non-compact case,  we have to replace it by a finite product of exponentials.  The  element  in  $G({\cal M})$  associated to \eqref{eq:g} is given by
\beqa
\hat g(m) = e^{\ii \,\theta^a(m) T_a} \ , \nn
\eeqa
where now $\theta^a(m)$ are square-integrable functions of ${\cal M}$.
 In a neighbourhood of  the identity  the following approximation holds
\beqa
\hat g(m) \sim 1 + \ii\; \theta^a(m) T_a = 1 + \ii \;\theta^{a I} \rho_I(m) T_a \ , \nn
\eeqa
where $\rho_I(m)\in {\cal B}$. In particular, the set of functions from ${\cal M} \to G$ leads,  at the infinitesimal level, to the Lie algebra  $\mathfrak g \big({\cal M} \big)$ with
basis
\beqa
\mathfrak g \big({\cal M} \big) = \Big\{T_{a I}(m) = T_a \rho_I(m), a=1,\cdots, \dim \mathfrak g, I \in {\cal I} \Big\} \ , \nn
\eeqa
and Lie brackets
\beqa
\label{eq:KM}
[T_{a I},T_{b J}]=\ii\; f_{ab}{}^c c_{IJ}{}^K T_{ c K} \ . 
\eeqa
If $\mathfrak g$ is a real Lie algebra, then the generalised Kac-Moody algebra, denoted by $\mathfrak{g}({\cal M})$, will be real, because the manifold ${\cal M}$ is  real. Clearly $\mathfrak{g}({\cal M})$ constitutes a generalisation of the usual notion of Kac-Moody algebras, but restricted hereafter to the context of
compact manifolds ${\cal M}$. The algebra \eqref{eq:KM} can be further enlarged introducing central charges and additional operators.
Actually, the possible central extensions of \eqref{eq:KM} were fully classified in \cite{ps}. It is worthy to be mentioned that the construction 
can be naturally adapted to Lie supergroups and Lie superalgebras, resulting in the notion of generalised super-Kac-Moody algebras \cite{Har,Azam,CFRS,RS}. 

 \medskip 
An alternative physical motivation for  considering generalised Kac-Moody algebras is related to Kaluza-Klein theories \cite{ss,bl,dpn}
 (and references therein) 
  and current algebras \cite{ad,gjt}. Indeed if we consider a $(4+n)$-dimensional compactified space-time of the form
\beqa
K=\mathbb R^{1,3} \times{\cal M} \ , \nn
\eeqa
where $\mathbb R^{1,3}$ is the four-dimensional space-time and ${\cal M}$ a compact  $n-$dimensional  real manifold, it follows from the N\oe ther theorem
that the conserved charges can be expressed in terms of the fields belonging to the $(4+n)-$dimensional space-time. If we denote
 by $T_a$ the conserved
charge associated to a Lie algebra $\mathfrak g$, and by $y^A$ ($A = 1 \cdots, n$) the coordinates on ${\cal M}$, then integration over the space
part of $\mathbb R^{1,3}$ but  not over the internal space ${\cal M}$ and the equal-time commutation relations lead to the current algebra
\beqa
\label{eq:CA0}
\big[T_a(y),T_{a'}(y')\big] =\ii\;  f_{a a'}{}^b T_b(y) \delta^n(y-y') \ ,
\eeqa
where the $\delta-$distribution is defined in Appendix \ref{ap:ID},  equation (\ref{eq:del}). Now consider $G_c$, a compact Lie group and $H \subset G_c$. Let us introduce a Hilbert basis of $L^2({\cal M})$ as above, and set
\beqa
T_a(y) =  T_{a I} \; \bar \rho^I(y) \ , \nn
\eeqa
then upon integration by $\int \d^n y \int \d^n y'$ (see Appendix \ref{ap:ID}) where ${\cal M}=G_c$ or  ${\cal M}=G_c/H$, gives rise to
\beqa\label{eq:CA1}
\big[T_{a I}, T_{a'I}\big]= \ii\; f_{a a'}{}^b c_{I I'}{}^J T_{b J} \ . 
\eeqa
We thus obtain a generalised Kac-Moody algebra as defined in \eqref{eq:KM}.  If we add a  Schwinger term to \eqref{eq:CA0}, we can define possible central extensions  in close analogy with the Pressley--Segal analysis of central extensions  of generalised Kac-Moody algebras \cite{ps,Scw}.
Now, considering the Lie algebra of vector fields on ${\cal M}$ generated by $L_{AI } = -\ii\;  \rho_I \partial_A$ (where
$\partial_A = \frac{\partial}{\partial y^A}$), the algebra \eqref{eq:CA1} extends to

\beqa
\label{eq:KM-vect}
\big[T_{aI},T_{bJ}\big]&=& \ii\; f_{a b}{}^c c_{IJ}{}^K  T_{cK}\ , \nn\\
\big[L_{AI},L_ {BJ}\big]&=& - \ii \Big( (\partial_A  \rho_J) L_{BI} - (\partial_B \rho_I) L_{AJ}\Big) \ ,\\
\big[L_{AI},T_{a J}\big]&=&   \rho_I \partial_A  \rho_J T_{a J} = d_{AI,J}{}^K T_{aK}\ ,  \nn
\eeqa
where the summation over repeated indices is implicit and $\rho_I \partial_A  \rho_J =d_{A I,J}{}^K \rho_K$.

The authors of Ref. \cite{dd}  analysed the symmetries induced by the massive modes appearing in the Fourier expansion for a five dimensional compactified space-time  $\mathbb R^{1,3} \times \mathbb S^1$ by using an algebra of the type \eqref{eq:KM-vect}; however in the
context of  centreless  (usual) Kac-Moody and Virasoro algebras. It was further mentioned that these results  can potentially be extrapolated to higher dimensional space-times.


\section{Fourier expansion on compact manifolds\label{Sec3}}

In this section we briefly discuss the Fourier expansion on compact manifolds. 
 Indeed, the usual Fourier analysis on the circle $\mathbb S^1$ can be extended   to compact manifolds.
Specifically, we consider two types of manifolds:  compact Lie groups and coset spaces of  compact Lie groups. For the two situations, we obtain the basic functions appearing in the Fourier analysis by group theoretical arguments.

\subsection{The compact manifold as a Lie group}\label{sec:PWtheo}

Let $G_c$ be a simple real compact Lie group and $\hat {\cal R} =\{{\cal R}_k, k \in \hat G_c\}$ be the set of all irreducible unitary representations of $G_c$, and $\hat G_c$ the set of labels of such representations.\footnote{{The notation $\hat G_c$ used here for the set of labels should not be confused with $\hat G$, which is often used in the literature to denote the affine algebras.}  }
 As  $G_c$ is compact, each of such representations is finite-dimensional; we denote the corresponding dimension of ${\cal R}_k$   by $d_k$.
For a matrix representation $D_{(k)}(g)\in {\cal R}_k$,   $g \in G_c$, we denote the matrix elements by $D_{(k)}{}^i{}_j(g)$. As  $G_c$ is a group, for two matrix representations $D_{(k)}(g), D_{(k)}(g')\in {\cal R}_k$, the matrix product  $D_{(k)}(g'')=D_{(k)}(g'g)=D_{(k)}(g) D_{(k)}(g')$ is again a representation and therefore belongs  to ${\cal R}_k$. In other words, each column (resp. each line)
of the matrix elements $D_{(k)}{}^i{}_j(g), j =1 \,\cdots, d_k$  (resp.  $D_{(k)}{}^i{}_j(g), i =1 \,\cdots, d_k$) is a $G_c-$representation.

Let $\d \mu(G_c)$ be the Haar measure of $G_c$ and consider the space of square integrable functions, $L^2(G_c)$, defined on the manifold $G_c$ and normalised
as 
\beqa
\int_{G_c} \d \mu(G_c)= 1 \ . \nn
\eeqa
This allows us to state the following theorem.

\begin{theorem}[Peter-Weyl \cite{PW}]\label{theo:PW}
Let  $\hat {\cal R}= \{{\cal R}_k, k \in \hat G_c\}$ be the set of all unitary irreducible representations of $G_c$, and 
$D_{(k)}(g)\in {\cal R}_k$, for $g\in G_c$. Then the set of functions on $G_c$,
\beqa
\psi_{(k)}{}^i{}_{j}(g) = \sqrt{d_k} D_{(k)}{}^i{}_j(g), \ \ k \in \hat G_c\ , i,j=1,\cdots,d_k, g \in G_c,
\eeqa
forms a complete Hilbert  basis of $L^2(G_c)$ with  inner product
\beqa
(\psi_{(k)}{}^i{}_{j},\psi'_{(k')}{}^{i'}{}_{j'})=\int _{G_c} \d \mu(G_c) \overline{\psi}^{(k)}{}_{i}{}^{j} (g) \psi_{(k')}{}^{ i'}{}_{j'}(g) =
\delta^{k}_{k'}
\delta_{i}^{i'}\delta^{j}_{j'} \ . \nn
\eeqa
\end{theorem}

\medskip

For any function $\Phi$ in $L^2(G_c)$, we have
\beqa
\Phi(g)= \sum_{k \in \hat G_c} \sum_{i,j=1}^{d_k} \phi^{k}{}_{i}{}^{j} \psi_{(k)}{}^{i}{}_{j}(g)
\equiv \phi^{k}{}_{i}{}^{j} \psi_{(k)}{}^{i}{}_{j}(g) \ ,\nn
\eeqa
where the coefficients $ \phi^{k}{}_{i}{}^{j}$ are given by
\beqa
\phi^{k}{}_{i}{}^{j} = \int_{G_c} \d\mu(G_c)  \overline  \psi^{(k)}{}_{i}{}^{j}(g)  \Phi(g) \ . \nn
\eeqa
As the representation is unitary, it follows that 
\beqa
\overline \psi^{(k)}{}_i{}^{j}(g) = \psi_{(k)}{}^i{}_{j}(g^{-1}) \ . \nn
\eeqa

In the following, we describe left-coset manifolds ${\cal M} = G_c/H$ for compact groups $G_c$.  As mentioned earlier, we do not discuss manifolds based on non-compact groups in detail.  Albeit formally feasible, as we shall illustrate with some examples, the construction involves various subtleties that go beyond the scope of this paper. Essentially, instead of using the Peter-Weyl theorem for the particular compact groups discussed here, the construction in the non-compact case requires the more general Plancherel theorem (see, e.g. \cite{HC,Schmid, barut}), compounded by the infinite-dimensional unitary representations of non-compact groups. Instead of the direct sums used in the context of square-integrable functions, the corresponding expression in the non-compact case uses the Plancherel integral.

\subsection{Compact manifolds as a coset space $G_c/H$}\label{sec:coset}
 
In this section, we consider the  manifold to be a left coset ${\cal M} = G_c/H$ with respect to a closed subgroup  $H$ of $G_c$.  Denote  by $\mathfrak h$ the Lie algebra associated to $H$.  In general, $G_c/H$, which is the set of equivalence classes
\beqa
g_1\sim  g_2  \ \  {\it iff} \ \ \exists\; h \in H \ \ \text{such\ that} \ \ g_2 = g_1 h \ ,\nn
\eeqa
does not form a group unless $H$ is normal in $G_c$. 
The elements of the coset space $G_c/H$ are denoted by $[g]$ and we have $[g_1]=[g_2]$ {\it iff} $\exists \;h\in H$ such that $g_2=g_1h$. Thus
if $r \in [r]$ then
\beqa
[gr]=[r'] \ \ \Leftrightarrow \ \ gr=r'h \ \ \Leftrightarrow \ \ r'=grh^{-1} \ , \nn
\eeqa
 which defines the left action of $G_c$ on $G_c/H$.

At the Lie algebra level, we write the generators of $\mathfrak g_c$,  namely $T_a$ (with $a=1, \cdots ,\dim \mathfrak g_c$), as follows:
$U_i$ with $i=1,\cdots ,\dim \mathfrak h$,  and $V_p$  with  $p=1,\cdots ,\dim \mathfrak g_c -\dim \mathfrak h$. The elements  $V_p$ belong to the space $\mathfrak g_c/\mathfrak h$, which is not  generally a Lie algebra. The commutations relations take the form
\beqa
\begin{array}{ll}
(a)&[U_j,U_k]=\ii\;g_{jk}{}^\ell U_\ell \ ,\\
(b)&[U_j,V_p]=\ii\; (R_j)_p{}^q V_q \ ,\\
(c)&[V_p,V_q]= \ii \;g_{pq}{}^j U_j + \ii \;g_{pq}{}^r V_r \ .
\end{array}\nn
\eeqa
The relations (a) are trivially satisfied as $\mathfrak h$ is a Lie subalgebra of $\mathfrak g_{c}$, whereas the
relations (b)  imply that $\mathfrak g_{c}/\mathfrak h$ is a representation of $\mathfrak g_{c}$.  We remark that   if 
$g_{pq}{}^r=0$ holds in (c),  then the manifold $G_c/H$ is said to be a symmetric space.

Now let us extend the harmonic expansion of Theorem \ref{theo:PW} to a coset space  $G_c/H$  seen  as a manifold \cite{ss}.  As the elements of $G_c/H$ are equivalence classes, we  consider now a function $\Phi \in L^2(G_c/H)$, with components $\Phi^i$, which belongs to a certain representation  $ {\cal R}_H$ of $H$  and, for a given $h\in H$, we denote by $D^i{}_j(h)$ its matrix elements  in ${\cal R}_H$. Thus, we have
\beqa
\Phi^i(gh)=D^i{}_j(h)\Phi^j(g) \ , \ \ h \in H,\ g \in G_c/H \ . \nn
\eeqa

In order to apply the harmonic expansion to a coset manifold $G_c/H$, we do not need to consider all the representations of $G_c$.  Let us consider the representation ${\cal R}_{(k)} \in {\cal R}, k\in  \hat G_c$  with matrix representative $D_{(k)}(g)$ for $g \in G_c$,  such that
\beqa
D_{(k)}(gh) = D_{(k)}(g) D(h)  \ . \nn
\eeqa
This relation is possible if and only if, in the embedding $H \subset G_c$, we have
\beqa
{\cal R}_{(k)} = m_{k}  {\cal R}_H \oplus \cdots \ . \nn
\eeqa
 In other words, ${\cal R}_H$ is contained  $m_k$ times (with $m_k>0$)  in ${\cal R}_{(k)}$, so that $m_{k}$ is the multiplicity of  ${\cal R}_H$  in the decomposition $H\subset G_c$. We denote by $\hat {\cal R} |_{{\cal R}_H}$ the set of representations of $G_c$ satisfying this property, while $\hat G_c|_{{\cal R}_H}$ denotes the set of corresponding labels.

The harmonic expansion takes the form
\beqa
\Phi^i(g) = \sum\limits_{k \in \hat G_c|_{{\cal R}_H}} \sum \limits_{n=1}^{m_{k} }\sum \limits_{j=1}^{d_k}\sqrt{\frac{d_k}{d_D} }
\Phi^{(k)jn} D_{(k)}{}^i{}_{j,n} (g) \ , \nn
\eeqa
where $d_D$ is the dimension of the representation ${\cal R}_H$.
Let $g=rh$, with $r$  a representative of the equivalence class $[r]\in G_c/H$. From the identities

\beqa
\int \d\mu(G_c) \sqrt{\frac{d_k}{d_D}} D_{(k)}{}^j{}_{i,n}(g^{-1})\Phi^i(g) &=&\int \d\mu(G_c) \sqrt{\frac{d_k}{d_D}} D_{(k)}{}^j{}_{i,n}(h^{-1}r^{-1})\Phi^i(rh)\nn\\
&=&\int \d\mu(G_c) \sqrt{\frac{d_k}{d_D}} D_{(k)}{}^{j}{}_{\ell',n}(r^{-1})  D^{\ell'}{}_{i}(h^{-1})D^i{}_\ell(h)\Phi^\ell(r)\nn\\
&=&
\int \d \mu(H) D^{\ell'}{}_{i}(h^{-1})D^i{}_\ell(h)\nn\\
&&\times
\int \d\mu(G_c/H) \sqrt{\frac{d_k}{d_D}} D_{(k)}{}^{j}{}_{\ell'}(r^{-1}) \Phi^\ell(r)\nn\\
&=&\int \d\mu(G_c/H) \sqrt{\frac{d_k}{d_D}} D_{(k)}{}^{j}{}_{\ell',n}(r^{-1}) \Phi^\ell(r)\ ,\nn
\eeqa
we conclude that the coefficients of the expansion are given by
\beqa
\Phi^{(k)jn}=\int \d\mu(G_c/H) \sqrt{\frac{d_k}{d_D}} D_{(k)}{}^{j}{}_{i,n}(r^{-1}) \Phi^i(r)\nn \ .
\eeqa

 Here, we shall be interested only   in functions $\Phi$ in the trivial representation of $H$, so that the expansion simplifies to
\beqa
 \Phi(g) = \sum\limits_{k \in \hat G_c|_{{\cal R}_0}} \sum \limits_{n=1}^{m_{k} }\sum \limits_{j=1}^{d_k}\sqrt{\frac{d_k}{d_D} }\Phi^{(k)jn} D_{(k)}{}^{i_0}{}_{j,n} (g) \ , \nn  
\eeqa
with ${\cal R}_0$  and $ D_{(k)}{}^{i_0}{}_{j,n} (g)$ the trivial representation of $H$. In all examples that will be presented in Section \ref{sec:ap}, we have $m_k=1$.

Unitary representations of compact Lie algebras are classified  either by their Dynkin labels or their Young tableaux  which correspond to tensors of a certain type,  notably for the classical series $\mathfrak{a}_n=\mathfrak{su}(n+1),\mathfrak{b}_n=\mathfrak{so}(2n+1),
\mathfrak{c}_n=\mathfrak{usp}(2n), \mathfrak{d}_n=\mathfrak{so}(2n)$.
The representations which contain the scalar representation can then be deduced from either the Dynkin label or from the Young tableau.
For instance, the only representations that lead to a scalar representation for the embedding  $SO(n-1)\subset SO(n)$ are the representations  with dominant
weight $|n,0,\cdots,0\rangle$ corresponding to $n$-th order symmetric traceless tensors, {\it i.e.}, the $n^{th}$-order symmetric power of the fundamental representation $\left[1,0^{n-1}\right]$.\\

\subsection{Labeling functions in the Peter-Weyl theorem}\label{sec:MLO}

The functions appearing in harmonic analysis on ${\cal M} =G_c$ or  ${\cal M} =G_c/H$ are  associated to
all  the finite-dimensional  unitary representations of $G_c$. It is however well known that, within a given representation,  
a weight vector is generally not uniquely defined by its eigenvalues with respect to a given Cartan subalgebra. The purpose
of this section is to identify a minimal set of   operators to characterise
unambiguously all weight vectors in an arbitrary representation.
 In essence, the main properties of the labelling problem  for semisimple Lie algebras are deduced from a theorem 
due to Racah \cite{Ra}, which in modern terminology can be stated as follows: 

\bepro \label{RAC} Let $\frak{g}$ be a simple (compact) Lie algebra of rank $\ell$. Then the following conditions hold:

\begin{enumerate}

\item  $\frak{g}$ admits $\ell$ independent primitive Casimir operators $\left\{C_{d_1},\cdots ,C_{d_\ell}\right\}$.\footnote{
 By primitive Casimir operators we mean those of minimal degree in the generators \cite{Co,Ra}.}  

\item Each $C_{d_{k}}$ can be represented as a homogeneous polynomial  of degree $d_k$ in the generators.

\item The degrees $d_k$ of the invariants satisfy the following numerical identity:
\begin{equation}
\sum_{k=1}^{\ell} d_k = \frac{\dim\frak{g}+\ell}{2}.    \nn 
\end{equation}

\item Any irreducible representation $\cal D$ of $\frak{g}$ is completely determined by $ \frac{\dim\frak{g}+\ell}{2}$ labels, from which
\begin{enumerate}
\item $\ell$ labels characterise the representation $\cal D$ as eigenvalues of the Casimir operators  $\left\{C_{d_1},\cdots ,C_{d_\ell}\right\}$.

\item A number of $\displaystyle \frac{\dim\frak{g}-\ell}{2}$ internal labels are required to separate the states within the multiplet
$\cal D$. 
\end{enumerate}
\end{enumerate}
\enpro

Although $\cal D$ can be distinguished from other non-equivalent representations by means of the eigenvalues of the Casimir operators or, alternatively, the highest weight with respect to a given Cartan subalgebra $\frak{h}$, the choice of internal labels is far from being unique, and  usually depends on a specific chain of proper subalgebras 
\begin{equation*}
\frak{g}_1\subset \frak{g}_2 \subset \dots \subset \frak{g}
\end{equation*}
such that, in each step, the Casimir operators of the subalgebra  are used to separate states \cite{Bdh,Lou}.
We give in Appendix \ref{sec:ML} some details on the construction of internal labels beyond the Cartan subalgebra.

\medskip
In fact, when ${\cal M} =G_c$,  we understand the action of the group $G_c$ on a
matrix $M$  in a representation ${\cal D}$  as a right action and left action, {\it i.e.}, considering $g\in  G_c$
and its corresponding matrix $D(g)$ in the representation ${\cal D}$
we have
\beqa
M \to M' = D(g) M D(g)^t \ . \nn
\eeqa
This means that the matrix elements of $M$ are labeled by three types of indices
associated to their corresponding operators:
\begin{enumerate}
\item $\ell$ labels which specify the representation: they  can be the eigenvalues of the  Casimir operators
or the eigenvalues of the Cartan generators on the highest weight; 
\item  $\frac{\dim \mathfrak g -\ell} 2$ labels which characterise the lines and are
associated to internal labels and Cartan subalgebra for the right action. 
\item $\frac{\dim \mathfrak g -\ell} 2$ labels  which characterise  the columns and are
associated to internal labels and Cartan subalgebra for the left action. 
\end{enumerate}
Therefore   $\dim \mathfrak{g}$ operators are needed to label unambiguously all states when ${\cal M} =G_c$,
whilst for  ${\cal M} =G_c/H$ the number of operators needed is $\ell$ to label all representations and $\frac{\dim \mathfrak g -\ell} 2$
internal labels.\footnote{Actually the full set of labels is only necessary for the generic case. For representations of $G_c$ exhibiting some kind of symmetry, the number of internal labels needed is usually smaller.}

 
\section{Generalised Kac-Moody algebras\label{Sec4}}

In this section we build explicitly the generalised Kac-Moody algebra associated to the manifold ${\cal M}$, where
${\cal M}$ is either a  compact Lie group $G_c$  or a coset space $G_c/H$ with respect to a  closed subgroup $H \subset G_c$.
This construction  proceeds in several steps.

\subsection{Construction of the  algebra}

Let $\mathfrak g$ be a simple real (or complex) Lie algebra with basis $\{T_a, a=1,\cdots, \dim \mathfrak g\}$. The Lie
brackets take the form
\beqa
\big[T_a,T_b\big] = \ii \;f_{ab}{}^c T_c \ . \nn
\eeqa
Further, denote the Killing form by 
\beqa
\Big<T_a,T_b\Big>_0 = g_{ab} \equiv\  \text{Tr}\Big(\text{ad}(T_a) \text{ad}(T_b)\Big)  \ . \nn
\eeqa

Let ${\cal M}$ be  a compact $n=(p+q)-$dimensional manifold of volume $V$,  isomorphic to either $G_c$ or $G_c/H$, and suppose that we have a parameterisation
$ y^A = (\varphi^i, u^r)= (\varphi^1,\cdots,\varphi^p, u^1,\cdots, u^{  q})$  such that
\beqa
\int _{\cal M} \d \mu({\cal M}) = \frac 1 V \int_{\cal M} \d ^p \varphi \; d^q u = 1 \ .\nn
\eeqa
Consider    the set of square integrable functions on ${\cal M}$ which are periodic in all $\varphi-$directions, but not in
the $u-$directions.  As done in Sections \ref{sec:PWtheo} and \ref{sec:coset}, we introduce  a Hilbert basis  of $L^2({\cal M})$
 identified with a minimal set of labels (see Section \ref{sec:MLO})
\beqa
{\cal B} = \Big\{\rho_ I(\varphi,u)\ , \ \ I \in {\cal I} \Big\} \ ,\nn
\eeqa
where ${\cal I}$ denotes the set of all labels needed to identify the states unambiguously.
Let $\mathfrak g({\cal M})$ be the set of smooth maps from ${\cal M}$ into $\mathfrak g$:
\beqa
\mathfrak g({\cal M}) = \Big\{T_{a I} = T_a \rho_I(\varphi,u) \ , a = 1, \dots, \dim \mathfrak g\ , I \in  {\cal I} \Big\} \ . \nn
\eeqa
In this case (see Appendix \ref{ap:ID}), the Lie brackets take the form
\beqa
\label{eq:KM-def}
\big[T_{aI}, T_{bJ}\big] =\ii\; f_{ab}{}^c c_{IJ}{}^K T_{cK} \ .
\eeqa
The precise form of the coefficients $ c_{IJ}{}^K $ defined in \eqref{eq:rhorho}  is irrelevant at this stage.  We shall present several explicit examples  in Section \ref{sec:ap}.
Finally,  the Killing form in $\mathfrak g({\cal M})$ is given by
\beqa
\label{eq:killKM}
\Big<X,Y \Big>_1 = \int _{\cal M} \d \mu({\cal M}) \Big<X,Y \Big>_0 \ , 
\eeqa
for $X,Y \in \mathfrak g({\cal M})$.
It follows (see Appendix \ref{ap:ID}) that 
\beqa 
\rho_I  (\varphi, u) = \eta_ {IJ} \overline{\rho}^ J (\varphi, u) \ , \nn
\eeqa
so that 
\beqa
\Big<T_{aI} ,T_{bJ}  \Big>_1 = g_{ab} \eta_{IJ} \ . \nn
\eeqa

\paragraph{ Central extensions of the generalised algebra.}

One natural question is whether the algebra determined by \eqref{eq:KM-def} admits central extensions. This problem
was completely solved by Pressley and Segal in \cite{ps} (see Proposition 4.28  therein). Given a one-chain $C$ ({\it i.e.}, a closed one-dimensional
 piecewise smooth curve), the central extension is given by the two-cocycle
\beqa
\label{eq:2Cocy}
\omega_C(X,Y) = \oint_C \big<X, \d Y\big>_0  \ , 
\eeqa
where $\d Y =\partial_A Y\ \d y^A= \partial_i Y\ \d \varphi^i + \partial_s Y\ \d u^s$ is the exterior derivative of $Y$.
We observe that the two-cocycle $\omega_C$ is non-trivial even if the manifold ${\cal M}$ has a trivial  fundamental group. In fact this result
extends  also for non-compact differentiable manifolds. Furthermore, as
pointed out by
Pressley and Segal, this result is somewhat  disappointing,  as central extensions are characterised by maps
from $C\to {\cal M}$, where $C$ is one-dimensional. Stated differently, there is no central extension  built up from maps
${\cal N} \to {\cal M}$ when $\dim {\cal N} >1$. The two-cocycle can be written in alternative form. Indeed we have  \cite{bt}
\beqa
\label{eq:2Cocy-gamm}
\omega_C(X,Y) = \int_{{\cal M}}  \big<X, \d Y\big>_0 \wedge \gamma \ , 
\eeqa
where $\gamma$ is a closed $(n-1)-$current (a distribution)  associated to $C$. In particular if
\beqa
\gamma = \sum \limits _{A=1}^n (-1)^A \gamma_A \;\d y^1 \wedge \cdots \wedge  \d y^{A-1} \wedge  \d y^{A+ 1} \wedge \cdots \wedge\d y^n , \nn
\eeqa
then
\beqa
\d \gamma = \sum \limits _{A=1}^n \partial_A \gamma_A \d y^1 \wedge \cdots \wedge  \d y^{n}=0 \ . \nn 
\eeqa

 Using the topological properties of the manifolds ${\mathbb S}^2$ and ${\mathbb S}^1 \times {\mathbb S}^1$, the authors of Ref. \cite{Frap} classified all possible
central extensions for the case in which $\gamma$ is defined only by functions and not distributions.
For our purposes, in order to have some contact with the current algebra \eqref{eq:CA}, hereafter, we will  consider $n$ specific $(n-1)-$forms :
\beqa
\gamma_{(A)} =    (-1)^A k_A  \d y^1 \wedge \cdots \wedge \d y^{A-1} \wedge  \d y^{A+ 1}\wedge \cdots \d y^n \ , \ A=1,\cdots ,n \nn
\eeqa
where $k_A \in \mathbb R$. Thus
\beqa
\omega_{(A)}(T_{aI}, T_{bJ})&=&  k_A g_{ab} \int _{{\cal M}} \d \mu({\cal M}) \; \rho_I(\varphi,u) \partial_A \rho_J(\varphi,u) \nn\\
                         &=& k_A g_{ab} d_{AIJ} \ .\nn
\eeqa
The brackets of the centrally  extended algebra $\mathfrak{g}({\cal M})$ take the form
\beqa
\label{eq:KM-ce}
\big[T_{aI},T_{bJ}\big] = \ii \; f_{ab}{}^c c_{IJ}{}^K T_{cK} +  g_{ab} \sum \limits_{A=1}^n k_A d_{AIJ} \ . 
\eeqa

The algebra constructed by this procedure is closely related to a current algebra with Schwinger terms
\beqa
\label{eq:CA}
\big[T_a(y),T_{a'}(y')\big] = \ii \;f_{a a'}{}^b T_b(y) \delta^n(y-y') -\ii\; \sum \limits_{A=1}^n k_A \partial_A \delta^n(y-y') \ . 
\eeqa
Indeed, upon integration by $\int \d^n y \int \d^n y'$ (see Appendix \ref{ap:ID}, in particular,  equation \eqref{eq:del})   equation \eqref{eq:CA}
leads to equation \eqref{eq:KM-ce}. In   Ref. \cite{Bars},   Bars constructed centrally extended extensions of the generalised Kac-Moody algebras
$\mathfrak{g}(\mathbb S^2)$ and $\mathfrak{g}(\mathbb S^1 \times \mathbb S^1)$   by using a current algebra
approach.

As already mentioned, a generalisation of Kac-Moody algebras to the case of non-compact manifolds can be considered, hence the natural question whether  these algebras admit central extensions arises. This question was briefly studied in \cite{RS}, where it was shown that \eqref{eq:2Cocy} still defines a two-cocycle as ${\cal C}$ has no boundary, but that the cocycle reformulation \eqref{eq:2Cocy-gamm} must be treated with care because the manifold  ${\cal M}$ is non-compact and divergence problems for the integrals may appear. A generic ansatz to circumvent this technical difficulty has not yet been found. 

\paragraph{ Derivations of the generalised algebra.}
The last step in the construction of the generalised Kac-Moody algebra associated to the manifold ${\cal M}$ is to introduce
the derivations $\partial_A$. However, due to the specific parametrisation of ${\cal M}$, the variables $\varphi$ and $u$
have different periodicity properties. The former are periodic whereas the latter are not. This in particular means
that the operators $d_j=-\ii\partial_{\varphi^j}$ associated to the variables $\varphi^j$ are Hermitian whilst, due to the boundary term in the integration
by parts, the operators $d_s= -\ii \partial_{u^s}$ associated to the variables $u^s$ are not Hermitian.
However, as we shall see, additional  Hermitian operators beyond  $d_j, j=1,\cdots, p$ can be considered.
The existence of these additional operators  follows from the relation between the manifold ${\cal M}$ and
the Lie group $G_c$.   We  can thus identify a maximal set
of commuting Hermitian operators. Of course, the operators $d_j,  j=1,\cdots,p$ are  commuting Hermitian operators.
  As just mentioned, since the manifolds that we consider are of the form $G_c$ or $G_c/H$, there exists a largest Lie
algebra  $\mathfrak{g}_m$ such that $\mathfrak g_c \subseteq \mathfrak g_m$,  with  $\mathfrak g_c$ the Lie algebra of $G_c$ (see examples below),
such that the basic functions $\rho_I$ belong to some unitary irreducible representation of $\mathfrak{g}_m$.
Furthermore the generators of the Lie algebra  $\mathfrak{g}_m$ can be realised as differential Hermitian operators acting on
${\cal M}$. Thus, among those generators we can extract the generators of the Cartan subalgebra $H_1,\cdots, H_k$, where
$k$ is the rank of $\mathfrak g_m$.  We express these operators as
\beqa
H_j= -\ii f_j^A(y) \partial_A \ . \nn
\eeqa
The Hermicity condition translates into
\beqa
\label{eq:herm}
\partial_A f_j^A (y) =0 \ \  \text{and}\ \  f_j^r| =0\ , \ r=1,\cdots, p,
\eeqa
where $f_j^r|=0$ means that the boundary term associated to all $u-$directions vanishes.
Now we identify among the generators $d_1, \cdots, d_p, H_1,\cdots, H_k$ the maximal set of commuting operators that we denote
$D_1,\cdots, D_r$.
 These generators are easily seen to adopt the form 
\beqa
D_j=-\ii f_j^A(y) \partial_A \ , \ j=1,\cdots, r \nn
\eeqa
and satisfy \eqref{eq:herm}.
Naturally, the functions $\rho_I$ are eigenfunctions of $H_j$ and we note
\beqa
H_j(\rho_I(y))= I(j) \rho_I(y) \ , \nn
\eeqa
with $I(j)$ the corresponding eigenvalue.

 It is worthy to be observed that there exists some  kind of duality between the Hermitian operators $D_j$ and central extensions. Indeed, one can easily show that
the $(n-1)-$forms
\beqa
\label{eq:form}
\gamma_j =  k_ j\sum \limits_{A=1}^n (-1)^A f_j^A(y)\; \d y^1 \wedge \cdots  \wedge \d y^{A-1} \wedge \d y^{A+1} \wedge \cdots \wedge \d y^n \ ,\ 
 j=1,\dots,
r \
\eeqa
$k_j\in \mathbb R$ are closed because of the condition \eqref{eq:herm}, and the corresponding two-cocycles are given by
\beqa
\label{eq:cc}
\omega_k(T_{aI},T_{bJ}) =  k_k  J(k) g_{ab} \eta_{IJ}\ . 
\eeqa

The generalised Kac-Moody algebra is thus generated by
\begin{enumerate}
\item  $T_{aI}$ which belong to $\mathfrak{g}({\cal M})$;
\item  the Hermitian operators $D_1,\cdots, D_r$;
\item the central charges $k_1,\cdots, k_r$ associated to the Hermitian operators.
\end{enumerate}

The non-vanishing brackets of the generalised Kac-Moody algebra
associated to ${\cal M}$  have the form
\beqa
\label{eq:KM-gen}
\big[T_{aI},T_{bJ}\big] &=&\ii\;  f_{ab}{}^c c_{IJ}{}^K T_{cK} +    g_{ab}  \eta_{IJ}  \sum \limits_{j=1}^r k_j I(j)
 \ , \nn\\
\big[D_j, T_{aI}\big] &=&   I(j)  T_{aI}\ ,
\eeqa
where $I(j)$ is the eigenvalue of $D_j$.
The authors of Ref. \cite{KT} defined  generalised Kac-Moody algebras associated to the torus $\mathbb T^n$, which coincides with
our construction for $G_c = U(1) ^n$,  and showed that
these algebras correspond to  specific examples of what they called `quasi-simple Lie algebras'. It should be observed that all their operators $d_A$ are Hermitian, hence they did not encounter the problem mentioned above.
 
\subsection{Root  system of generalised Kac-Moody algebras}\label{sec:root}

The purpose of this section is to identify  a root structure for the generalised algebras defined by equation \eqref{eq:KM-gen}.
We begin with the roots of the finite-dimensional  simple Lie algebra $\mathfrak g$. Suppose that $\mathfrak g$ is of rank $\ell$.
Let $H^i, i =1,\cdots, \ell$, be the  generators of the Cartan subalgebra of $\mathfrak g$ and let $\Sigma$ be the root system
of  $\mathfrak g$. We consider the corresponding operators $E_\alpha, \alpha \in \Sigma$,  in the usual Cartan-Weyl basis.
If we introduce
\beqa
\label{eq:hatg}
\hat{ \mathfrak g}({\cal M})= \text{Span}\Big\{ T_{aI}, D_j, k_j, a=1,\cdots, \dim \mathfrak g, I \in {\cal I}, j=1,\cdots, r \Big\} \ ,
\eeqa
then we observe from the algebra \eqref{eq:KM-gen} that, in addition to the elements of the Cartan subalgebra
of $\mathfrak g$, the operators $D_j$ and $k_j$ commute  with each other for $ j=1,\dots, r$.
Thus the Cartan subalgebra  of $\hat {\mathfrak{g}}({\cal M})$ is  then generated by $H^i$, $D_j$ and $k_j$, where  $i=1,\dots, \ell,\; j = 1,\cdots, r$, and the Cartan-Weyl basis takes the form $H^i_I$   and $E_{\alpha I}$,  where the non-vanishing brackets read
\beqa
\label{eq:CW}
\big[H^i_{I}, H^{i'}_{I '}\big] &=&  \eta_{ II'} h^{ii'}\sum _{p=1}^r I'(k) k_p   \ , \nn\\
\big[H^i_{I}, E_{\alpha J} \big]&=& c_{I J}{}^K  \alpha^i \; E_{\alpha K }\  ,
\nn\\
\big[E_{\alpha_ I}, E_{\beta_ J}\big] &=&
\left\{
\begin{array}{ll}
{\cal N}_{\alpha,\beta} \; c_{I J }{}^K \; E_{\alpha+\beta K }\ ,
& \alpha+\beta \in \Sigma ,\\[4pt]
c_{I J }{}^K\;  \alpha\cdot H_{K}
+ \eta_{IJ}\sum\limits _{p=1}^r J(k)  k_p\ ,   & \alpha + \beta=0,\\[4pt]
0,&\left\{\begin{array}{l}\alpha + \beta \ne 0\ ,\\ \alpha+\beta \not \in \Sigma , \end{array} \right.\ 
\end{array}
\right.\\
\big[D_i,E_{\alpha J}\big]&=& J(i) E_{\alpha_ J}\ , \nn\\
\big[D_i ,H^j_{J}]&=& J(i)  H^i_{J}\nn \ ,
\eeqa
where
\beqa
h^{ij} = \big<H^i,H^j\big>_0 \ , \nn
\eeqa
with the Killing form $ \big<\cdot,  \cdot\big>_0$ defined at the beginning of Section \ref{Sec4},
and the operators associated to roots of $\mathfrak g$ are normalised as
\beqa
\big<E_\alpha, E_\beta\big>_0 = \delta_{\alpha,-\beta} \ . \nn
\eeqa
Proceeding along the same lines as for usual Kac-Moody algebras (see {\it e.g.} \cite{go}, p. 343-344), 
we have for the Killing form of $\hat{ \mathfrak g}({\cal M})$
\beqa
\label{eq:CSA}
\Big<T_{aI},T_{bJ}\Big>&=&\eta_{IJ} g_{ab}\ ,\nn\\
\Big<D_j,T_{aI}\Big>&=&\Big<k_ j ,T_{a I }\Big> \ =\ 0\ , \\
\Big<k_i ,k_j\Big> &=&\Big<D_i ,D_j\Big>\ =\ 0\ ,\nn\\
\Big<D_i ,k_j\Big>&=&\delta_{j}^{i} \ . \nn
\eeqa

The root spaces are given by
\beqa
\label{eq:root-s}
\mathfrak g_{(\alpha, n_1,\cdots, n_r)} &=& \Big\{E_{\alpha I } \ \text{with} \ I(1) = n_1 ,\cdots , I(r)=n_r \Big\} \ ,\alpha \in \Sigma,
n_1,\cdots, n_r \in \mathbb Z \ , \nn\\ 
\mathfrak g_{(0, n_1,\cdots, n_r )} &=& \Big\{H^i_{I}\  \text{with} \ I(1) = n_1 ,\cdots , I(r)=n_r \Big\} \ , n_1,\cdots, n_r \in \mathbb Z \ . 
\eeqa
 Unlike the usual Kac-Moody algebras, the root  spaces associated to roots are infinite dimensional and we have
\beqa
\big[\mathfrak g_{(0,{\bf n})}, \mathfrak g_{(\alpha,{\bf m})}\big]&\subset& \mathfrak  g_{(\alpha,{\bf m+n})}, \nn\\
\big[\mathfrak g_{(\alpha,{\bf m})}, \mathfrak g_{(\beta,{\bf n})}\big]&\subset& \mathfrak  g_{(\alpha+\beta,{\bf m+n})}, \ \ \alpha+ \beta \in \Sigma\nn
\eeqa
with ${\bf n } = (n_1,\cdots,n_r)$. Introduce also ${\bf 0}=(0,\cdots,0)$.
 It is important to observe that  
the Lie bracket between two elements involves not only the
root structure, but also the representation theory of $G_c$, in the form of the  Clebsch-Gordan coefficients  $c_{IJ}{}^K$ (see \eqref{eq:CW}).\\

\noindent 

To define the set of positive roots, we use the lexicographic order:
\beqa
\label{eq:ord}
(\alpha,0,\cdots,0,n_1,\cdots,n_r) >0 \ \ \text{if} \left\{
\begin{array}{cc} \text{either}&\left\{\begin{array}{l}
\exists\; k \in \{1,\cdots, r\} \ \ \text{s.t.} \\
n_r=\cdots =n_{k+1} =0\ \  \text{and} \ \ n_k >0
\end{array}\right.\\
\text{or}& n_r=\cdots=n _1=0 \ \ \text{ and} \ \ \alpha>0 \ . 
\end{array}\right. 
\eeqa
 By \eqref{eq:CSA}, we can  endow the weight space with
a scalar product. Indeed
\beqa
(\alpha,c_1,\cdots,c_r,n_1,\cdots,n_r) \cdot  (\alpha',c'_1,\cdots,c'_r,n'_1,\cdots,n'_r) =\alpha \cdot \alpha'+
\sum_{j=1}^r\big(n_j c'_j + n'_J c_j\big) \ . \nn
\eeqa

  We further observe that, as happens for usual Kac-Moody algebras \cite{Kac2}, we have two types of roots.
The set of roots $(\alpha, {\bf 0}, {\bf n})$ of $\mathfrak {g}_{(\alpha, {\bf n})}$ with $\alpha \in \Sigma, {\bf n} \in \mathbb Z^r$ satisfy
\beqa
(\alpha, {\bf 0}, {\bf n}) \cdot (\alpha, {\bf 0}, {\bf n}) = \alpha\cdot \alpha >0 \ , \nn
\eeqa
and are called {\it real roots}, whilst the set   
$(0, {\bf 0}, {\bf n})$  of $\mathfrak{g}_{(0, {\bf n})}$ with $ {\bf n} \in \mathbb Z^r$ and satisfying
\beqa
(0, {\bf 0}, {\bf n}) \cdot (0, {\bf 0}, {\bf n'}) = 0 \ , \nn
\eeqa
is called the set of  {\it  imaginary roots}.

\medskip
Recall that  $r$ denotes  the number of central charges (see \eqref{eq:hatg}), that we called the order of centrality.
We now show that unless $r=1$,  we cannot find a system of simple roots for $\hat{\mathfrak g}$.
 To this extent, introduce $\alpha_i, i=1,\cdots,\ell$ the simple roots of $\mathfrak g$.  If $r=1$, and we denote  by $\psi$ the highest root of $\mathfrak{g}$, it is easy to see that
\beqa
\label{eq:s1}
\hat \alpha_i =(\alpha_i,0,0) \ , \ \ i=1,\cdots, \ell\ , \ \ \hat \alpha_{\ell+1}=(-\psi,0,1) \ 
\eeqa
is a system of simple roots of $\hat{\mathfrak g}$. Now, if we suppose that $r= 2$, as the positive roots are given by
(i) $(\alpha,0,0,0,0)$ with $\alpha>0$, or (ii) $(\alpha,0,0,n_1,0)$ with $\alpha \in \Sigma, n_1>0$, or (iii)
$(\alpha,0,0,n_1,n_2), \alpha \in \Sigma, n_1\in \mathbb Z, n_2>0$ and since the roots $(\alpha,0,0,n_1,0)$ are neither bounded from below
nor from above because $n_1 \in \mathbb Z$,  we cannot define a simple root of the form $(-\psi,0,0,-n_{\mathrm{max}},1)$, where $n_{\max}$
 corresponds to the highest possible value of $n_1$ (or $-n_{\max}$ the lowest possible value of $n_1$). This means that for $r\ge 2$ we cannot  construct a system of simple roots.
 In other words, the only generalised Kac-Moody algebras that admit simple roots are (obviously) the usual Kac-Moody algebras,
but also the Kac-Moody algebras associated to $SU(2)/U(1)$ studied in Section \ref{sec:su2-u1}.
In the latter case, the  Dynkin diagram of $\hat{\mathfrak g}(SU(2)/U(1))$  is analogous to the Dynkin diagram of
the corresponding usual Kac-Moody algebra $\hat{\mathfrak g}(U(1))$, but is  dressed by the representation theory of $SO(3)$.
Indeed, in this case the root  space  is infinite dimensional (see  \eqref{eq:root-s}).

 We have seen that
for a generalised Kac-Moody algebra of centrality order $r>1$, the set of imaginary roots is $r$-dimensional.
We may then wonder whether the algebra associated to the manifold ${\cal M}$ has some relationship with a degenerate Kac-Moody algebra
with Cartan matrix of co-rank $r$. In fact,  the algebra associated to the manifold ${\cal M}$ does not belong to the general classification of Kac-Moody algebras as given by Kac in \cite{Kac2}. Indeed, Kac-Moody algebras are defined by a (symmetrisable) Cartan matrix and thus admit
a Chevalley-Serre presentation. Differently, the algebras considered in this paper represent generalisations of affine Lie algebras.
 In particular, we have seen that for centrality orders strictly higher than one,  there does not exist a system of simple roots, and hence no Cartan matrix or Chevalley-Serre basis exist. Moreover,  we can construct all the generators of our generalised algebra, whilst this is not the case for the  Kac-Moody algebras (different from affine Lie algebras).  Observe moreover that for a centrality order $r=1$ corresponding to the algebra $\hat{\mathfrak g}(U(2)/U(1))$, we have a system of simple roots and a Chevalley-Serre presentation of the algebra. Even if the Dynkin diagram of $\hat{\mathfrak g}(U(2)/U(1))$ coincides
with the Dynkin diagram of $\hat{\mathfrak g}(U(1))$, the former is dressed by the representation theory of $SO(3)$.
\\

To finish this section, we briefly show that the generalised Kac-Moody algebras constructed so far share some properties with the
so-called Lorentzian Kac-Moody algebras \cite{gow,west}. Lorentzian Kac-Moody algebras appear in $M-$theory or in eleven-dimensional
supergravity compactifed on tori. Such algebras  are defined by a Cartan matrix (or a Dynkin diagram) subjected to some constraints. 

To this extent, introduce the root-lattice of $\hat{ \mathfrak g}({\cal M})$. We begin with the root lattice of $ \mathfrak g$
\beqa
\Lambda_R(\mathfrak g) = \Big\{ \sum_{i=1}^\ell n^i \alpha_i\ , \ n^i \in \mathbb Z\Big\} \ , \nn
\eeqa
with $\alpha_1,\cdots,\alpha_\ell$ the simple roots of $\mathfrak g$, supposed of rank $\ell$.
We then introduce the two-dimensional Lorentzian even self-dual lattice \cite{gow, west}
\beqa
\Pi^{1,1} = \Big\{(m,n)\ , \ m,n \in \mathbb Z\ \Big\} \ , \nn 
\eeqa
endowed with the Lorentzian scalar product
\beqa
(m,n) \cdot (m',n') = m n'+ n m' \ . \nn
\eeqa
Let   $(e=(0,1), \bar e = (1,0))$ satisfying $e\cdot \bar e=1, e\cdot e = \bar e \cdot \bar e =0$ be  a basis of $\Pi^{1,1}$.
Now assume that ${\cal M}=\mathbb T^r$ and 
 introduce  $r-$copies of  $\Pi^{1,1}$, and the corresponding basis $(e_i, \bar e_i), i=1,\cdots, r$.
 Therefore  we have (see \eqref{eq:root-s})
\beqa
\Lambda_R(\hat{ \mathfrak g}(\mathbb T^r)) \subset \Lambda_R(\mathfrak g) \oplus\underbrace{\Pi^{1,1} \oplus \cdots \oplus \Pi^{1,1}}_{r-\mathrm{times}} \ , \nn
\eeqa
{\it i.e.}, the root lattice of $\hat{ \mathfrak g}(\mathbb T^r)$ is a sublattice of
 $\Lambda_R(\mathfrak g) \oplus\Pi^{1,1} \oplus \cdots \oplus \Pi^{1,1}$.
 More precisely,  $\alpha \in \Lambda_R(\hat{ \mathfrak g}({\cal M}))$ if $\alpha\cdot e_i=0, i=1,\cdots,r$.
Thus the root system,  as well as all generators of $\hat{ \mathfrak g}(\mathbb T^r)$,  are known whereas, as we have  seen previously, for $r>1$ it is not possible to identify a system of simple roots. As a consequence, a Chevalley-Serre basis is not available.

In the same manner, some types of Lorentzian Kac-Moody algebras can be obtained from any semisimple Lie algebra $\mathfrak{g}$.
 For instance, the so-called `very extended Lie algebra' $\mathfrak{g}^{+++}$ is a rank $\ell+3$ Lorentzian Lie algebra
 associated to the semisimple Lie algebra $\mathfrak{g}$  (of rank $\ell$), where the simple roots
 are constructed from the simple roots of $\mathfrak{g}$ and two copies of $\Pi^{1,1}$ \cite{gow,west}. Therefore,
\beqa
\Lambda_R( {\mathfrak g}^{+++}) \subset \Lambda_R( \mathfrak g) \oplus \Pi^{1,1} \oplus \Pi^{1,1} \ . \nn
\eeqa
The simple roots of the very extended Lie algebra   ${\mathfrak g}^{+++}$ are  known. This means that one can introduce
a Chevalley-Serre basis for these algebras. However,  in this case explicit formulae for all generators of $\mathfrak{g}^{+++}$  
 are not available.

Thus even if the root lattices of  $\hat{ \mathfrak g}(\mathbb T^2)$ and ${\mathfrak g}^{+++}$ are both  sub-lattices of
$ \Lambda_R( \mathfrak g) \oplus \Pi^{1,1} \oplus \Pi^{1,1}$ those two Lie algebras have different properties.

\subsection{Representations of generalised Kac-Moody algebras}

 In this paragraph we outline some relevant points concerning the representations of generalised Kac-Moody algebras, 
with special emphasis on the existence of central charges in connection with the unitarity of representations. In the following, we   consider the quasi-simple Lie algebras as introduced in \cite{KT}, and assume that $\mathfrak g$ is a compact real  Lie algebra.
In the Cartan-Weyl basis, the algebra is generated by 
\beqa
\label{eq:QS}
\hat {\mathfrak g}(U(1)^r) = \big\{H^i_{\bf m}, E_{\alpha, \bf m}\ ,\;  \alpha \in \Sigma, \; {\bf m}\in \mathbb Z^n, \; d_i, k_i, i=1,\cdots, r \big\} \ ,\nn 
\eeqa
and the Lie brackets  are given by (${\bf m}=(m_1,\cdots,m_r)$)
\beqa
\big[H^i_{\bf m}, H^{i'}_{\bf m '}\big] &=&  \delta_{{\bf m} + {\bf m'}}  h^{ii'} \sum\limits_{i=1}^r  m_i k_i   \ , \nn\\
\big[H^i_{\bf m}, E_{\alpha \bf n} \big]&=&   \alpha^i \; E_{\alpha {\bf m} + {\bf n} }\  ,
\nn\\
\big[E_{\alpha {\bf m}}, E_{\beta  {\bf n}}\big] &=&
\left\{
\begin{array}{ll}
{\cal N}_{\alpha,\beta} \; \; E_{\alpha+\beta {\bf m} + {\bf n} }\ ,
& \alpha+\beta \in \Sigma ,\\[4pt]
  \alpha\cdot H_{ {\bf m} + {\bf n}  }
+ \delta_{{\bf m} + {\bf m'}} \sum\limits_{i=1}^r  m_i k_i ,   & \alpha + \beta=0,\\[4pt]
0,&\left\{\begin{array}{l}\alpha + \beta \ne 0\ ,\\ \alpha+\beta \not \in \Sigma , \end{array} \right.\ 
\end{array}
\right.\\
\big[d_i,E_{\alpha {\bf m}}\big]&=& m_i E_{\alpha \bf m}\ , \nn\\
\big[d_i ,H^j_{\bf m}]&=& m_i H^i_{\bf m}\nn \ .
\eeqa

 In the following, in order to increase the readability of some long formulae, we sometimes use the convention that $\delta_{a+ b}=\delta_{a+ b}^{0}=\delta_{a}^{-b}$.

As before,  we consider the set $\alpha_i, i=1,\cdots ,\ell$ of simple roots  and $\psi$ the highest root  of $\mathfrak g$.
Consider also the fundamental weights $\mu^i, i=1,\cdots, \ell$ of $\mathfrak g$ satisfying
\beqa
2 \mu^i \cdot \frac{\alpha_j}{\alpha_j\cdot \alpha_j} = \delta^i_j \ . \nn
\eeqa

We also suppose to be given a representation of $\hat{\mathfrak g}(U(1)^r)$
with highest weight
\beqa
\left| \hat \mu_0\right>=\left|  \mu_0, {\bf c}, {\bf m} \right>  \ , \ \ \text{where} \ \ \mu_0 = p_i \mu^i\ , \ \  (p_1,\cdots,p_r) \in \mathbb N ^r \nn
\eeqa
 such that the relations
\beqa
H^i \left| \hat \mu_0\right>&=&\mu_0^i \left| \hat \mu_0\right> \ , \nn\\
k_i \left| \hat \mu_0\right>&=&c_i \left| \hat \mu_0\right>  \ , \nn\\
d_i  \left| \hat \mu_0\right>&=&m_i \left| \hat \mu_0\right> ,  \nn
\eeqa
and
\beqa
E_{\alpha {\bf m}} \left| \hat \mu_0\right> &=& 0\ , \ \ (\alpha, {\bf 0}, {\bf m})>0 \ , \nn\\
H^i_{\bf m}  \left| \hat \mu_0\right> &=&0 \ , \ \ {\bf m} >0 \ . \nn
\eeqa
are satisfied. For any positive real root $\hat \alpha=(-\alpha, {\bf 0}, {\bf m})$, the generators
\beqa
X^\pm_{ \alpha,   {\bf m}} = \sqrt{\frac 2 {\alpha\cdot\alpha}} E_{\mp \alpha,\pm  {\bf m}}\ ,\ \ h_\alpha =\frac 2{\alpha\cdot \alpha}
\Big(- \alpha_i H^i_{\bf 0} + \sum\limits_{i=1}^r m_i k_i \Big) \ , 
\eeqa
 span an $\mathfrak{su}(2)-$subalgebra. The unitarity condition implies the constraints 
\beqa
\label{eq:uni}
&\frac 2{\alpha \cdot \alpha}\Big( -\alpha\cdot \mu_0 + \sum \limits _{i=1}^r  c_i m_i \Big)\in \mathbb Z\ , \nn\\
&\sum \limits _{i=1}^r  c_i m_i\ge \alpha \cdot \mu_0\ , \ 
\eeqa
 with the latter identity being a consequence of the relation
\beqa
||X^{-}_{\alpha,   {\bf m}}\, | \hat{\mu} \rangle||^2 =\langle \hat{\mu} \;|\; X^{+}_{ \alpha,   {\bf m}} X^{-}_{\alpha,   {\bf m}} |\hat \mu\rangle&=&\langle \hat{\mu} \;|\; \left[X^{+}_{ \alpha,   {\bf m}},X^{-}_{ \alpha,   {\bf m}}\right]\; |\hat \mu\rangle=\langle \hat\mu \;|\; h_{\alpha}\;  |\; \hat\mu\rangle \nn\\
&=& \frac 2{\alpha \cdot \alpha}\Big( -\alpha\cdot \mu_0 + \sum \limits _{i=1}^r  c_i m_i \Big)\ge 0 \ . \nn
\eeqa
If $\alpha>0$ then $\alpha \cdot \mu_0>0$, whereas for $\alpha<0$ we obtain $\alpha \cdot \mu_0<0$. We thus suppose that $\alpha>0$. In this case
the second relation in \eqref{eq:uni} is very strong. Indeed if $(-\alpha,{\bf 0},{\bf m})>0$, this means that
${\bf m} = (m_1,\cdots,m_{k-1},m_k,0,\cdots,0)$
with $m_k>0$ and $m_1,\cdots,m_{k-1} \in \mathbb Z$.
The second condition of \eqref{eq:uni},  which must be satisfied for any $k=1,\cdots,r$, is equivalent to impose that only one central charge is non-vanishing. Therefore, without loss of generality we can  suppose
$c_r =c \ne0 $ and $c_i =0, i=1,\cdots, r-1$. 

 Next, prior to analyse unitary representations, we observe that, since $c_1=\cdots=c_{r-1}=0$, the algebra  $\hat {\mathfrak g}(U(1)^r)$ does not admit highest weight unitary representations.
For this purpose, 
consider now $\hat\alpha=\left(-\alpha,{\bf 0},m_1,\dots ,m_{k},0,\dots 0\right)>0$ with $1\leq k\leq r$, $m_k>0$ and  $(m_1,\dots ,m_{k-1})\in \mathbb{Z}^{k-1}$. The  operators (with ${\bf m}=(m_1,\cdots,m_k,0,\cdots,0)$) 
\beqa
Y^\pm_{\hat \alpha} = \sqrt{\frac 2 {\alpha\cdot\alpha}} E_{\mp \alpha,\pm  {\bf m}}\ ,\ \ h_\alpha^{\prime} =\frac 2{\alpha\cdot \alpha}
\Big(- \alpha_i H^i_{\bf 0}  \Big) \ , 
\eeqa
also generate an $\mathfrak{su}(2)-$subalgebra. As before, the condition $||Y^{\pm}_{\hat\alpha}\, | \hat{\mu} \rangle||^2\geq 0$ holds, from which we deduce that 
\beqa 
\sum_{p=1}^{k} m_ic_i\geq \alpha\cdot\mu \ .
\eeqa
However, as $c_1=\dots =c_{r-1}=0$, this cannot be satisfied. 
This contradiction arises from the fact that the central charges vanish for $k_1,\dots ,k_{r-1}$, 
 and is in direct agreement with the 
 well known result of classical Kac-Moody algebras  that states that  the unique highest weight unitary representation for $k=0$ is the trivial one (the adjoint representation, for which the central charge vanishes, is not a highest weight representation). 

 With these preliminaries,  we will construct unitary representations  in two steps.
As there is only one non-vanishing central charge $k_r$, in the first step  we consider  the usual Kac-Moody algebra
\beqa
\label{eq:KM1}
\tilde{\mathfrak{g}}=\mathfrak{g}\left(U(1)\right)= \left\{ T_{a m}, k_r, d_r, \alpha\in\Sigma, m\in\mathbb{Z}\right\}. 
\eeqa
Representations of the latter are well known and correspond to the unitary representations of $\tilde{\mathfrak{g}}$ with highest weight $|\tilde\mu\rangle = |\mu, c_r,m_r\rangle$.
In this case, with $\psi$ being the highest root of $\mathfrak{g}$, we obtain:
\beqa
\label{eq:psi}
\frac{\psi}{\psi\cdot \psi}= \sum \limits_{i=1}^\ell  q^i \frac{\alpha_i}{\alpha_i \cdot \alpha_i} \ ,  \ \ q^i \in \mathbb N, 
\eeqa
and the second condition  of \eqref{eq:uni} translates to
\beqa
\label{eq:urep}
 x \ge  p_i q^i \ ,
\eeqa
where $x=2 \frac{c}{\psi\cdot \psi}$ is the level of the representation.
We now recall some known results on unitary representations of $\tilde{\mathfrak{g}}$  (see {\it e.g.} \cite{go}).

The simple  roots of $\tilde{\mathfrak g}$ are given by
\beqa
\hat \alpha_0=( -\psi,0,1)\ ,  \ \ \hat \alpha_i=(\alpha_i,0,0) \ , i=1,\cdots,\ell \nn
\eeqa
where the second entry corresponds to the eigenvalue of the  non-vanishing central charge and the third one to the eigenvalue of the corresponding  Hermitean operator, say $k_r$ and $d_r$ respectively.  Introduce  also
\beqa
\hat \Sigma =\Big\{\hat \alpha = (\alpha,0,n)\ , \ \  (0,0,n)\ , \ \   \alpha \in \Sigma, n \in \mathbb Z \Big\}\ . \nn
\eeqa
The  fundamental weights are defined by
\beqa
\hat \mu^0 = (0,\frac12 q^0 \psi\cdot \psi,0)  \ \ \text{with} \ \ q^0 =1 \ , \
\hat \mu^i = (\mu^i,\frac12 q^i \psi\cdot \psi,0) \ , \ i=1,\cdots, \ell  \ ,\nn
\eeqa
with the $q^i, i=1,\cdots,r$ defined by \eqref{eq:psi}.
We obviously have
\beqa
2 \hat \mu^i \cdot \frac{\hat \alpha_j}{\hat \alpha_j \cdot \hat \alpha_j} = \delta^i_j \ . \nn 
\eeqa
A highest weight is then specified by
\beqa
\hat \mu_ 0 = p_ i \hat \mu^i , \ \ \text{with} \  p_i \in \mathbb N \ , i=0,\cdots, \ell \ , \nn
\eeqa
and
the level of the representation is then given by
\beqa
x= \sum \limits _{i=0}^\ell p_i q^i\ge \sum \limits _{i=1}^\ell  p_i q^i \ . \nn
\eeqa
We denote the corresponding representation space as ${\cal D}_{\hat \mu_0}$.

 In a second step, let $\hat{\mathfrak{g}}=\tilde{\mathfrak{g}}\left(U(1)^{r-1}\right)$, which corresponds
to the set of smooth maps from $U(1)^{r-1}$ into $\widetilde{\mathfrak {g}}$. 
We deduce that (see \eqref{eq:KM1}) 
\beqa
\label{eq:tg}
 \hat {\mathfrak g}(U(1)^r)&=& \widetilde {\mathfrak g}(U(1)^{r-1})\\
 &=&\Big\{ T_{am_r} e^{ \ii \sum \limits_{k=1}^{n-1} m_k \varphi_k  } , \ \  {\bf m} \in \mathbb Z^{r-1}\ , \ d_r \ , k_r \ \ \text{and}  \ \  d_j=-\ii \partial_j\ ,\ \ j=1,\cdots,n-1\Big\} \ .\nn
\eeqa
As seen in \eqref{eq:uni}, the central charges associated to $d_j, j=1,\cdots, r-1$ vanish. Consider now

\beqa
\label{eq:R}
\mathcal{R}=\left\{ |{\bf m}\rangle, {\bf m}\in \mathbb{Z}^{r-1}\right\} \ ,
\eeqa
 the set of all unitary representations of $U(1)^{r-1}$.
We have
\beqa 
d_j |{\bf m}\rangle = m_j  |{\bf m}\rangle\nn \ . 
\eeqa
Using  the harmonic expansion on $U(1)^{r-1}$
\beqa 
\langle (\varphi_1,\cdots,\varphi_{r-1})|{\bf m}\rangle = {\bf e}^{\rm{i}\left(m_1\varphi_1+\dots m_{r-1}\varphi_{r-1}\right)} \ ,  \nn
\eeqa
 unitary representations of $\hat{\mathfrak{g}}$ are given by the tensor product
\beqa
\label{eq:uni}
\widehat{\mathcal{D}_{\tilde{\mu}}}=\mathcal{D}_{\tilde{\mu}}\otimes \mathcal{R}
\eeqa
and correspond to a harmonic expansion of the unitary representation $\mathcal{D}_{\tilde{\mu}}$ of $\tilde{\mathfrak{g}}$ on the manifold $U(1)^{r-1}$. 

A unitary representation of $\hat {\mathfrak g}(U(1)^r)$ follows directly from unitary representations of $\widetilde {\mathfrak g}$.
As the space \eqref{eq:R} is neither  bounded  from above nor bounded from below, unitary representations of  $\hat {\mathfrak g}(U(1)^r)$ 
are not highest weight representations.
As we have just seen, unitarity of representations implies only {\it one} non-vanishing central charge. This result 
seems to be contradictory at a first sight, in particular when considering 
the tensor product of two-representations for two different non-vanishing central charges. As an illustration, consider
for instance the generalised Kac-Moody algebra associated to the manifold  ${\cal M}=\mathbb T^2$.
Denote with ${\bf  k}=(k_1,k_2)$ the central charges ($k_1,k_2 \ne 0$) of $\mathfrak g_{12} =\hat {\mathfrak{g}}(U(1)^2)$, without any unitarity constraints.
 From the previous result, two cases can be considered if one wants unitary highest weight representations:
 (1) ${\bf k}=(k_1,0)$  or (2) ${\bf k}=(0,k_2)$. These two choices lead to two possible isomorphic ({\it but different}) algebras that we denote
 $\mathfrak g_1$ respectively  $\mathfrak g_2$ (see Eq.[\ref{eq:tg}]). We can now consider unitary representations of the first
 or the second algebra. Let ${\cal D}_1 \otimes {\cal R}_2$   (resp. ${\cal R}_ 1 \otimes {\cal D}_2$) be a unitary representation
 of $\mathfrak g_1$ (resp. $\mathfrak g_2$) with the notations of \eqref{eq:uni}.  While 
 $\Big({\cal D}_1 \otimes {\cal R}_2\Big) \otimes \Big({\cal R}_ 1 \otimes {\cal D}_2\Big)$ is certainly a representation of the algebra
 $\mathfrak g_1 \times \mathfrak g_2$,  {\it it is  not} a representation of the algebra $\mathfrak{g}_{12}$, which has {\it two}
 non-vanishing central charges. This shows that the contradiction is only apparent, with no conflict emerging from the construction.

Now we can extend part of the results to the general case, {\it i.e.}, when ${\cal M} = G_c$ or $G_c/H$.
Let $\alpha$ be a root of the compact Lie algebra $\mathfrak g$.   Next, introduce 
\beqa
X^+_{\alpha, {\bf m}} \in {\mathfrak g}_{(-\alpha, {\bf m})} \ , \ \
X^-_{\alpha, {\bf m}} \in {\mathfrak g}_{(\alpha, -{\bf m})}  \ , \ \
h_\alpha \in  {\mathfrak g}_{(0, {\bf 0})} \ , \nn
\eeqa
with ${\bf m} =(m_1,\cdots,m_r) >0$.
Note that in this case, with the notations of  equation (\ref{eq:CW}),  for $X^+_{\alpha, {\bf m}}
\in \mathfrak{g}_{(-\alpha, {\bf m})}$,
 $X^-_{\alpha, {\bf m}} \in \mathfrak{g}_{(\alpha, -{\bf m})}$  we have ${\bf m}(k)=m_k$ and $ -{\bf m}(k)=-m_k$ respectively. Furthermore,
 the coefficient $\eta_{{\bf m} {\bf n}}$ appearing in the bracket $\big[X^+_{\alpha,{\bf m}},X^-_{\alpha, {\bf m}}\big]$ simplifies in this case to $\eta_{\bf m}$,
which is a sign (see the examples given in the next section). Assume further that the operators
$X^\pm_{\alpha, {\bf m}},  h_\alpha - \frac 2{\alpha \cdot \alpha}\eta_{{\bf m}} \sum \limits_{i=1}^r m_i k_i$ are chosen
in such way  that they generate an $\mathfrak {su}(2)-$subalgebra:
\beqa
\label{eq:gene-su2}
\big[h_\alpha, X^\pm_{\alpha, {\bf m}}\big] = \pm  X^\pm_{\alpha, {\bf m}} \ , \ \
\big[X^+_{\alpha,{\bf m}},X^-_{\alpha, {\bf m}}\big]= h_\alpha - \frac 2{\alpha \cdot \alpha}\eta_{{\bf m}} \sum \limits_{i=1}^r m_i k_i  \ .
\eeqa
Then, in analogy to the previous discussion of the unitarity of representations, it follows that all
central charges except one must be equal to zero.  It is important to observe that, in absence of symmetries between the generators $D_i$, we can have $r$ different possibilities given by (eventually reordering 
the eigenvalues of the operators $D_i$ to define positive roots, see equation (\ref{eq:ord})) 
\beqa
{\bf c} = (0,\cdots,0, c_p,0,\cdots, 0) \ \ \text{or} \ \ {\bf k} = (0,\cdots,0, k_p,0,\cdots, 0) \ \ p\in\{1,\cdots,r\} \ . \nn
\eeqa
In this situation, it remains to
identify the precise form of the operators in \eqref{eq:gene-su2} and
to establish a condition analogous to \eqref{eq:urep} to characterise unitary
representations. Independently of these conditions,
unitarity leads to only one non-vanishing central charge. Consequently,
there are no obstructions to introduce a system of simple roots, as seen in Section \ref{sec:root}.
The resulting Dynkin diagram of $\hat{\mathfrak g}({\cal M})$  is analogous to the Dynkin diagram of
the corresponding usual Kac-Moody algebra  $\hat{\mathfrak g}(U(1))$, but dressed with the representation theory of $G_c$.

Let us emphasise again that, given the Lie algebra $\hat {\mathfrak g}({\cal M})$, all central charges except one must be equal to zero in order to  guarantee the unitarity of a representation. This result can be compared  to that of the previous section, where we proved that
only when the order of centrality, {\it i.e.}, the number of non-vanishing central charges is equal to one, the algebra admits a system of simple roots.


\section{Explicit construction in low rank}\label{sec:ap}

\noindent In  this section, we  illustrate the procedure previously developed  with some physically relevant  examples in which we shall identify the coefficients $c_{IJ}{}^K$ together with the Hermitian operators $D_i$ and their corresponding central extensions $k_i$. The brackets will  always be given by equation \eqref{eq:KM-gen}.
We begin naturally with the Lie group $SU(2)$, then we turn  to the coset spaces $SU(2)/U(1)$, $SO(4)/SO(3)$, $SU(3)/SU(2)$ and $G_2/SU(3)$. For elementary definitions see {\it e.g.} \cite{GT}.

\subsection{ Real Lie group $SU(2)$}\label{sec:su2}

The group $SU(2)$ is defined as the set of special unitary $2\times 2$ complex matrices:
\beqa
SU(2) = \Big\{U \in {\cal M}_2(\mathbb C) \ \ \text{such~that~}\ U^\dag U = 1, \ \det U=1 \Big\}\ , \nn
\eeqa
that can be written in the form
\beqa
U =\begin{pmatrix} \phantom{-}\alpha & \beta \\ -\overline \beta & \overline \alpha \end{pmatrix} \ , \ \ \alpha,\beta \in \mathbb C \ , \ \  |\alpha|^2+|\beta|^2 =1 \ , \nn 
\eeqa
showing that the isomorphism of manifolds $SU(2) \cong \mathbb S^3$ holds.   Indeed, setting  (see \cite{GT})
\beqa 
\left\{\begin{array}{lll}
\alpha &=& \cos \theta e^{i \varphi_1} \ , \\ 
\beta& =& \sin \theta  e^{i \varphi_2} \ ,
\end{array} \right.
\ \ 0\le \theta \le \frac{\pi}2  \ ,\ \  0 \le \varphi_1 <2 \pi\  ,\  \   0 \le \varphi_2 <2 \pi\label{eq_alphabeta}
\eeqa
we obtain a parameterisation of the sphere $\mathbb S^3$.
In the language of Appendix \ref{ap:ID}, this leads to  the parameterisation of ${\mathbb S}^3$ given  by
\beqa
0\le \varphi_1,\varphi_2 \le 2 \pi\ , 0\le u = \frac 12 \sin^2 \theta \le \frac 12 \ .
\eeqa
This parameterisation is a bijection on a dense subset of $SU(2)$, namely  when $\theta\ne \left\{0,\frac{\pi}2\right\}$. We observe that this parameterisation is not a  homeomorphism  from $[0,\frac{\pi}2]\times [0,2 \pi)\times [0,2 \pi)$  onto $\mathbb S^3$,  as the interval is non-compact. The   scalar product on $SU(2)$ is given by  (see \cite{GT})
\beqa
(f,g) = \frac 1{2 \pi^2}\int \limits_{0}^{\frac{\pi}2} \sin\theta\cos \theta \d \theta \int\limits_0^{2 \pi} \d \varphi_1  \int\limits_0^{2 \pi} \d \varphi_2 \;\overline{f(\theta,\varphi_1,\varphi_2)} \; {g(\theta,\varphi_1,\varphi_2)}\ .\nn 
\eeqa

It is easy to observe that the functions
\beqa
\psi_{a,b}(\theta,\varphi_1,\varphi_2) = \sqrt{\frac{(a+b+1)!}{a!\; b!} }\alpha^a (-\overline \beta)^b\ , a,b \in \mathbb N \ , \nn
\eeqa
satisfy the relations
\beqa
\label{eq:ab}
(\psi_{a,b},\psi_{a',b'})  = \delta^a_{a'}\delta_{b'}^{b}\ . 
\eeqa
Furthermore,  we can see  from equation \eqref{eq:ab} that the  functions $\alpha, \beta \in \mathbb S^3$  defined in equation \eqref{eq_alphabeta} enable us to obtain all the matrix elements  introduced in  Section  \ref{sec:PWtheo} in  a ready manner.
To this extent,  we introduce a differential realisation of the generators of the Lie algebra $\mathfrak{su}(2)$ in the above  parameterisation of $SU(2)$:   
\beqa
\label{eq:su2}
J_+ &=& J_1 + \ii J_2 = \frac12 e^{\ii(\varphi_1 -\varphi_2)}\Big(-\ii \tan \theta \frac{\partial}{\partial \varphi_1} -\ii \cot \theta  \frac{\partial}{\partial \varphi_2} + \frac{\partial}{\partial \theta} \Big) \ ,\nn\\
J_- &=& J_1 - \ii J_2 = \frac12 e^{-\ii(\varphi_1 -\varphi_2)}\Big(-\ii \tan \theta \frac{\partial}{\partial \varphi_1} -\ii \cot \theta  \frac{\partial}{\partial \varphi_2} - \frac{\partial}{\partial \theta} \Big)\ , \\
J_3&=& -\frac \ii 2 \big(\frac{\partial}{\partial \varphi_1}-\frac{\partial}{\partial \varphi_2}\big) \ , \nn
\eeqa
with Lie  brackets
\beqa
\big[J_3,J_\pm\big] = \pm J_{\pm} \ , \ \
\big[J_+,J_-\big] = 2 J_{3}  \ . \nn
\eeqa
This differential realisation acts on the  rows of matrices and thus corresponds to a right action. Similarly we can define a left action
  acting on the columns. The generators are the same as in \eqref{eq:su2}, except that we have to replace $\varphi_2$ by
  $-\varphi_2$. We do not give the form of the generators except for the last one
  \beqa
  \label{eq:JL}
J_3'= -\frac \ii 2\left(\frac{\partial}{\partial \varphi_1}+\frac{\partial}{\partial \varphi_2}\right) \ ,
\eeqa
but it can be explicitly checked that right and left actions commute.
   
The space of irreducible unitary representations is given by $\widehat {\cal R} = \Big\{{\cal R}_\ell,\ \ell \in \frac12 \mathbb N \ \Big\}$ with
the representation ${\cal R}_\ell$ of dimension $d_\ell =2 \ell +1$. Thus, for each $\ell$, we have to identify $2\ell+1$ equivalent representations associated to the right action.
The key  observation for this identification is given by the two complex-conjugate  two-dimensional spinor representations  defined as
\beqa
{\cal D}_{\frac12,\frac12}&=& \Big\{ \Phi_{\frac 12 ,\frac12,\frac12}= \sqrt{2} \alpha \ , \Phi_{\frac12,\frac12,-\frac12}= \sqrt{2} \beta \Big\} \ , \nn\\
{\cal D}_{-\frac12,\frac12}&=& \Big\{ \Phi_{-\frac12,\frac12,\frac12}= -\sqrt{2} \overline \beta \ , \Phi_{-\frac12,\frac12,-\frac12}= \sqrt{2} \overline \alpha \Big\}  \nn \ , 
\eeqa
 such that ${\cal D}_{-\frac12,\frac12} = \overline{{\cal D}}_{\frac12,\frac12}  \cong {\cal D}_{\frac12,\frac12}$, as expected.
In the notation above, the first index corresponds to the eigenvalue of the Cartan generator of the left action or $J_3'$ (see \eqref{eq:JL}), the last index
to the eigenvalue of the Cartan generator of the right action or $J_3$ (see \eqref{eq:su2}), whereas the second index corresponds to the eigenvalue of the Casimir operator
of the spinor representation. This identification is in accordance with Section \ref{sec:MLO} and the labelling  problem.

This can be extended easily to  an arbitrary representation. Indeed, for any $\ell \in \frac12 \mathbb N$, define the $2\ell +1$ equivalent representation spaces corresponding to the right action ${\cal D}_{m',\ell}$, $-\ell \le m'\le \ell$.
Each space  admits the highest weight vector 
\beqa
\label{eq:hw}
\Phi_{m',\ell,\ell}&=& \sqrt{\frac{(2\ell +1)!}{(\ell +   m')!(\ell-   m')!}} \alpha^{\ell +m'} (-\overline \beta)^{\ell -m'}\\
&=&(-1)^{\ell-m'} \sqrt{\frac{(2\ell +1)!}{(\ell +  m')!(\ell-  m')!}}
e^{\ii(\ell+ m') \varphi_1-\ii(\ell -m') \varphi_2} \cos^{\ell+m'}(\theta) \sin^{\ell -m'}(\theta) \ . \nn
\eeqa
 In order to obtain the remaining vectors of the representation space   ${\cal D}_{m',\ell}$ we use the relation 
\beqa
L_\pm^k\Big(e^{\ii m_1 \varphi_1 -\ii m_2 \varphi_2} F(\theta)\Big)= \frac{(\mp )^k}{2^k} e^{\ii(m_1 \pm k)\varphi_1 - \ii(m_2 \pm k)\varphi_2} \sin^{k \pm m_2}\theta \cos^{\pm m_1}\theta\nn\\
\hskip 2.truecm \frac{\d^k}{\d (\cos \theta)^k}\Big[ \sin^{\mp m_2}\theta \;\cos^{\mp m_1}\theta\; F(\theta)\Big]\ , \nn
\eeqa
which   can be proved by induction.
It follows that 
\beqa
\Phi_{m',\ell,m}&=& \sqrt{\frac{(\ell +m)!}{(2\ell)!(\ell -m)!}}\; \Big(J_-\Big)^{\ell-m} \Bigg[(-1)^{\ell-m'} \sqrt{\frac{(2\ell +1)!}{(\ell +  m')!(\ell-  m')!}}\nn\\
&&\hskip 3.truecm
\times  e^{\ii(\ell+m') \varphi_1-\ii (\ell -m') \varphi_2} \cos^{\ell+m} \theta \sin^{\ell-m'}\theta \Bigg] \nn\\
&=&(-1)^{\ell -m'}\frac 1 {2^{\ell -m}}\sqrt{\frac{2\ell+1}{(\ell +m')!(\ell-m')!}\frac{(\ell+m)!}{(\ell-m)!}}\; e^{\ii(m+m')\varphi_1-\ii(m-m')\varphi_2}\nn\\
&& \times \sin^{-m+m'}\theta \ \cos^{-\ell-m'} \theta \frac{\d}{\d (\cos\theta)^{\ell -m}}\Big[\sin^{2\ell -2m'} \theta \cos^{2\ell+2m'} \theta\Big] \ . \nn
\eeqa

When $\ell$ is an integer number and $m'\ge 0$,   the formula simplifies.
 If we define $u=\cos 2 \theta$ and the polynomials
\beqa
P_{m', \ell,m}(u)&=& (-1)^{\ell-m'} \frac1 {2^\ell} \sqrt{\frac {1}{(\ell+m')!(\ell -m')!}} \frac{\d^{\ell-m}}{\d u^{\ell -m}} \Big[(1-u^2)^{\ell-m'}(1+u)^{2m'}\Big] , \nn
\eeqa
we have
\beqa
\Phi_{m',l,m}&=& \sqrt{\frac{(\ell +m)!}{(\ell-m)!}} e^{\ii(m+m')\varphi_1-\ii(m-m')\varphi_2} \sin ^{-m+m'} 2\theta (1+\cos 2 \theta)^{-m'} P_{m',\ell,m}(u) \ . \nn
\eeqa
In particular
\beqa
\Phi_{m',\ell,0}&= e^{\ii m'(\varphi_1-\varphi_2)} \sin ^{m'} 2\theta (1+\cos 2 \theta)^{-m'} P_{m',\ell,0}(u) \ . \nn
\eeqa
 There are analogous formul\ae \ for $m'\le 0$. 

 We conclude that the $\Phi$-functions are orthonormal
\beqa
(\Phi_{m_1',\ell_1,m_1}, \Phi_{m_2',\ell_2,m_2})=\delta^{m_1'}_{m_2'} \delta^{\ell_1}_{ \ell_2} \delta^{m_1}_{m_2} \ , \nn
\eeqa
and constitute an  orthonormal Hilbert  basis   which is well adapted to the Peter-Weyl theorem  applied to  $SU(2)$:
\beqa
\label{eq:H-S3}
{\cal B} = \Big\{\Phi_{m',\ell,m}, \  \ell \in \frac12 \mathbb N, -\ell \le m,m' \le \ell  \Big\} \ . 
\eeqa
From the highest weight \eqref{eq:hw} we have the conjugacy property
\beqa
\label{eq:conjsu2}
\bar \Phi^{m',\ell,m}(\theta,\varphi_1,\varphi_2)= (-1)^{m-m'}  \Phi_{-m',\ell,-m}(\theta,\varphi_1,\varphi_2) \ . 
\eeqa

 In order to define the Lie algebra $\hat{ \mathfrak g}(SU(2))$,  we  introduce $T_{a, m',\ell,m} = T_a  \Phi_{m',\ell,m}(\theta,\varphi_1,\varphi_2)$, the
two Hermitian operators $J_3', J_3$ and their associated central charges $k,k'$ (see \eqref{eq:form} and \eqref{eq:cc}) .
The Lie brackets take then the form (see \eqref{eq:KM-gen})
\beqa
\label{eq:KM-S3}
\big[T_{a_1,m_1',\ell_1,m_1},T_{a_2,m_2',\ell_2,m_2}\big]&=& \ii f_{a_1 a_2}{}^{a_ 3} c_{m_1',\ell_1,m_1,m_2',\ell_2,m_2}{}^{m_3',\ell_3,m_3}
T_{a_3,m_3',\ell_3,m_3} \nn\\
&&+ (-1)^{m_1-m_1'} g_{a_1 a_2} \delta_{\ell_1,\ell_2}  \delta_{m_1+ m_2}\delta_{m'_1 + m'_2,}(k m_2+k' m_2') \ , \\
\big[J_3',T_{a,m',\ell_1,m}\big]&=& m'T_{a,m',\ell_1,m} \ , \nn\\
\big[J_3,T_{a,m',\ell_1,m}\big]&=& mT_{a,m',\ell_1,m} \ . \nn
\eeqa

We now proceed with the evaluation of the $c_{IJ}{}^K$ coefficients.
 By using well-known results from the coupling theory of angular momenta (see {\it e.g.} \cite{Edm,Van}) we obtain
\beqa
\label{eq:PhiPhi}
&\Phi_{m'_1,\ell_1,m_1}(\theta,\varphi_1,\varphi_2) \Phi_{m'_2,\ell_2,m_2}(\theta,\varphi_1,\varphi_2)\nn\\
&=
 \sum \limits _{\ell=|\ell_1 -\ell_2|}^{\ell_1 +\ell_2} \lambda(m'_1,m'_2,\ell,\ell_1,\ell_2)  \Big({}^{\ell_1}_{m_1} \ \ {}^{\ell_2}_{m_2} \Big| {}^{\hskip .5truecm \ell}_{m_1+m_2}  \Big)
 \Phi_{m'_1+m'_2,\ell,m_1+m_2}(\theta,\varphi_1,\varphi_2),
\eeqa
where $\Big({}^{\ell_1}_{m_1} \ \ {}^{\ell_2}_{m_2} \Big| {}^{\hskip .5truecm \ell}_{m_1+m_2}  \Big)$  are the Clebsch-Gordan  coefficients associated to the right action.
In this expansion,  we consider only the allowed values of $m'_1 + m'_2$,  such that $ - \ell\le m'_1+m'_2\le \ell$.
There is no generic closed expression  to compute the coefficients $\lambda(m_1',m_2',\ell,\ell_1,\ell_2)$, although they can be computed recursively. For instance, for the highest value of $m'_1, m'_2,\ell,\ell_1,\ell_2$ we obtain : 
\beqa
\lambda(m'_1,m'_2,\ell_1+\ell_2,\ell_1,\ell_2) = \sqrt{\frac{(2\ell_1+1)!(2\ell_2+1)!}{(2(\ell_1+\ell_2)+1)!}\,
\frac{(\ell_1+\ell_2+m'_1+m'_2)!(\ell_1+\ell_2-(m'_1+m'_2))!}
{(\ell_1+m'_1)!(\ell_1-m'_1)!(\ell_2+m'_2)!(\ell_2- m'_2)!}} \ . \nn
\eeqa

\subsection{ Coset space $SU(2)/U(1)$}\label{sec:su2-u1}

Assume that  $Q \in U(1) \subset SU(2)$  is given by
\beqa
Q=  e^{2 \ii \;\theta J_3'}= \begin{pmatrix} e^{ \ii\theta}&0\\ 0&e^{-\ii\theta} \end{pmatrix} \ . \nn
\eeqa
This means that $\alpha, \beta$ have a $U(1)-$charge equal to 1 and   $\overline \alpha, \overline \beta$ have a $U(1)-$charge  $-1$. More generally, the functions
$\Phi_{m',\ell,m}(\theta,\varphi_1,\varphi_2)$ have a charge $2m'$.
From Section \ref{sec:coset}, we  just need to consider functions that are neutral, {\it i.e}, $\Phi_{0,\ell,m}$.
This is possible if $\ell$ is an integer number. For such functions, we have
\beqa
\Phi_{0,\ell,m}&=&\frac{(-1)^{\ell}}{ 2^\ell\ell !}\sqrt{(2\ell+1)\frac{(\ell+m)!}{(\ell-m)!}}\; e^{\ii m(\varphi_1-\varphi_2)}
\sin^{-m}2\theta \frac{\d^{\ell -m}}{\d (\cos 2 \theta)^{\ell-m} }\Big((1-\cos^2 2 \theta)^\ell\Big) \  \nn\\
&=&Y_{\ell m}(2\theta, \varphi_1 -\varphi_2) \ , \nn
\eeqa
where $Y_{\ell  m}$ are the usual spherical  harmonics on the sphere ${\mathbb S}^2$. (Note, however, the unconventional normalisation factor for the $Y$-functions.)
 If we perform the change of  coordinates 
 \beqa
0 \le \psi=2 \theta \le \pi\ , \ \
0 \le \varphi= \varphi_1-\varphi_2< 2\pi \ , \ \
0 \le \widetilde{\varphi}= \varphi_1+\varphi_2< 4\pi  \ , \nn
\eeqa
 then the points $m=(\psi,\varphi,  \widetilde{\varphi} = \text{cons.})$ parameterise points on the manifold $\mathbb  S^2 \cong SU(2)/U(1) \subset \mathbb  S^3 \cong  SU(2).$
 With this parameterisation, for the level surfaces $\varphi=$ \text{cons.} we have,  on the one hand
\beqa
L_\pm&=& e^{\pm \ii\varphi} \Big(\ii\cot \psi \frac{\partial}{\partial \varphi} \pm \frac{\partial}{\partial \psi} - \ii\frac1 { \sin  \psi }  \frac{\partial}{\partial \widetilde \varphi}\Big)\nn\\
&=&\Big|_{\widetilde \varphi = \text{cons.}} \; e^{\pm \ii\varphi} \Big(\ii\cot \psi \frac{\partial}{\partial \varphi} \pm  \frac{\partial}{\partial \psi} \Big)\nn \ ,\\
L_3&=& - \ii \frac{\partial}{\partial \varphi} \ , \nn
\eeqa
as well as the relation 
\beqa
(f,g) &=& \frac 1{2 \pi^2}\int \limits_{0}^{\frac{\pi}2} \sin\theta\cos \theta\; \d \theta \int\limits_0^{2 \pi} \d \varphi_1  \int\limits_0^{2 \pi} \d \varphi_2 \; \overline{f(\theta,\varphi_1,\varphi_2)} \; {g(\theta,\varphi_1,\varphi_2)}\nn\\
&=&\Big|_{\widetilde \varphi = \text{cons.}}\; \frac1{4\pi} \int \limits_{0}^{\pi} \sin \psi \;\d \psi \int\limits_0^{2 \pi} \d \varphi
\overline{f(\psi,\varphi,\text{cte})} \; {g(\psi,\varphi,\text{cons.})} \ , \nn
\eeqa
 thus reducing to the generators of $\mathfrak{so}(3)$ (resp. to the Hilbert  scalar product on $\mathbb S^2$). In the language of Appendix \ref{ap:ID}, ${\mathbb S}^2$ is parameterised by
 \beqa
0\le \varphi \le 2 \pi\ , \ \ -1\le u = \cos \psi \le 1 \ . \nn
\eeqa

It follows that the adapted Hilbert basis for $SU(2)/U(1)$ is given by the  usual spherical harmonics,
\beqa
{\cal B} =\Big\{ Y_{\ell m}, \ell \in \mathbb N, \ -\ell \le m \le \ell \Big\} \ , \nn
\eeqa
with the conjugacy relation
\beqa
\bar Y^{\ell m}(\psi,\varphi)=(-1)^m Y_{\ell,-m}(\psi,\varphi) \ . \nn
\eeqa
 In order to define the Lie algebra   $\hat{ \mathfrak g}\big(SU(2)/U(1)\big)$ we introduce $T_{a,\ell, m}= T_a Y_{\ell  m}(\psi,\varphi)$, the Hermitian  operator
$L_3$ and its associated central charge $k$ (see \eqref{eq:form} and \eqref{eq:cc}). The Lie brackets take the form  (see \eqref{eq:KM-gen})
\beqa
\label{eq:KM-S2}
\big[T_{a_1,\ell_1,m_1},T_{a_2,\ell_2,m_2}\big]&=& \ii f_{a_1 a_2}{}^{a_ 3} c_{\ell_1,m_1,\ell_2,m_2}{}^{\ell_3,m_3}
T_{a_3,\ell_3,m_3} \nn\\
&&+ (-1)^{m_1} k m_2\; g_{a_1 a_2} \delta_{\ell_1,\ell_2}\delta_{m_1+ m_2}  \ , \\
\big[J_3,T_{a,\ell,m}\big]&=& mT_{a,\ell,m} \ . \nn
\eeqa
For the spherical harmonics, it is  well known that
\beqa
\label{eq:YY}
Y_{\ell_1 m_1} Y_{\ell_2 m_2} = \sum \limits _{\ell=|\ell_1 -\ell_2|}^{\ell_1 +\ell_2}
\sqrt{\frac  {(2\ell_1 +1)(2\ell_2 +1)} {2\ell +1}}  \Big({}^{\ell_1}_{0} \ \ {}^{\ell_2}_{0} \Big| {}^{\ell}_{0}  \Big)
\Big({}^{\ell_1}_{m_1} \ \ {}^{\ell_2}_{m_2} \Big| {}^{\hskip .5truecm \ell}_{m_1+m_2}  \Big)  Y_{\ell m_1+m_2} \ ,
\eeqa
 which leads to the $c_{IJ}{}^K$ coefficients in \eqref{eq:KM-S2}.
 
Potential applications can {\it e.g.} be conceived in Supergravity \cite{dewit1, dewit2}, using the space $SL(2,\mathbb{R})/U(1)$ related to the non-compact group $SL(2,\mathbb{R})$, the unitary representations of which are known \cite{barg}, and correspond to the discrete series (either lower or upper bounded) and the continuous series (principal and supplementary). Discrete (respectively continuous) series are characterised by a discrete (respectively continuous)  spectrum of the Casimir operator of $\mathfrak{sl}(2,\mathbb R)$, from which we conclude that the discrete series are
 normalisable, whereas the continuous series are not.
They can be found, for instance, in Section 2 of Ref. \cite{Schmid}, and can be related to  Eqs. (5.26)  and  (5.27) of Ref. \cite{GT};  (the former corresponding to the discrete series, while the latter, within the continuous  series, requires to distinguish between ``bosons'' ($n$ even in \cite{Schmid}, p. 197) and ``fermions'' ($n$ odd in \cite{Schmid}), with the correspondence: $2\mu=E_0-\frac12+s$ and $2\lambda=-E_0-\frac12+s$ ($\mu$, $\lambda$ from Ref. \cite{GT}) with $E_0=0$ for bosons and $E_0=\frac12$ for fermions and with $s \in\ii \mathbb R$
(resp.  $s \in\mathbb R$) for the principal (respectively complementary) series. In the context of harmonic functions, the discussion of non-compact groups rapidly shows to be considerably intrincate 
\cite{Campoamor-Stursberg:2014ffa}.  Indeed, as the number  $s$ for the continuous series  (see \cite{Campoamor-Stursberg:2014ffa}) is either real or purely
imaginary, the expressions in Section 4.3 of this reference \cite{Campoamor-Stursberg:2014ffa} are meaningless for the continuous series, whereas for discrete series, convergence occurs for spin greater than 1/2, whereas here, $s=0$, $1/2$.
Harmonic analysis of homogeneous spaces $G/H$, with $G$ a non-compact Lie group and $H$ a compact closed subgroup, of the same type as
$SL(2,\mathbb R)/U(1)$, has been considered to some extent in \cite{barut}, Chapter 15.


\subsection{Coset space $SO(4)/SO(3)$}\label{sec:so4-so3}

The manifold $SO(4)/SO(3)$ is well-known to be isomorphic to the three-sphere $ \mathbb S^3$.
 Its interest in our context is that it gives rise
 to an equivalent realisation of the algebra \eqref{eq:KM-S3},  but with a different Hilbert basis.
 This construction can furthermore be extended to the coset spaces $SO(n)/SO(n-1) \cong \mathbb S^{n-1}$ for values $n>4$.

The sphere $\mathbb S^3$ can be parameterised by
\beqa
\label{eq:S3}
\left.
\begin{array}{lll}
x_1&=&\sin \psi \sin \theta \cos\varphi \ , \\
x_2&=&\sin \psi\sin \theta \sin \varphi  \ ,\\
x_3&=&\sin \psi\cos \theta \ , \\
x_4&=&\cos \psi \ ,
\end{array}
\right\}
\begin{array}{l}
0\le \psi\le \pi\ ,\\
0\le \theta\le \pi\ ,\\
0\le \varphi <  2 \pi\ ,
\end{array} \
\eeqa
and endowed with a scalar product  defined by
\beqa
(f,g)= \frac 1 {2\pi^2} \int \limits_0^\pi \sin^2 \psi \d \psi \int \limits _0^\pi \sin \theta \d \theta \int \limits_0^{2\pi} \d \varphi
\overline{f(\theta,\varphi,\psi)}\; {g(\theta,\varphi,\psi)} \ .
\nn\eeqa
Again, using the terminology of Appendix \ref{ap:ID}, the sphere ${\mathbb S}^3$ is parameterised by
 \beqa
0\le \varphi \le 2 \pi\ , \ \ -1\le u_1 = \cos \theta \le 1 \ , 0\le u_2 =\frac12\psi -\frac12 \cos \psi \sin \psi \le \frac {\pi} 2 \ . \nn
\eeqa

Representations of $SO(4)$ are characterised by their Dynkin labels or by a Young tableau associated to a tensor with a certain symmetry.
Among  tensors, only  traceless $n^{\text{th}}$-order symmetric  tensors admit a scalar representation with respect to the  embedding $SO(3) \subset SO(4)$.
  Let us denote by ${\cal D}_{\frac n2, \frac n2}, n \in \mathbb N$ the representation corresponding to the set of traceless  $n^{\text{th}}$-order tensors, which is of dimension $(n+1)^2$.  If $D_{(n)}^i{}_j$ are the corresponding matrix
elements, from  the result of Section \ref{sec:coset}   and because ${\cal D}_{\frac n2,\frac n2}$  contains
the scalar representation in the decomposition through the embedding $SO(3)\subset SO(4)$,    for any Fourier  expansion  on  $\mathbb S^3$ we have to consider the indices $i=i_0$, where $D_{(n)}^{i_0}{}_j$  
are in the scalar representation of $SO(3)$. 

These matrix elements can be  easily obtained.  To that purpose, we introduce the generators of the Lie algebra
$\mathfrak{so}(4)$ in the usual  \{$N_0,N_\pm,N'_0,N'_\pm$\}  basis:
\beqa
\begin{array}{lll}
\big[N_0,N_\pm\big]&=&\pm N_\pm,\\
\big[N_+,N_-\big]&=&2 N_0,
\end{array} \ \ \ 
\begin{array}{lll}
\big[N'_0,N'_\pm\big]&=&\pm N'_\pm,\\
\big[N'_+,N'_-\big]&=&2 N'_0,
\end{array} \ \  \
\begin{array}{l}
\big[N_a,N'_b\big]=0
\end{array} \ . \nn
\eeqa

From the expression of the generators of the Lie algebra defined on $\mathbb S^3$ and the expansion on the sphere, we obtain:
\beqa
N_0&=&-\frac \ii 2 \Bigg[\frac{\partial}{\partial \varphi} + \cot \psi \sin \theta \frac{\partial}{\partial \theta} -\cos \theta \frac{\partial}{\partial \psi}\Bigg] \ , \nn\\
N_+&=&\frac12 e^{\ii \varphi} \Bigg[\Big(\ii\cot \theta -\frac{\cot \psi}{\sin \theta}\Big)\frac{\partial}{\partial \varphi} + \Big(1+\ii \cot \psi\cos \theta\Big)\frac{\partial}{\partial \theta}
+\ii \sin \theta \frac{\partial}{\partial \psi}\Bigg]\ , \nn\\
N_-&=&\frac12 e^{-\ii\varphi} \Bigg[\Big(\ii\cot \theta +\frac{\cot \psi}{\sin \theta}\Big)\frac{\partial}{\partial \varphi} + \Big(-1+\ii \cot \psi\cos \theta\Big)\frac{\partial}{\partial \theta}
+\ii \sin \theta \frac{\partial}{\partial \psi}\Bigg]\ , \nn
\eeqa
\beqa
N'_0&=&-\frac \ii2 \Bigg[\frac{\partial}{\partial \varphi} - \cot \psi \sin \theta \frac{\partial}{\partial \theta} +\cos \theta \frac{\partial}{\partial \psi}\Bigg] \ , \nn\\
N'_+&=&\frac12 e^{\ii\varphi} \Bigg[\Big(\ii\cot \theta +\frac{\cot \psi}{\sin \theta}\Big)\frac{\partial}{\partial \varphi} + \Big(1-\ii\cot \psi\cos \theta\Big)\frac{\partial}{\partial \theta}
-\ii \sin \theta \frac{\partial}{\partial \psi}\Bigg]\ , \nn\\
N'_-&=&\frac12 e^{\ii\varphi} \Bigg[\Big(i\cot \theta -\frac{\cot \psi}{\sin \theta}\Big)\frac{\partial}{\partial \varphi} - \Big(1+\ii \cot \psi\cos \theta\Big)\frac{\partial}{\partial \theta}
-\ii \sin \theta \frac{\partial}{\partial \psi}\Bigg]\ . \nn 
\eeqa
We can construct spherical harmonics, since we have
\beqa
\Phi_{\frac12\frac12;\frac12,\frac12}(\theta,\varphi,\psi) &=& \sqrt{2} e^{\ii\varphi} \sin \theta \sin \psi= \sqrt 2 \big(x_1 + \ii x_2)\nn\ , \\
\Phi_{\frac12,-\frac12;\frac12,\frac12}(\theta,\varphi,\psi) &=& \sqrt{2} (-\cos \theta\sin \psi  + \ii \cos \psi )=\sqrt 2 (-x_3+\ii x_4)\nn\ , \\
\Phi_{\frac12,\frac12;\frac12,-\frac12}(\theta,\varphi,\psi) &=& -\sqrt{2} (\cos \theta\sin \psi  + \ii \cos \psi )=-\sqrt 2 (x_3+\ii x_4)\nn\ , \\
\Phi_{\frac12,-\frac12;\frac12,-\frac12}(\theta,\varphi,\psi) &=& -\sqrt{2} e^{-\ii\varphi} \sin \theta \sin \psi= -\sqrt 2 \big(x_1 -\ii x_2)\nn \ ,
\eeqa
  denoting this representation by ${\cal D}_{\frac 12,\frac 12}$, and whose highest weight is given by $\Phi_{\frac12\frac12;\frac12,\frac12}$.
 According to this prescription, the highest weight of the representation ${\cal D}_{\frac n 2,\frac n2}$ is given by
\beqa
\Phi_{\frac n2,\frac n2;\frac n2,\frac n2}(\theta,\varphi,\psi) =\sqrt{n+1} e^{\ii n \varphi} \sin^n \theta \sin^n \psi \ .\nn
\eeqa
The remaining vectors
$\Phi_{\frac n 2, m_1;\frac n 2,  m_2}(\theta,\varphi,\psi)$ are explicitly obtained  by the action of the operators $N_-, N'_-$, where $m_1$, $m_2$ indicate the eigenvalues of $N_0, N_0'$.
Moreover, we have the conjugacy properties
\beqa
\overline \Phi^{\frac n 2, m_1;\frac n 2,  m_2}(\theta,\varphi,\psi)=(-1)^{m_1+m_2} \Phi_{\frac n 2, -m_1;\frac n 2,  -m_2}(\theta,\varphi,\psi) \ . \nn
\eeqa
Since the first and third indices are redundant, we set
\beqa
\Phi_{\frac n 2, m_1;\frac n 2,  m_2} \to \Phi_{n , m_1, m_2} \ .\nn
\eeqa

The representation space ${\cal D}_{\frac n2,\frac n2}$ can also be obtained in another way. Introduce $y_i = r x_i$ (see equation
\eqref{eq:S3}) in spherical coordinates, as well as the Laplacian $\nabla^2$ in this system of coordinates (to be defined below, in equation \eqref{eq:lap}) and   the space of polynomials of degree $n$, $\mathbb R_n[y_1,y_2,y_3,y_4]$, so that we have
\beqa
\label{eq:traceless}
{\cal D}_{\frac n 2 ,\frac n2}= \Big\{P(y_1,y_2,y_3,y_4)|_{r=1} \ \ \text{where} \ \ P \in \mathbb R_n[y_1,y_2,y_3,y_4] \ \ \text{s.t.} \ \ \nabla^2 P(y_1,y_2,y_3,y_4) = 0 \Big\} \ .
\eeqa
In  equation \eqref{eq:traceless}, the Laplacian constraint simply projects on traceless polynomials.
Then the Hilbert basis on $\mathbb S^3$ relevant
for our purpose is defined by
\beqa
\label{eq:H-S3p}
{\cal B} = \Big\{ \Phi_{ n ;m_1, m_2}, \ \ n \in \mathbb N, - \frac n 2 \le m_1,m_2 \le \frac n 2 \Big\}\  .
\eeqa

We further have the orthonormality relations
\beqa
\Big(\Phi_{ n , m_1, m_2},\Phi_{ {n'} , m'_1,  m'_2}\Big) = \delta^n_{n'} \delta^{m_1}_{ m_1'}\delta^{m_2}_{m_2'} \ . \nn
\eeqa
In the notations above for $\Phi_{ n ,m_1, m_2}$, we have taken the following conventions: $n$ labels the representation space,  here ${\cal D}_{\frac n2, \frac n2}$, and $m_1$ (resp. $m_2$) labels the
eigenvalue of $N_0$ (resp. $N'_0$).
From the results of Section \ref{sec:MLO}, we would expect to need four labels to classify all representations of $\mathfrak{so}(4)$.
However,  as only certain types of representations appear in the Fourier expansion,
it will turn out that three labels are sufficient.
In order to define the Lie algebra $\hat{ \mathfrak g}\big(SO(4)/SO(3)\big)$, we introduce
$T_{a,n,m_1,m_2}= T_a \Phi_{ n , m_1, m_2}(\theta,\varphi,\psi)$
 the Hermitian  operator $N_0, N_0'$. The  associated central charges $k_0$ and $k'_0$ are more involved and are obtained using the
two-forms dual to $N_0,N_0'$ given by
\beqa
\gamma_0&=&-\frac \ii 2 k_0 \Bigg[\d \theta \wedge \d \psi  -\cot \psi \sin \theta \d \varphi \wedge\d \psi
-\cos \theta \d \varphi \wedge \d \theta\Bigg] \sin^2 \psi \sin \theta \ , \nn\\
\gamma'_0&=&-\frac \ii 2 k_0' \Bigg[\d \theta \wedge \d \psi  +\cot \psi \sin \theta \d \varphi \wedge\d \psi
+\cos \theta \d \varphi \wedge \d \theta\Bigg] \sin^2 \psi \sin \theta \ . \nn
\eeqa
The Lie brackets take the form (see \eqref{eq:KM-gen})
\beqa
\label{eq:KM-S3p}
\big[T_{a,n,m_1,m_2},T_{a',n',m'_1,m'_2}\big]&=& \ii f_{a a'}{}^{a'' } c_{n,m_1,m_2,n',m'_1,m'_2}{}^{n'',m''_1,m''_2}
T_{a'',n'',m''_1,m''_2,} \nn\\
&& + (-1)^{m_1+m_2} g_{a'a'}  \delta_{n n'} \delta_{m_1+ m_1'}\delta_{m_1 + m'_2}(k_0 m'_1+k'_0 m_2') \ , \\
\big[N_0,T_{a,n,m_1,m_2}\big]&=& m_1 T_{a,n,m_1,m_2} \ , \nn\\
\big[N_0',T_{a,n,m_1,m_2}\big]&=& m_2 T_{a,n,m_1,m_2} \ . \nn
\eeqa
 
 \noindent  Since the two Hilbert bases \eqref{eq:H-S3} and \eqref{eq:H-S3p}
are admissible  bases for the space $L^2(\mathbb S^3)$, the algebra \eqref{eq:KM-S3p} is isomorphic to the algebra \eqref{eq:KM-S3}.
The former is presented  using the representation theory of $\mathbb S^3 \cong SU(2)$, whilst the latter is described in terms of the representation
theory of $\mathbb S^3 = SO(4)/SU(3)$.\\

The coefficients $c_{IJ}{}^K$ in \eqref{eq:KM-S3p} can be obtained from the relation
\beqa
\Phi_{n,m_1,m_2}(\theta,\varphi,\psi) \Phi_{n',m'_1,m'} (\theta,\varphi,\psi)&=& \sum \limits_{N=|n-n'|}^{n+n'}
\lambda(N,n,n') \Big({}^{\hskip .35truecm n}_{m_1,m_2} \ \ {}^{\hskip .4truecm n'}_{m'_1,m'_2} \Big|{}^{\hskip .9truecm N}_{m_1+m_1',m_2+m'_2}\Big)\nn\\
&&
\hskip 2.truecm \times \Phi_{N,m_1+m'_1,m_2+m'_2}(\theta,\varphi,\psi),
\eeqa
where $\Big({}^{\hskip .35truecm n}_{m_1,m_2} \ \ {}^{\hskip .4truecm n'}_{m'_1,m'_2} \Big|{}^{\hskip .9truecm N}_{m_1+m_1',m_2+m'_2}\Big)$ are the Clebsch-Gordan coefficients of the decomposition  
\beqa
{\cal D}_{\frac n 2,\frac n 2 }\otimes {\cal D}_{\frac {n'} 2,\frac {n'} 2} = \bigoplus \limits_{N=|n-n'|}^{n+n'} {\cal D}_{\frac N2, \frac N 2} \ , \nn
\eeqa
and $\lambda(N,n,n')$ are  coefficients which can be computed recursively.

There is a third presentation of the algebra  given in  \eqref{eq:KM-S3p}. The two Casimir operators of the $\mathfrak{so}(4)$ algebra are given by
\beqa
Q=N_0^2 + \frac12 (N_+ N_- + N_-N_+) \ , \ \
Q' = N'_0{}^2 + \frac12 ( N'_+  N'_- +  N'_- N'_+) \ , \nn
\eeqa
or by
\beqa
C = 2 ( Q  + Q ') \ , \ \
C' = 2(  Q -  Q ')\ . \nn
\eeqa
For the representation ${\cal D}_{\frac n2,\frac n2}$, we have
\beqa
\label{eq:cas}
C=n(n+2) \ , \ \ C'=0 \ . 
\eeqa
However, the Laplacian in spherical coordinates $(r,\phi,\theta,\psi)$  takes the form
\beqa
\label{eq:lap}
\nabla ^2 = \frac 1{\sqrt g} \partial_i(\sqrt{g} g^{ij} \partial_j)=\frac {\partial^2}{\partial r^2} + \frac 3 r \frac{\partial}{\partial r} - \frac C {r^2} \ , 
\eeqa
with the metric given by
\beqa
\d s^2 = \d r^2 + r^2\big(\d \psi^2 +\sin^2  \psi\; \d \theta^2 + \sin^2 \psi \sin^2 \theta \;\d \phi^2\big) \ . 
\eeqa
Hence, the condition \eqref{eq:cas} for functions of the sphere ${\mathbb S}^3$ in the representation ${\cal D}_{\frac n2,\frac n2}$ is equivalent to impose the constraint  
\beqa
\label{eq:lap3}
\nabla^2 Y(\theta,\phi,\psi)=-\frac{n(n+2)}{r^2}  Y(\theta,\phi,\psi) \ . 
\eeqa
This definition is obviously equivalent to equation \eqref{eq:traceless},  as 
\beqa
\nabla^2 \big(r^n Y(\theta,\phi,\psi)\big)=0 \ . \nn
\eeqa
If we set
\beqa
Y(\theta,\phi,\psi)\equiv Y_{n\ell m}(\theta,\phi,\psi) = H_{n \ell}(\psi) Y_{\ell m}(\theta,\phi) \ , \ \ n \in \mathbb N\  ,\ \  0\le \ell \le n\ , \ \ -\ell \le m \le \ell \ ,\nn
\eeqa
then the condition \eqref{eq:lap3} is equivalent to  the second-order linear homogeneous ordinary differential equation
\beqa
\frac{\d^2 H_{n\ell}(\psi)}{\d \psi^2} + 2 \cot \psi \frac{\d H_{n\ell}(\psi)}{\d \psi} + \Big(n(n+2) -\frac{\ell(\ell+1)}{\sin^2 \psi}\Big) H_{n\ell}(\psi)= 0 \ .
\eeqa
Indeed, we can express the functions $H_{n \ell}(\psi)$ as follows
\beqa
H_{n\ell}(\psi) = N_{n\ell} \sin^\ell \psi\; C_{n-\ell}^{1+\ell}(\cos \psi) \ ,\nn
\eeqa
where $C_{n-\ell}^{1+\ell}$ are the  Gegenbauer polynomials defined by (see {\it e.g.} \cite{Kam, Ge})
\beqa
\frac1 {(1 + x^2 - 2 x \cos \psi)^\alpha} = \sum \limits_{n=0}^{\infty} C_n^\alpha(\cos \psi) x^n \ , \ \ \text{for}\ \ |x|<1 \ , \nn
\eeqa
and the normalisation coefficients are given by
\beqa
N_{n\ell}=(-1)^\ell (2 \ell)! \sqrt{\frac{(n+1)(n-\ell)!}{(n+\ell+1)!}} \ . \nn
\eeqa
The harmonic functions $Y_{n\ell m}$ are characterised by the eigenvalues of the Casimir operators in the embedding chain  $SO(4) \supset SO(3) \supset SO(2)$, with $n$ being associated to $SO(4)$, $\ell$  to $SO(3)$ and $m$ to $SO(2)$.  This is another illustration of the missing label problem studied in Section \ref{sec:MLO}.
The functions $Y_{n\ell m}$ in \eqref{eq:lap3}  can be expressed in terms of the functions $\Phi_{n,m_1,m_2}$,  by means of
the decomposition 
\beqa
{\cal D}_{\frac n2, \frac n2} = \bigoplus\limits_{k=0}^n {\cal D}_n \ , \nn
\eeqa
 valid for subduced representations in the embedding $SO(3) \subset SO(4)$.
These functions can be considered as an alternative Hilbert basis on the $3$-sphere, with
\beqa
(Y_{n\ell m},Y_{n' \ell' m'}) = \delta^n_{n'} \delta^\ell_{ \ell'} \delta^m_{m'} \ , \nn
\eeqa
and constitute the standard basis used in the harmonic analysis on $\mathbb S^3$.

It is possible to  define the Lie algebra $\hat {\mathfrak{g}}(SO(4)/SO(3))$  by means of the generators $T_{a,n,\ell, m} = T_a Y_{n\ell m}$,
 $N_0+N_0', N_0 -N_0'$ and $k_0+k_0', k_0-k_0'$.
However, the functions $Y_{n \ell m }$ are not  simultaneous  eigenfunctions of
 the operators $N_0$ and $N'_0$, but of  their sum $N_0 + N_0'$.  This follows at once from the identities
\beqa
Y_{11\pm 1} = \Phi_{1,\frac12,\pm \frac12}\ ,
Y_{110}=\frac1 {\sqrt 2}\Big( \Phi_{1,-\frac12, \frac12}+\Phi_{1,\frac12,-\frac12}\Big) \ , \ \
Y_{100}=\frac  \ii  {\sqrt 2}\Big( \Phi_{1,-\frac12, \frac12}-\Phi_{1,\frac12,-\frac12}\Big) \ . \nn
\eeqa
Thus, in this new basis, the commutation relations are more involved. In particular, the fact that the $Y$s are not
eigenfunctions of $N_0-N_0'$ in the second line of \eqref{eq:KM-S3p} implies that the coefficient of $k_0-k_0'$
together with the commutator with $N_0 -N_0'$ have to be consequently modified.

The construction described in this section can be further extended to the coset  spaces $SO(n)/SO(n-1) \cong \mathbb S^{n-1}$. As a matter of fact, the only representations of  $SO(n)$ that contain a scalar representation with respect to the embedding $SO(n-1) \subset SO(n)$ are the traceless symmetric $n^{\text{th}}$-order tensors. These representations are obtained from the symmetric powers of the fundamental representation $\left[1,0^{n-1}\right]$ after  extracting the trace, and correspond to representations with Dynkin  labels $\left[n,0,\cdots,0\right]$, in the notation of \cite{Pat}.

\medskip

Similarly to what happened with the results of Section  \ref{sec:su2-u1}, the construction above could be potentially of interest for applications to either Supergravity or Einstein-Maxwell scalar theories in $4+1$ dimensions, with a non-compact Riemannian version $SO(3,1)/SO(3)$ of our manifold $SO(4)/SO(3)$, that can be seen as the special case for $n=3$ of the sequence $SO(n,1)/SO(n)$ of symmetric real manifolds in supergravity theories, which yield non-symmetric homogeneous special K\"ahler (respectively special quaternionic K\"ahler) manifolds \cite{dewit1,dewit2}. Unitary representations of $SL(2,\mathbb C)$ were originally obtained by Gel'fand \cite{gms}, while the unitary representations of $SO(1,n)$ were studied in \cite{otto}. 


\subsection{Coset space $SU(3)/SU(2)$\label{ssSU3SU2}}

In order to construct the generalised Kac-Moody algebra associated to the coset space $SU(3)/SU(2)$, we  must first conveniently parameterise  the manifold $SU(3)$.  To this extent, we proceed in four steps: 
\begin{enumerate}
\item If $w_1, w_2,w_3$ are three orthonormal vectors of $\mathbb C^3$, {\it i.e.}, satisfying $w_i^\dag w_j =\delta_{j}^{i}$,  the matrix
$M_1 = \begin{pmatrix} w_1&w_2&w_3 \end{pmatrix}$  subjected to the additional constraint $\det M_1=1$, is such that $M_1^\dag M_1 =1$, and therefore belongs to $SU(3)$. We choose  the matrix $M_1$ as
\beqa
M_1=
\begin{pmatrix}\phantom{-} \cos   \theta  e^{\ii  \omega_1}&\sin \theta \cos  \xi e^{\ii  \omega_2}&\sin  \theta   \sin \xi e^{-\ii  ( \omega_1 +\omega_2)}\\
-\sin  \theta e^{\ii  \omega_1}&\cos \theta \cos \xi e^{\ii  \omega_2}&\cos \theta \sin \xi e ^{-\ii ( \omega_1+ \omega_2)}\\
0&-\sin \xi e^{\ii \omega_2}&\cos \xi e^{-\ii  ( \omega_1+\omega_2)}\end{pmatrix} \ , \nn
\eeqa
with $0\le \omega_1,\omega_2  <  2\pi ,0\le \xi,\theta \le \frac{\pi} 2 $ .
\item We introduce the diagonal matrix
\beqa
U_1=\begin{pmatrix} e^{ \ii \lambda}&0&0\\0&e^{\ii \lambda}&0\\0&0&e^{-2 \ii \lambda} \end{pmatrix}\ , \ 0\le \lambda < 2 \pi \nn
\eeqa
which belongs to $U(1) \subset SU(3)$.
\item We further consider the matrix 
\beqa
U_2 = \begin{pmatrix}1&0&0\\
                     0&\phantom{-1} \alpha&\beta\\
                     0&-\overline \beta&\overline \alpha\\
                     
      \end{pmatrix} \  , \nn
\eeqa
with
\beqa
\left\{
\begin{array}{lll}
\alpha&=& \cos \rho e^{ \ii\psi_1}\ ,\\
\beta&=& \sin \rho e^{ \ii\psi_2}\ ,
\end{array}\right.
\ \
\begin{array}{l}
0\le \rho\le \frac \pi 2\ ,\\
0\le \psi_1, \psi_2 < 2 \pi \ .
\end{array}\nn
\eeqa
This implies that $U_2 \in SU(2) \subset SU(3)$.
\item Finally, we define  the matrix  
\beqa
U\  \equiv\ U_2 U_1 M_1 \in SU(3) \ , \nn
\eeqa
which parameterises a point on the manifold $SU(3)$.
\end{enumerate}

If we define $\lambda+\omega_1 = \varphi_3$, $\lambda+\omega_2=\varphi_1$ and $\lambda -\omega_1 -\omega_2 = \varphi_2$,  then we have
   the following identities in the first row of the matrix $U$,
\beqa
U_{11}&=& \cos \theta  e^{\ii \varphi_3}\nn \ ,\\
U_{12}&=& \sin \theta \cos \xi e^{\ii  \varphi_1} \ ,\nn\\
U_{13}&=& \sin \theta\sin \xi e^{\ii \varphi_2} \ .\nn
\eeqa
Similarly, we have  
\beqa
U_{21}&=& -\alpha \sin \theta e^{\ii  \varphi_3}\ ,\nn\\
U_{22}&=&\alpha \cos\theta \cos \xi e^{\ii  \varphi_1}  -\beta \sin \xi e^{-\ii (\varphi_2+\varphi_3)}\ ,\nn\\
U_{23}&=&\alpha \cos \theta \sin\xi e^{\ii  \varphi_2} + \beta \cos \xi e^{-\ii (\varphi_1 + \varphi_3)}\nn \ ,
\eeqa
 for the second row of $U$,  and for the third row,
\beqa
U_{31}&=& \overline \beta \sin \theta e^{\ii  \varphi_3}\nn\ ,\\
U_{32}&=&-\overline \beta \cos\theta \cos \xi e^{\ii  \varphi_1}  -\overline \alpha \sin \xi e^{-\ii (\varphi_2+ \varphi_3)}\nn \ ,\\
U_{33}&=&-\overline \beta \cos \theta \sin\xi e^{\ii \varphi_2}+ \overline \alpha \cos \xi e^{-\ii (\varphi_1+ \varphi_3)}\nn \ .
\eeqa
The three  rows $\{U_{i1},U_{i2},U_{i3}\}, i=1,2,3$ span the fundamental three-dimensional representation  $[1,0]$ of $SU(3)$, whereas
$\{\overline U_{i1},\overline U_{i2},\overline U_{i3}\}, i=1,2,3$   span the anti-fundamental three-dimensional representation  $[0,1]$ of $SU(3)$.
It is well known that any representation of $SU(3)$ can be obtained from appropriate tensor products of the fundamental and anti-fundamental representations,  from which we conclude that the
Hilbert basis for the manifold $SU(3)$ can be deduced from the $U_{ij}$ and $\overline U_{ij}$ functions.

 From the  Peter-Weyl theorem, for a given representation ${\cal D}$ of dimension $d$,  we obtain, with the functions $U_{ij}$ and $\overline U_{ij}$, $d$ copies of the representation ${\cal D}$.  If we denote by ${\cal D}_{n,m}$ the representation of highest weight $n \mu_1 + m \mu_2$ (see below for the notations), then it is straightforward   to obtain the correct number of copies of  the representations ${\cal D}_{n,0}$ and ${\cal D}_{0,n}$, say $(n+2)!/(2!n!)$. For  general representations ${\cal D}_{n,m}$, however, it is more involved.  For instance, we have eight copies  of the adjoint representation ${\cal D}_{1,1}$, rather than nine, as it would seem at a first glance. Indeed, the highest weight of the adjoint representations are $\overline U_{11} U_{12}$, $ \overline U_{11} U_{22}$, $ \overline U_{11} U_{32}$, $ \overline U_{21} U_{12}$, $ \overline U_{21} U_{22}$, $ \overline U_{21} U_{32}$, $\overline U_{31} U_{12}$, $ \overline U_{31} U_{22}$, $\overline U_{31} U_{32}$, but  because of the relation $\overline U_{11} U_{12} + \overline U_{21} U_{12} +\overline U_{31} U_{32}=0$, there are actually eight independent copies of the adjoint representation. Thus, having constructed all representations with the correct multiplicity,  an appropriate Kac-Moody algebra $\hat{\mathfrak g}(SU(3))$ can be defined. 

It is worth noticing that the embedding of Kostant's principal subalgebra $\mathfrak{su}(2)_P$, isomorphic to ${\mathfrak{so}}(3)$, into the Lie algebra ${\mathfrak{su}}(3)$, is actually maximal, constituting the unique symmetric embedding of Kostant's ${\mathfrak{su}}(2)_P$ into any simple, compact Lie algebra \cite{kostant}. 

 However, we are mainly interested in the Fourier expansion on the coset space $$SU(3)/SU(2) \cong \mathbb S^5\ .$$  This isomorphism is related to the possible description of the corresponding tangent space as the 5-dimensional irreducible representation space of either 
 $\mathfrak{so}(5)\simeq \mathfrak{usp}(4)$  or  $\mathfrak{su}(2)$. Since this representation of $\mathfrak{su}(2)$ is self-conjugate,
 this implies, by Theorem 1.5 of \cite{dynkin},
the maximal embedding $\mathfrak{su}(2) \subset \mathfrak{usp}(4)$, so that the  subduced representation remains irreducible. It should be 
mentioned in this respect that the action of $SU(2)$ on its 5-dimensional irreducible representation is an example of a $\theta$-group \cite{vkac1,vkac2,Vin}, a
remarkable class of linear groups of transformations related to symmetric, (pseudo-)Riemannian coset spaces that has shown to be of current interest in several physical applications.
  According to the results of Section \ref{sec:coset}, the only functions that must be considered are
\beqa
\left\{
\begin{array}{lll}
z_3=U_{11}&=& \cos \theta  e^{\ii\varphi_3}\ , \\
z_2=U_{13}&=& \sin \theta \sin \xi e^{\ii \varphi_2}\ , \\
z_1=U_{12}&=& \sin \theta\cos \xi e^{\ii\varphi_1} \ ,
\end{array}\right.\ \ 
\begin{array}{l}
0\le \theta \le \frac \pi 2 \ , \\
0\le \xi \le \frac \pi 2 \ ,\\
0 \le \varphi_i < 2 \pi \ .
\end{array}\ \nn
\eeqa
They parameterise the five-sphere as a consequence of $|z_1|^2+|z_2|^2+|z_3|^2=1$. In the language of Appendix \ref{ap:ID}, the parameterisation of ${\mathbb S}^5$ is given by \cite{rb}
 \beqa
0\le \varphi_1, \varphi_2 ,\varphi_3  < 2 \pi\ , \ \ 0\le u_1  = \frac12 \sin^2 \xi  \le \frac12 \ \ , \ 
0\le u_2  = \frac14 \sin^4 \theta  \le \frac14
\ . \nn
\eeqa
The scalar product on the five-sphere is taken  as
\beqa
(f,g)&=& \frac1 {\pi^3}\int\limits_0^\frac{\pi}2 \sin^3 \theta \cos \theta \d  \theta \int \limits_0^\frac{\pi}2 \sin \xi \cos \xi \d \xi
\int \limits_0^{2\pi} \d \varphi_1
\int \limits_0^{2\pi} \d \varphi_2
\int \limits_0^{2\pi} \d \varphi_3
\nn\\&&
\times
\overline{f(\theta,\xi,\varphi_1,\varphi_2,\varphi_3)}\;{g(\theta,\xi,\varphi_1,\varphi_2,\varphi_3)}  \ , \nn
\eeqa
while the generators of the Lie algebra $\mathfrak{su}(3)$ are 
\beqa
\label{eq:su3-S5}
E_1^+&=&\frac12 e^{\ii (\varphi_1-\varphi_2)} \Big(\frac{\partial}{\partial\xi} -\ii  \tan \xi \frac{\partial}{\partial \varphi_1} -\ii  \cot \xi \frac{\partial}{\partial \varphi_2}\Big) \ , \nn \\
E_2^+&=&\frac12 e^{\ii (\varphi_2-\varphi_3)}\Big(-\sin \xi \frac {\partial}{\partial \theta} -\cot \theta \cos \xi \frac{\partial}{\partial \xi} - \ii \frac{\cot \theta}{\sin \xi}
\frac{ \partial}{\partial \varphi_2} - \ii\tan \theta \sin \xi \frac{\partial}{\partial \varphi_3}\Big) \ , \nn \\
E_3^+&=& \frac12  e^{\ii (\varphi_1 -\varphi_3)}\Big(-\cos\xi \frac{\partial}{\partial \theta} +\cot \theta \sin \xi \frac{\partial}{\partial \xi}
-\ii  \frac{\cot\theta}{\cos \xi}\frac{\partial}{\partial\varphi_1} -\ii  \tan \theta\cos\xi \frac{\partial}{\partial\varphi_3} \Big) \ , \nn\\
E_1^-&=&\frac12 e^{\ii (-\varphi_1+\varphi_2)} \Big(-\frac{\partial}{\partial\xi} -\ii  \tan \xi \frac{\partial}{\partial\varphi_1} -\ii  \cot \xi \frac{\partial}{\partial\varphi_2}\Big) \ ,  \\
E_2^-&=&\frac12 e^{\ii (-\varphi_2+\varphi_3)}\Big(\sin \xi \frac{\partial}{\partial \theta} +\cot \theta \cos \xi\frac{\partial}{ \partial \xi} - \ii  \frac{\cot \theta}{\sin \xi}
 \frac{\partial}{\partial\varphi_2} - \ii \tan \theta \sin \xi \frac{\partial}{\partial\varphi_3}\Big) \ , \nn \\
E_3^-&=& \frac12  e^{\ii (-\varphi_1 +\varphi_3)}\Big(\cos\xi \frac{\partial}{\partial\theta} -\cot \theta \sin \xi \frac{\partial}{\partial\xi}
-\ii  \frac{\cot\theta}{\cos \xi}\frac{\partial}{\partial\varphi_1} -\ii  \tan \theta\cos\xi\frac{\partial}{ \partial\varphi_3} \Big) \ , \nn\\
h_1&=& -\ii  \Big(\frac{\partial}{\partial\varphi_1} -\frac{\partial}{\partial\varphi_2}\Big) \ , \nn \\
h_2&=& -\ii  \Big(\frac{\partial}{\partial\varphi_2}-\frac{\partial}{\partial\varphi_3}\Big) \ . \nn
\eeqa
Here $(E_i^\pm,h_i), i=1,2$, are the generators associated  with the two simple roots $\alpha_i,i=1,2$ of the complexification of $\mathfrak{su}(3)$,  and $E^\pm_3=\pm[E^\pm_1,E^\pm_2]$. We denote the fundamental weights by $\mu_i$   with $i=1,2$.

A direct computation  with equation \eqref{eq:su3-S5} shows that the functions 
\beqa
\psi_{n_1,n_2,n_3}^n(\theta,\xi,\varphi)&=& \sqrt{\frac{(n+2)!}{2 n_1! n_2 ! n_3!}}
\sin^{n_1+n_2} \theta \cos^{n_3} \theta
\cos^{n_1} \xi \sin^{n_2} \xi\; e^{\ii(n_1 \varphi_1 + n_2 \varphi_2+n_3 \varphi_3)} \ ,\nn\\
&& \hskip 1.truecm n_1+n_2+n_3 =n \nn
\eeqa
span the representation with highest weight $\big|n \mu_1\big>$, while the functions $\overline{\psi}_{ n}^{n_1,n_2,n_3}$ span
the representation
with highest weight $\big|n \mu_2\big>$. These highest weights are explicitly given by
\beqa
\big<\theta,\xi,\varphi_1,\varphi_2,\varphi_3\big|n \mu_1\big>&=&\psi^n_{n,0,0}(\theta,\xi,\varphi_1,\varphi_2,\varphi_3)=\sqrt{\frac{(n+2)!}{2 n!}}\sin^n \theta \cos^n \xi e^{\ii n \varphi_1}\ , \nn\\
\big<\theta,\xi,\varphi_1,\varphi_2,\varphi_3\big|n \mu_2\big>&=&\overline{\psi}_n^{0,0,n}(\theta,\xi,\varphi_1,\varphi_2,\varphi_3)=\sqrt{\frac{(n+2)!}{2 n!}}\cos\theta^n  e^{-\ii n\varphi_3}\ . \nn
\eeqa
Furthermore,
\beqa
(\psi_{n_1,n_2,n_3}^n,\psi_{m_1,m_2,m_3}^m)=\delta^m_n \delta^{n_1} _{m_1}  \delta^{n_2}_{ m_2}  \delta^{n_3}_{ m_3}  \ , \ \ 
(\overline{\psi}_n^{n_1,n_2,n_3},\psi_{m_1,m_2,m_3}^m)= 0 \ . \nn
\eeqa

The representations of highest weight $\big|n \mu_1 + m \mu_2\big>$, denoted by ${\cal D}_{n,m}$, are obtained from 
the highest weight 
\beqa
 \psi^{n,m}_{n,0,0;0,0,m}=\sqrt{\frac{(n+m+2)!}{2 n! m!}}\sin^n \theta \cos^n \xi \cos^m\theta e^{\ii(n \varphi_1 -m \varphi_2) }\ , \nn
\eeqa
which enables us to obtain the full representation  from the operators $E_i^-$, where $i=1,2$. The functions 
$\psi_{n_1,n_2,n_3;m_1,m_2,m_3}^{n,m}$ that span the representation
${\cal D}_{n,m}$ (which can alternatively be obtained  by combining   $\psi^n_{n_1,n_2,n_3}$ with $\overline{\psi}_m^{m_1,m_2,m_3}$ and substracting the  trace).
We now identify the minimal set of indices to characterise any function.  From Section \ref{sec:MLO}, each representation is characterised by two numbers. We denote by ${\cal D}_{n,m}$ the representation of highest weight $n \mu_1 + m \mu_2$. Then three internal labels are needed to distinguished elements inside each representation space. The inner states are distinguished by the eigenvalues of $h_1,h_2$ as well as the value of the Casimir operator of the $\mathfrak{su}(2)$-subalgebra  generated by $E_1^\pm, h_1$ that we take equal to
\beqa
Q= \frac14 h_1^2 + \frac12(E_1^+ E_1^- + E_1^- E_1^+) \ . \nn
\eeqa 
Then for $n,m \in \mathbb N$, the representation space reduces to
\beqa
{\cal D}_{n,m} = \Big\{\psi_{n, m, n_1,n_2,\ell}, \ \ n_1,n_2 \ \ \text{s.t.} \ \ n_1 \mu_1 + n_2 \mu_2 \text{ is a  weight,}\ 0\le \ell \le\frac12(n+m)\Big\}\nn
\eeqa
and the Hilbert basis of   $\mathbb S^5$ adapted to our construction is defined by
\beqa
\label{eq:H-S5}
{\cal B} = \Big\{\psi_{n, m, n_1,n_2,\ell}, n,m \in \mathbb N, n_1,n_2, \ \ \text{s.t.} \ \ n_1 \mu_1 + n_2 \mu_2 \text{ is a  weight,}\ 0\le \ell \le\frac12(n+m)\Big\} \ . 
\eeqa
In the notation above, $n$, $m$ correspond to the representation $D_{n,m}$, $n_1,n_2$ are the eigenvalues of $h_1,h_2$ and $\ell$ is the eigenvalue of
the additional internal label $Q$. In order to obtain the conjugacy relation, we observe that we have the following normalisation for the fundamental and anti-fundamental
representations
\beqa
{\cal D}_{1,0}=\big\{\sqrt{3} z_1, \sqrt{3} z_2, \sqrt{3} z_3\big\}\ , \ \
{\cal D}_{1,0}=\overline{{\cal D}}_{1,0}=\big\{\sqrt{3} \bar z_3, -\sqrt{3} \bar z_2, \sqrt{3} \bar z_3\big\}\ , \ \
\eeqa
and thus
\beqa
 \overline\psi^{n, m, n_1,n_2,\ell}(\theta,\xi,\varphi_1,\varphi_2,\varphi_3)=(-1)^{\frac13(n-m-n_1+n_2)} \psi_{m, n, -n_1,-n_2,\ell}(\theta,\xi,\varphi_1,\varphi_2,\varphi_3) \ . \nn
\eeqa
Then we define the Lie algebra   $\hat{ \mathfrak g}\big(SU(3)/SU(2)\big)$ by introducing $T_{a,n, m, n_1,n_2,\ell}=
T_a \psi_{n, m, n_1,n_2,\ell}$.  We identify the maximal set of commuting operators by observing that relations \eqref{eq:su3-S5} can be extended
to define a differential realisation of the Lie algebra $\mathfrak{so}(6) \supset \mathfrak{su}(3)$. Within this differential realisation, all   the representations
$D_{n,m}$  are rearranged into representations of $\mathfrak{so}(6)$ corresponding to symmetric traceless  tensors.
The two constructions based on $\mathbb S^5 = SU(3)/SU(2)$ or on $\mathbb S^5 = SO(6)/SO(5)$ lead to isomorphic algebras
in straight analogy with Sections \ref{sec:su2} and \ref{sec:so4-so3}.
We will not use the representations of $\mathfrak{so}(6)$ to build the generalised Kac-Moody algebra.

The Hermitian  operators  are  taken to be $h_1,h_2$ and
\beqa
h=-\ii\left(\frac{\partial}{\partial \varphi_1} + \frac{\partial}{\partial \varphi_2} +\frac{\partial}{\partial \varphi_3} \right),
\eeqa
 the latter being associated to the  Cartan subalgebra of $\mathfrak{so}(6)$, and we have
 \beqa
h \psi_{n,m,n_1,n_2,\ell}=(n-m) \psi_{n,m,n_1,n_2,\ell} \ . \nn
\eeqa
 We finally introduce the associated central charge $k_1,k_2,k$ (see \eqref{eq:form} and \eqref{eq:cc}). The Lie brackets take the form (see \eqref{eq:KM-gen})
\beqa
\label{eq:KM-S5}
\big[T_{a,n, m, n_1,n_2,\ell},T_{a',n', m', n'_1,n'_2,\ell'}\big]&=& \ii f_{a a'}{}^{a''}
c_{n, m, n_1,n_2,\ell,n', m', n'_1,n'_2,\ell'}{}^{n'', m'', n''_1,n''_2,\ell''}
T_{a'',n'', m'', n''_1,n''_2,\ell''} \nn\\
&&+ (-1)^{\frac13(n-m-n_1+n_2)}(k_1n_2  + k_2 n_1 ' + k(n'-m'))\times\nn\\
&& \hskip .5truecm  g_{a a'} \delta_{n m'} \delta_{m n'} \delta_{n_1+n_2 '} \delta_{n_2 +n_1'} \delta_{\ell\ell'}  \ ,\nn \\
\big[h_1,T_{a,n, m, n_1,n_2,\ell}\big]&=& n_1 T_{a,n, m, n_1,n_2,\ell} \ , \\
\big[h_2,T_{a,n, m, n_1,n_2,\ell}\big]&=& n_2 T_{a,n, m, n_1,n_2,\ell} \ , \nn\\
\big[h,T_{a,n, m, n_1,n_2,\ell}\big]&=& (n-m) T_{a,n, m, n_1,n_2,\ell} \ , \nn
\eeqa
 Observe that, because of the identity $\overline{\cal D}_{n,m}= {\cal D}_{m,n}$, the structure constants involve the normalisation term $\delta_{m n'} \delta_{n m'} \delta_{n_1+n_2'}\delta_{n_2+n_1'}$, and
not $\delta_{m m'} \delta_{n n'}\delta_{n_1+n_1'}\delta_{n_2+n_2'}$ as may be naively expected. Such a subtlety is only encountered when the Lie algebra $\mathfrak g$ admits complex ({\it i.e.}, neither real nor pseudo-real) representations, and is not present
for the remaining examples studied in this article.

The $c_{IJ}{}^K$ coefficients are obtained from
\beqa
\psi_{n, m, n_1,n_2,\ell}\psi_{n', m', n'_1,n'_2,\ell'}&=&
\sum
\lambda(N,M,n,n',m,m',\ell,\ell',L) \Big({}^{\hskip .15truecm n}_{\; n_1} {}^{\hskip .15cm  m}_{\; \; n_2 }{}^{\; \ell}  \ \
{}^{\hskip .15truecm n'}_{\; n'_1} {}^{\hskip .15truecm m'}_{\; \; n_2'}{}^{\;\ell'} \Big|
{}^{\hskip .35cm N \hskip .8truecm M}_{\; n_1+n_1' \; \;n_2+n_2'} {}^{\;L} {}^{\hskip .35cm }_{}\Big)\nn\\
&& \hskip 2.truecm \times\psi_{N,M,n_1+n'_1,n_2+n'_2,L}. \nn
\eeqa
Here
$\Big({}^{\hskip .15truecm n}_{\; n_1} {}^{\hskip .15cm  m}_{\; \; n_2 }{}^{\; \ell}  \ \
{}^{\hskip .15truecm n'}_{\; n'_1} {}^{\hskip .15truecm m'}_{\; \; n_2'}{}^{\;\ell'} \Big|
{}^{\hskip .35cm N \hskip .8truecm M}_{\; n_1+n_1' \; \;n_2+n_2'} {}^{\;L} {}^{\hskip .35cm }_{}\Big)$
are the Clebsch-Gordan coefficients of the decomposition  
\beqa
{\cal D}_{n ,m }\otimes {\cal D}_{n',m'}  = \bigoplus \limits_{N,M} D_{N, M} \ . \nn
\eeqa
As before, the  coefficients $\lambda(N,M,n,n',m,m',\ell,\ell',L)$  can be computed recursively.

This construction can be extended naturally, along the same lines, to the generic coset space $SU(n+1)/SU(n) \cong \mathbb S^{2n+1}$.
The only representations that contain the scalar representations 
with respect to the embedding $SU(n-1)\subset SU(n)$ are ${\cal D}_{n,0,\cdots,0,m}$ (in the notations of \cite{GT})
and correspond to traceless tensor products of the fundamental and the anti-fundamental
representations.  It should be observed that the generalised Kac-Moody algebra that we obtain  from  $SU(n-1)\subset SU(n)$
is isomorphic to the construction from the
coset space $SO(2n+2)/SO(2n+1) \cong \mathbb S^{2n+1}$.


\subsection{ Coset space $G_2/SU(3)$}\label{sec:su3}

Prior to the construction of parameterisations of the manifolds $G_2$ and $G_2/SU(3)$, we briefly recall some fundamental properties  of the  exceptional Lie algebra $\mathfrak{g_2}$ (see {\it e.g.} \cite{Wybo}).
The Cartan matrix of $\mathfrak g_2$ is given by
\beqa
\begin{pmatrix}\phantom{-}2&-1\\ -3&\phantom{-}2 \end{pmatrix} \  \nn
\eeqa
 and its simple roots  and fundamental weights are given respectively by
\beqa
\begin{array}{llllll}
\alpha_1 &=& \big|2,-3\big> \ , &
\alpha_2&=& \big|-1,2\big> \ , \\
\mu_1 &=& \big|1,0\big> \ , &
\mu_2&=& \big|0,1\big> \ .
\end{array}\nn
\eeqa

The representation  of highest weight $\mu_2$ is seven-dimensional and real, explicitly: 
\beqa
&\big<z\big| 1,-2\big>=z_{[1,-2]}\ ,\ \ \big<z\big| -1,1\big>=z_{[-1,1]}\ ,\ \ \big<z\big| 0,1\big>=z_{[0,1]}\ ,\nn\\
&\big<z\big| -1,2\big>=\overline z_{[-1,2]}\ , \ \ \big<z\big| 1,-1\big>=\overline z_{[1,-1]}\ , \ \  \big<z\big| 0,-1\big>= \overline z_{[0,-1]}\ ,\nn\\
&\big<z\big| 0,0\big>=x_0\ . \nn
\eeqa
 Because the representation is real, $x_0$ is a real number, while  $z_{[1,-2]}, z_{[-1,1]},z_{[0,1]}$ are complex and conjugate to $\overline z_{[-1,2]},\overline z_{[1,-1]}, \overline z_{[0,-1]}$, respectively. A differential realisation of the algebra deduced from this representation is given by (see Chapter 10 in  \cite{zz})
 \begin{allowdisplaybreaks}
\beqa
\label{eq:diff}
E_{\alpha_1[2,-3]} &=& z_{[1,-2]}\partial_{[-1,1]} -\overline z_{[1,-1]}\overline \partial_{[-1,2]} \ , \nn\\
E_{-\alpha_1[-2,3]} &=& z_{[-1,1]}\partial_{[1,-2]} -\overline z_{[-1,2]}\overline \partial_{[1,-1]} \ ,\nn\\
E_{\alpha_2[-1,2]}&=&z_{[0,1]} \overline \partial_{[1,-1]}-z_{[-1,1]}\overline \partial_{[0,-1]} + \sqrt 2 \Big(x_0 \partial_{[1,-2]} - \overline z_{[-1,2]} \partial_0\Big) \ , \nn\\
E_{-\alpha_2[1,-2]}&=&\overline z_{[0,-1]}  \partial_{[-1,1]}-\overline z_{[1,-1]} \partial_{[0,1]} + \sqrt 2 \Big(x_0 \overline \partial_{[-1,2]} - z_{[1,-2]} \partial_0\Big) \ , \nn\\
E_{\alpha_1+\alpha_2[1,-1]}&=&z_{[1,-2]} \overline \partial_{[0,-1]}-z_{[0,1]}\overline \partial_{[-1,2]} + \sqrt 2 \Big(x_0 \partial_{[-1,1]} - \overline z_{[1,-1]} \partial_0\Big) \ , \nn\\
E_{-\alpha_1-\alpha_2[-1,1]}&=&\overline z_{[-1,2]}  \partial_{[0,1]}-\overline z_{[0,-1]} \partial_{[1,-2]} + \sqrt 2 \Big(x_0 \overline \partial_{[1,-1]} -  z_{[-1,1]} \partial_0\Big) \ , \nn\\
E_{\alpha_1+2\alpha_2[0,1]}&=&\overline z_{[1,-1]}  \partial_{[1,-2]}-\overline z_{[-1,2]} \partial_{[-1,1]} + \sqrt 2 \Big(x_0 \overline \partial_{[0,-1]} -  z_{[0,1]} \partial_0\Big) \ , \\
E_{-\alpha_1-2\alpha_2[0,1]}&=& z_{[-1,1]}  \overline\partial_{[-1,2]}- z_{[1,-2]} \overline\partial_{[1,-1]} + \sqrt 2 \Big(x_0  \partial_{[0,1]} -  \overline z_{[0,-1]} \partial_0\Big) \ , \nn\\
E_{\alpha_1+3\alpha_2[-1,3]} &=& z_{[0,1]}\partial_{[1,-2]} -\overline z_{[-1,2]}\overline \partial_{[0,-1]} \ , \nn\\
E_{-\alpha_1-3\alpha_2[1,-3]} &=& z_{[1,-2]}\partial_{[0,1]} -\overline z_{[0,-1]}\overline \partial_{[-1,2]} \ , \nn\\
E_{2\alpha_1+3\alpha_2[1,0]} &=& z_{[0,1]}\partial_{[-1,1]} -\overline z_{[1,-1]}\overline \partial_{[0,-1]} \ , \nn\\
E_{-2\alpha_1-3\alpha_2[1,0]} &=& z_{[-1,1]}\partial_{[0,1]} -\overline z_{[0,-1]}\overline \partial_{[1,-1]} \ , \nn\\
h_1&=&z_{[1,-2]}\partial_{[1,-2]}-z_{[-1,1]}\partial_{[-1,1]}-\overline z_{[-1,2]}\overline \partial_{[-1,2]}+\overline z_{[1,-1]}\overline \partial_{[1,-1]} \ , \nn\\
h_2&=&-2z_{[1,-2]}\partial_{[1,-2]}+z_{[-1,1]}\partial_{[-1,1]}+ z_{[0,1]}\partial_{[0,1]} \nn\\
&&+2 \overline z_{[-1,2]}\overline \partial_{[-1,2]}-\overline z_{[1,-1]}\overline \partial_{[1,-1]}
-\overline z_{[0,-1]} \overline \partial_{[0,-1]}\ . \nn
\eeqa
 \end{allowdisplaybreaks}

\medskip
\noindent Due to the embedding ${\mathfrak g_2} \subset \mathfrak{so}(7)$, the quadratic form
\beqa
q(z)=x_0^2 + 2 z_{[1,-2]}\overline z_{[-1,2]} + 2 z_{[-1,1]}\overline z_{[1,-1]} + 2 z_{[0,1]}\overline z_{[0,-1]} \ , \nn
\eeqa
is preserved {\it i.e.}, for any element $\xi$ in the realisation (\ref{eq:diff}) of $\mathfrak{g}_2$, we have $\xi(q(z))=0$.
As a consequence, the representations with highest weight $n \mu_2$ can be described in terms of $n^{\text{th}}$-order polynomials in the variables $z$. In order to factor out those terms proportional to $q(z)$, we establish that 
\beqa
\label{eq:D0n}
{\cal D}_{0,n}= \{ P \in \mathbb R_n[x_0,  z_{[1,-2]},  z_{[-1,1]},  z_{[0,1]}, \overline z_{[-1,2]}, \overline z_{[1,-1]},\overline z_{[0,-1]}] \ \ \text{s.t.}\  \nabla^2(P)=0\} \ , 
\eeqa
where
\beqa
\nabla^2= \partial_0^2 + 2 \partial_{[1,-2]}\overline \partial_{[-1,2]} + 2 \partial_{[-1,1}\overline \partial_{[1,-1]} + 2 \partial_{[0,1]}\overline \partial_{[0,-1]} \ , \nn
\eeqa
and $\mathbb R_n[x_0,  z_{[1,-2]},  z_{[-1,1]},  z_{[0,1]}, \overline z_{[-1,2]}, \overline z_{[1,-1]},\overline z_{[0,-1]}]$ denotes the space of $n^{\text{th}}$-order polynomials.

\smallskip
\noindent For the embedding $\mathfrak{su}(3) \subset \mathfrak g_2$, the branching rule for the adjoint representation (with highest weight $\mu_1$) of $\mathfrak{g}_2$ is given by  
\beqa
\utilde {\bf 14} = \utilde {\bf 8} \oplus  \utilde {\bf 6}  \ . \nn
\eeqa
The adjoint representation is always real,  whereas  the two {\it complex conjugate}  fundamental and anti-fundamental representations regroup  into a six-dimensional {\it real} representation:
\beqa
\label{eq:633}
\utilde {\bf 3} \oplus  \utilde {\bf \overline 3}=  \utilde {\bf 6} \ .
\eeqa
The embedding $\mathfrak{su}(3) \subset \mathfrak g_2$ can be explicitly described as follows. Let
\beqa
\beta_1&=& 2 \alpha_1 + 3 \alpha_2 \ , \nn\\
\beta_2&=&-\alpha_1 \ , \nn
\eeqa
be the two simple roots of $ \mathfrak{su}(3)$,  so that
the generators of the subalgebra $\mathfrak{su}(3)\subset \mathfrak g_2$ are  expressed as
\beqa
\begin{array}{llllll}
E_{\beta_{1}}&=& E_{2 \alpha_1 + 3 \alpha_2} \ ,& E_{-\beta_{1}}&=& E_{-2 \alpha_1 -3 \alpha_2}\ , \\
E_{\beta_2} &=&E_{-\alpha_1} \ , &E_{-\beta_2} &=&E_{\alpha_1} \ , \nn\\
E_{\beta_{1}+\beta_2}&=&E_{ \alpha_1 + 3 \alpha_2} \ , &E_{-\beta_{1}-\beta_2}&=&E_{ -\alpha_1 - 3 \alpha_2} \, \nn\\
H_1&=&2 h_1 +h_2 \ , &H_2&=&-h_1 \ .
\end{array}\nn
\eeqa
 By introducing also the fundamental weight,
\beqa
\mu_1 &=&\frac23 \beta_1 + \frac13 \beta_2 = \alpha_1 + 2 \alpha_2 \ ,\nn \\
\mu_2 &=&\frac13 \beta_1 + \frac23 \beta_2= \alpha_2 \ ,\nn
\eeqa
the generators of the coset $\mathfrak{g}_2/\mathfrak{su}(3)$ read as 
\beqa
\begin{array}{llllll}
&\utilde{\bf 3}&&&\utilde{\bf \overline 3}\\ \hline 
E_{\mu_1}&=& E_ {\alpha_1 + 2 \alpha_2}\ ,&E_{\mu_2}&=& E_ { \alpha_2}\ , \nn\\
E_{\mu_1 -\beta_1}&=& E_ {-\alpha_1 - \alpha_2}\ ,&E_{\mu_2-\beta_2}&=& E_ {\alpha_1+ \alpha_2}\ , \nn\\
E_{\mu_1 -\beta_1 -\beta_2}&=& E_ { - \alpha_2}\ ,&E_{\mu_2-\beta_2-\beta_1}&=& E_ {-\alpha_1-2 \alpha_2}\ .
\end{array}\nn
\eeqa 

The next step in the construction is to derive a matrix representation. This can be easily done  by means of the vectors
\beqa
Z&=&(z_{[-1,2]},z_{[-1,1]},z_{[0,1]},\overline z_{[1,-2]},\overline z_{[1,-1]},\overline z_{[0,-1]}, x_0) , \nn\\
\partial Z&=&(\partial_{[1,-2]},\partial_{[1,-1]},\partial_{[0,-1]},\overline \partial_{[-1,2]},
\overline \partial_{[-1,1]},\overline \partial_{[0,1]}, \partial_0)^t\ , \nn
\eeqa 
such that, to any (first-order) differential operator $D$ of $\mathfrak{g}_2$, we can associate the matrix $M$ defined by
\beqa
D= Z M \partial_Z \ . \nn
\eeqa
The matrices $M_\gamma$  associated to the roots $\gamma$  of $\mathfrak{g}_2$, as well as those $h_1, h_2$, corresponding to the Cartan subalgebra, can be constructed in  a straightforward manner  {\it via} this prescription.  For the subalgebra $\mathfrak{su}(3)$, these matrices reduce to
\beqa
T_i = \begin{pmatrix} \lambda_i&0&0\\
           0&-\overline\lambda_i&0\\
           0&\phantom{-}0&0
           \end{pmatrix} \ ,\ \ i=1,\dots, 8 \ , \nn
\eeqa
where $\lambda_i$ are the $3\times 3$ Gell-Mann matrices. For the coset $\mathfrak{g_2}/\mathfrak{su}(3)$
we have
\beqa
x^i U_i &=& \left(
\begin{array}{ccccc}
0&0&0&0&x^5+\ii  x^6\\
0&0&0&-x^5-\ii x^6&0\\
0&0&0&-x^3-\ii x^4&x^1+\ii x^2\\
0&-x^5+\ii x^6&-x^3+\ii x^4&0&0\\
x^5-\ii x^6&0&x^1-\ii x^2&0&0\\
x^3-\ii x^4&-x^1+\ii x^2&0&0&0\\
\sqrt{2}(x^1+\ii x^2)&\sqrt{2}(x^3+\ii x^4)&-\sqrt{2}(x^5+\ii x^6)&-\sqrt{2}(x^1-\ii x^2)&-\sqrt{2}(x^3-\ii x^4)
\end{array}
\right.\nn
\\ \nn\\
&&\hskip 7.truecm
\left.
\begin{array}{cc}
x^3+\ii x^4&\sqrt{2}(x^1-\ii x^2)\\
-x^1-\ii x^2&\sqrt{2}(x^3-\ii x^4)\\
0&-\sqrt{2}(x^5-\ii x^6)\\
0&-\sqrt{2}(x^1+\ii x^2)\\
0&-\sqrt{2}(x^3+\ii x^4)\\
0&\sqrt{2}(x^5+\ii x^6)\\
\sqrt{2}(x^5-\ii x^6)&0
\end{array}
\right)\ ,\nn
\eeqa
with the matrices $U_i$ defined by 
\beqa
\begin{array}{llllll}
U_1&=& M_{\alpha_2} -M_{-\alpha_2} \ , &U_2&=&{ {\rm i}}(M_{\alpha_2} +M_{-\alpha_2}) \ ,\\
U_3&=&M_{\alpha_1 + \alpha_2} -M_{-\alpha_1 - \alpha_2}\ , &U_4&=&{ {\rm i}} (M_{\alpha_1 + \alpha_2} +M_{-\alpha_1 - \alpha_2})\ ,\\
U_5&=&M_{\alpha_1 +2 \alpha_2} -M_{-\alpha_1 -2 \alpha_2}\ , &U_6&=&{ {\rm i}} (M_{\alpha_1 +2 \alpha_2} +M_{-\alpha_1 - 2\alpha_2})\ .
\end{array} \nn  
\eeqa
It should be observed that these matrices are not well adapted,  because the representation is real. 
 With equation \eqref{eq:633}, we consider the real basis
\beqa
\label{eq:realb}
X=\begin{pmatrix} x_{[-1,2]}=\frac1 {\sqrt{2}} (z_{[-1,2]}+\overline z_{[1,-2]})\\
                   x_{[-1,1]}=\frac1 {\sqrt{2}} (z_{[-1,1]}+\overline z_{[1,-1]})\\
                   x_{[0,1]}=\frac1 {\sqrt{2}} (z_{[0,1]}+\overline z_{[0,-1]})\\
                   y_{[-1,2]}=-\frac \ii  {\sqrt{2}} (z_{[-1,2]}-\overline z_{[1,-2]})\\
                   y_{[-1,1]}=-\frac \ii  {\sqrt{2}} (z_{[-1,1]}-\overline z_{[1,-1]})\\
                   y_{[0,1]}=-\frac \ii  {\sqrt{2}} (z_{[0,1]}-\overline z_{[0,-1]})\\
                   x_0
                   \end{pmatrix} \ .
                   \eeqa
Over this basis, the generators of the $\mathfrak{su}(3)$-subalgebra take the form
\beqa
S_i= \begin{pmatrix} \phantom{-}\frac12(\lambda_i - \overline\lambda_i)& \frac { {\rm i}}2(\lambda_i + \overline\lambda_i)&0\\ 
                     - \frac {  {\rm i}}2(\lambda_i + \overline\lambda_i )& \frac12(\lambda_i - \overline\lambda_i)&0\\
                     0&0&0
                     \end{pmatrix},\nn
\eeqa
while for the coset $\mathfrak{g}_2/\mathfrak{su}(3)$ the generators $V_i$ are given by
\beqa
\ii x^j V_j=\begin{pmatrix}
\phantom{-2}0&\phantom{2}-x^6&\phantom{2}-x^4&\phantom{-2}0&\phantom{2}-x^5&\phantom{2}-x^3&\phantom{-}2x^2\\
\phantom{-2}x^6&\phantom{-2}0&\phantom{-2}x^2&\phantom{2}-x^5&\phantom{-2}0&\phantom{-2}x^1&\phantom{-}2x^4\\
\phantom{-2}x^4&\phantom{2}-x^2&\phantom{-2}0&\phantom{2}-x^3&\phantom{2}-x^1&\phantom{-2}0&-2x^6\\
\phantom{-2}0&\phantom{-2}x^5&\phantom{-2}x^3&\phantom{-2}0&\phantom{2}-x^6&\phantom{2}-x^4&\phantom{-}2x^1\\
\phantom{-2}x^5&\phantom{-2}0&\phantom{-2}x^1&\phantom{-2}x^6&\phantom{-2}0&\phantom{2}-x^2&-2x^3\\
\phantom{-2}x^3&\phantom{2}-x^1&\phantom{-2}0&\phantom{-2}x^4&\phantom{-2}x^2&\phantom{-2}0&\phantom{-}2x^5\\
-2x^2&-2x^4&\phantom{-}2x^6&-2x^1&\phantom{-}2x^3&-2x^5&\phantom{-2}0
\end{pmatrix} \ . \nn
\eeqa

 Given an appropriate real basis, we can construct a parameterisation of the manifold $G_2$.  From the coset space structure, we rewrite a matrix for $G_2$ in the form 
\beqa
G= M_3 M_2 \nn
\eeqa
where $M_3$ is a matrix that parameterises $SU(3)$, and $M_2$ a matrix that  parameterises the factor space $G_2/SU(3)$.
The matrix $M_3$ can be directly obtained from Section \ref{ssSU3SU2}:
\beqa
M_3 = \begin{pmatrix} \phantom{-}\frac12(U+ \overline U)&\frac {  {\rm i}} 2(U-\overline U)&0\\
                     -\frac {  {\rm i}} 2 (U-\overline U)& \frac12 (U+\overline U)&0\\
                      0&0&1\end{pmatrix} \ .
\eeqa
The matrix $M_2$ is constructed as follows. Take
\beqa
U_1(\theta) = e^{{  {\rm i}} \theta V_1} \nn
\eeqa
and consider the specific point on $G_2/SU(3)$ given by $U_1(\pi)$.
 Next, introduce the matrix $P= R R_1 R_2 R_3 R_4 R_5$, with the $R-$matrices being appropriate rotations:
 $R$ angle $-\varphi$ in the plane $(x_{[-1,2]}, x_{[-1,1]})$, $R_1$ angle $-\theta_1$ in the plane  $(x_{[-1,1]},
 x_{[0,1]})$, 
  $R_2$ angle $-\theta_2$ in the plane $(x_{[0,1]},y_{[-1,,2]})$,
  $R_3$ angle $-\theta_3$ in the plane $(y_{[-1,,2]}, y_{[-1,1]})$,
   $R_4$ angle $-\theta_4$ in the plane $(y_{[-1,1]},y_{[0,1]})$, and
   $R_5$ angle $-1/2\theta_5$ in the plane $(y_{[0,1]},x_0)$ with the notations of  \eqref{eq:realb}
and define $M_2$ by
\beqa
M_2 = P^{-1} U_1(\pi) P \ . \nn
\eeqa
The parameterisation of the manifold is thus given by
\beqa
G =  M_3 P^{-1} U_1(\pi)   P\ . 
\eeqa
At a first glance, the matrix $M_2$ seems not to be very illuminating. Fortunately, however, as follows from Section \ref{sec:coset}, only the last column of the matrix $M_2$ will be relevant for the harmonic analysis on $G_2/SU(3)$. It reduces to
a very simple expression:
\beqa
\label{eq:S6-p}
M_{2,}{}_{71}&=& \sin \theta_1\sin \theta_2 \sin \theta_3 \sin \theta_4 \sin \theta_5 \sin \varphi \equiv- \frac { {\ii}} 2(z_1-\bar z_1)\ , \nn\\
M_{2,}{}_{72}&=& \sin \theta_1\sin \theta_2 \sin \theta_3 \sin \theta_4 \sin \theta_5 \cos \varphi  \equiv
\frac 1 2(z_1 +\bar z_1)\ , \nn\\
M_{2,}{}_{73}&=& \cos \theta_1\sin \theta_2 \sin \theta_3 \sin \theta_4 \sin \theta_5 \equiv\frac12(z_2+\bar z_2)\ \ , \nn\\
M_{2,}{}_{74}&=& \cos \theta_2 \sin \theta_3 \sin \theta_4 \sin \theta_5 \equiv -\frac {  {\rm i}} 2(z_2 -\bar z_2)  \ , \\
M_{2,}{}_{75}&=& \cos \theta_3 \sin \theta_4 \sin \theta_5 \equiv \frac 1 2(z_3 +\bar z_3) \ , \nn\\
M_{2,}{}_{76}&=& \cos \theta_4 \sin \theta_5 \equiv  -\frac {  {\rm i}} 2(z_3 -\bar z_3)\ , \nn\\
M_{2,}{}_{77}&=& \cos  \theta_5  \equiv x_0 \ , \nn
\eeqa
which is a parameterisation of the sphere $\mathbb S^6\cong G_2/SU(3)$   with $0\le \varphi < 2 \pi$,  $0\le \theta_i \le \pi$, $i=1,\cdots,5$,  that corresponds to the usual spherical coordinates.

\medskip

\noindent Comparing with the approach of Appendix \ref{ap:ID}, the sphere ${\mathbb S}^6$ is parameterised by
 \beqa
0\le \varphi \le 2 \pi\ , \ \ -1\le u_1  &=&\cos \theta_1  \le 1 \ \ ,\nn\\
0\le u_2  &=& \frac12(\theta_2 -\cos \theta_2 \sin \theta_2 )\le \frac \pi 2
\, \nn\\
-\frac23 \le u_3 &=& -\frac 1 3(\sin^2 \theta_3 \cos \theta_3 + 2 \cos\theta_3)\le \frac23 \ , \nn\\
0\le u_4 &=& \frac18((3 \theta_4 -3 \sin \theta_4 \cos \theta_4 -2 \sin^3 \theta_4 \cos \theta_4) \le \frac38 \pi. \nn\\
-\frac8{15}\le u_5 &=&-\frac8{15} \cos\theta_5 -\frac 4{15} \sin\theta^2_5 \cos \theta_5 -\frac15 \sin\theta_5^4 \cos \theta_5
\le \frac 8{15} \ .\nn
\eeqa

\medskip

If we define the scalar product on $\mathbb S^6$ by
\beqa
(f,g) &=& \frac{15}{16 \pi^3} \int \limits_0^\pi \d \theta_5 \sin^5 \theta_5 \int \limits_0^\pi \d \theta_4 \sin^4 \theta_4\int \limits_0^\pi \d \theta_3 \sin^3 \theta_3
\int \limits_0^\pi \d \theta_2 \sin^2 \theta_2 \int \limits_0^\pi \d \theta_1 \sin \theta_1 \int \limits_0^{2\pi} \d \varphi\nn\\
&&\hskip 1.5truecm \overline{f(\theta_1,\theta_2,\theta_3,\theta_4,\theta_5, \varphi)}\;{g(\theta_1,\theta_2,\theta_3,\theta_4,\theta_5, \varphi)},\nn
\eeqa
without loss of generality ({\it e.g.} after having conjugated the matrix $M_2$ by an appropriate permutation) we can introduce the harmonic functions
\begin{allowdisplaybreaks}
\beqa
\label{eq:harmg2}
\Phi_{1;1,-2}&=&\sqrt{\frac 7 4}\big(M_{2,}{}_{72}+ { {\rm i}}M_{2,}{}_{71}\big)=- \sqrt{\frac 7 2} z_{[-1,2]}\nn\\
&=&\sqrt{\frac 7 2}e^{{ {\rm i}} \varphi} \sin \theta_1 \sin \theta_2 \sin \theta_3 \sin \theta_4 \sin\theta_5
           \ , \nn\\
\Phi_{1;-1,2}&=&\sqrt{\frac 7 4} \big(M_{2,}{}_{72}-{  {\rm i}} M_{2,}{}_{71}\big) =\sqrt{\frac 7 2} \overline z_{[1,-2]} \nn\\
&=&\sqrt{\frac 7 2}e^{-{  {\rm i}} \varphi} \sin \theta_1 \sin \theta_2 \sin \theta_3 \sin \theta_4 \sin\theta_5\  \ , \nn\\
 \nn\\
\Phi_{1;-1,1}&=&\sqrt{\frac 7 4}\big(M_{2,}{}_{73}+{  {\rm i}} M_{2,}{}_{74}\big) =- \sqrt{\frac 7 2} z_{[-1,1]}  \nn\\
&=&
\sqrt{\frac 7 2}  \sin \theta_3\sin \theta_4 \sin\theta_5\big(\cos \theta_1 \sin \theta_2 + { {\rm i}} \cos \theta_2\Big) \ , \nn\\
\Phi_{1;1,-1}&=&\sqrt{\frac 7 4}\big(M_{2,}{}_{73}-{ {\rm i}} M_{2,}{}_{74}\big)
=- \sqrt{\frac 7 2} \overline z_{[1,-1]}\\
&=& \sqrt{\frac 7 2}\sin \theta_3\sin \theta_4 \sin\theta_5\big(\cos \theta_1 \sin \theta_2 - {  {\rm i}} \cos \theta_2\Big) \ , \nn\\
\Phi_{1;0,1}&=&\sqrt{\frac 7 4}\big(M_{2,}{}_{75}+{  {\rm i}} M_{2,}{}_{76}\big)
=\sqrt{\frac 7 2} z_{[0,1]}\nn\\
&=&\sqrt{\frac 7 2}\sin\theta_5\big( \cos \theta_3 \sin \theta_4 +{  {\rm i}} \cos \theta_4\big) \ , \nn\\
\Phi_{1;0,-1}&=&\sqrt{\frac 7 4}\big(M_{2,}{}_{75}-{  {\rm i}} M_{2,}{}_{76}\big)
=-\sqrt{\frac 7 2} \overline z_{[0,-1]}\nn\\
&=&\sqrt{\frac  72} \sin\theta_5\big( \cos \theta_3 \sin \theta_4 -{  {\rm i}} \cos \theta_4\big) \ , \nn\\
\Phi_{1;0,0}&=&\sqrt{7} M_{2,}{}_{77} = \sqrt{7} x_0 =\sqrt 7 \cos  \theta_5, \nn
\eeqa
\end{allowdisplaybreaks}

\noi
which  are orthonormal with respect to the scalar product on $\mathbb S^6$.
The precise signs are obtained from \eqref{eq:diff}.
These functions  parameterise the representation ${\cal D}_{0,1}$. The highest weight of the representation ${\cal D}_{0,n}$ (see  equation \eqref{eq:D0n}) is therefore given by
\beqa
\Phi_{n,0,n}^{\alpha} = \sqrt{\frac 1 {60} \frac 1{4^n} \frac{(2n+5)!}{n!(n+2)!}} z_{[0,1]}^n = \sqrt{\frac 1 {60} \frac 1{4^n} \frac{(2n+5)!}{n!(n+2)!}}\sin\theta^n_5\big( \cos \theta_3 \sin \theta_4 +{  {\rm i}} \cos \theta_4\big)^n \ ,\nn 
\eeqa
and ${\cal D}_{0,n}$ is constructed by the action of the operators given in  equation (\ref{eq:diff}). Only at the very end, we substitute  equation \eqref{eq:harmg2} into the $n^{\text{th}}$-order polynomials of ${\cal D}_{0,n}$, given in  equation \eqref{eq:D0n},
to obtain the corresponding harmonic functions.

 In contrast to the previous cases, for $G_2$ we have a degeneracy problem. According to Proposition \ref{RAC}, we need $6$ internal labels to separate states
within an irreducible representation of $G_2$, the Casimir operators of $G_2$ being used to characterise the representation. Considering the reduction chain 
\beqa
G_2 \supset SU(3)\supset SU(2)\supset U(1),\label{redcha}
\eeqa
provides us with five internal labels, namely the Casimir operators of $SU(3)$ and $SU(2)$, as well as the generators of the Cartan subalgebra. It is thus necessary 
to consider an additional label.  This operator can be constructed by the method of elementary multiplets (see \cite{Sh2}) 
observing that the adjoint representation of $G_2$ decomposes as the direct sum of an octet $T$ and two 
conjugate triplets $V,{\overline{V}}$ of $SU(3)$. The simplest labelling problem resulting from this method is a cubic operator $TV\overline{V}$ in the generators of $G_2$, such that in each monomial one 
generator belongs to the octet and each of the triples, respectively. This operator is Hermitian  and commutes with the elements of $SU(3)$ \cite{Sh3}. 

\medskip
 It is worthy to be observed that the sphere can be obtained in two different ways, either as the coset space $\mathbb S^6 = G_2/SU(3)$, or alternatively as   $\mathbb S^6 =SO(7)/SO(6)$. On the other hand, as the subduced representations ${\cal D}_{n,0,0}$ of $SO(7)$ are isomorphic to the representation representation ${\cal D}_{0,n}$ of $G_2$ (see {\it  e.g.} \cite{Pat2}), it follows that 
 the harmonic functions on $G_2/SU(3)$ are the same as the harmonic functions on $SO(7)/SO(6)$, except that the former are labeled with the quantum numbers of $G_2$, whereas the latter are labeled by the quantum numbers of $SO(7)$. This, in particular,  implies  that
we have the Lie algebra isomorphism
\beqa
\hat{\mathfrak g}\big(G_2/SU(3)\big)\cong \hat{\mathfrak g}\big(SO(7)/SO(6)\big) \ . \nn
\eeqa

 This enables us to construct harmonic
functions on the sphere  $\mathbb S^6$ using two alternative ways, either 
using the representation theory of $\mathfrak g_2$ or  the representation theory of $\mathfrak{so}(7)$.
In the second case, we can extend the differential realisation of $\mathfrak{g}_2$ given in \eqref{eq:diff} to a differential realisation
of $\mathfrak{so}(7)$. Moreover all representations ${\cal D}_{0,n}$ turn out to be  representations of $\mathfrak{so}(7)$
corresponding to symmetric traceless tensors.

The  differential realisation  of $\mathfrak{so}(7)$ is given on page 407 in \cite{GT}. It is not  useful to reproduce the expression for all
the generators of $\mathfrak{so}(7)$, but only for the Cartan subalgebra:
\beqa
\label{eq:herm-S6}
h_1 &=& -\ii\frac{\partial}{\partial \varphi} \ , \nn\\
h_2&=&-\ii\Big(\cot \theta_2 \sin \theta_1 \frac{\partial}{\partial \theta_1}-\cos \theta_1 \frac{\partial}{\partial \theta_2} \Big) \ , \\
h_3&=&-\ii\Big(\cot \theta_4 \sin \theta_3 \frac{\partial}{\partial \theta_3}-\cos \theta_3 \frac{\partial}{\partial \theta_4} \Big) \ .\nn
\eeqa

According to Proposition  \ref{RAC}, we need nine internal labels.  Actually, as we
merely consider symmetric traceless tensors in fact, only 6 labels are required \cite{Gir}. The reduction chain
\footnote{As in Section \ref{sec:so4-so3}, the set of  Gegenbauer polynomials obtained from the reduction chain 
$SO(7)\supset SO(6)\supset SO(5) \supset SO(4)\supset SO(3)  \supset SO(2)$  of $\mathbb S^6$ does not constitute an adapted set of harmonic
functions in our case. This will further hold for all $n-$spheres.}

\beqa
SO(7) \supset SO(5) \supset SO(3)\ , \nn
\eeqa
provides three additional operators, namely the two Casimir operators of $SO(5)$ and the Casimir operator
of $SO(3)$. Thus with the generators of the Cartan subalgebra we have identified six labels.
We introduce the vector representation (with the notations of \eqref{eq:S6-p})
\beqa
\label{eq:Dn00}
{\cal D}_{1,0,0} =\Big\{\Psi_{1,1,0,0}=\sqrt{\frac72}z_1, \Psi_{1,-1,0,0}= -\sqrt{\frac72}\bar z_1,
\Psi_{1,0,1,0} =\sqrt{\frac72}z_2, \Psi_{1,0,-1,0}= \sqrt{\frac72}\bar z_2,\nn\\
\Psi_{1,0,0,1}= \sqrt{\frac72}z_3, \Psi_{1,0,0,-1} = -\sqrt{\frac72}z_3 ,\Psi_{1,0,0,0}=\sqrt{7}x_0 \Big\}
\ , 
\eeqa 
 obtained explicitly (signs included)
from the differential realisation of $\mathfrak{so}(7)$. The first label is associated to the vector representation
${\cal D}_{1,0,0}$, whilst the last three indices correspond to the eigenvalues of the Cartan subalgebra. In a similar manner,
with the
highest weight vector of the representation ${\cal D}_{n,0,0},  n \in \mathbb N$ being given by 
\beqa
\Psi_{n,n,0,0} = \sqrt{\frac 1 {60} \frac 1{4^n} \frac{(2n+5)!}{n!(n+2)!}}  z_1^n
 = \sqrt{\frac 1 {60} \frac 1{4^n} \frac{(2n+5)!}{n!(n+2)!}}  e^{\ii n \varphi}
 \sin^n \theta_1\sin^n \theta_2 \sin^n \theta_3 \sin ^n\theta_4 \sin^n \theta_5 
\ , \nn
\eeqa
the representation ${\cal D}_{n,0,0}$ can be easily obtained.
  The labels introduced previously enables us to determine an adapted Hilbert basis:
\beqa
\label{eq:H-S7}
{\cal B} =\Big\{\Psi_{n,m_1,m_2,m_3,\ell_1,\ell_2,\ell_3} \ , \ n \in \mathbb N \Big\} \ . 
\eeqa
In this notation, the first index corresponds to the representation ${\cal D}_{n,0,0}$, the three last indices
to the eigenvalues of the Cartan subalgebra of $\mathfrak{so}(7)$
and the remaining  indices to the additional internal labels.
From \eqref{eq:Dn00}
we have the 
  conjugacy relation
\beqa
\bar \Psi^{n,m_1,m_2,m_3,\ell_1,\ell_2,\ell_3}=(-1)^{n_1+n_2} \Psi_{n,-m_1,-m_2,-m_3,\ell_1,\ell_2,\ell_3} \ . \nn
\eeqa

The generators of $\hat {\mathfrak g}(SO(7)/SO(6))$ are then given by $T_{a,n,m_1,m_2,m_3,\ell_1,\ell_2,\ell_3}
=T_a \Psi_{n,m_1,m_2,m_3,\ell_1,\ell_2,\ell_3}$, the Hermitian operators \eqref{eq:herm-S6}.
The corresponding $5-$forms  are thus
\beqa
\gamma_1 &=&- \ii k_1\;\d \theta_1 \wedge \d \theta_2\wedge \d \theta_3\wedge \d \theta_4\wedge \d \theta_5
 \sin^5\theta_5 \sin^4\theta_4\sin^3\theta_3\sin^2\theta_2\sin\theta_1\ , \nn\\
\gamma_2 &=&-\ii k_2  \Big(-\cot \theta_2 \sin \theta_1 \d \varphi \wedge \d \theta_2\wedge \d \theta_3\wedge \d \theta_4\wedge \d \theta_5
-\cos\theta_1  \d \varphi \wedge \d \theta_1\wedge \d \theta_3\wedge \d \theta_4\wedge \d \theta_5\Big)\nn\\
&&\hskip 2.truecm
\times \sin^5\theta_5 \sin^4\theta_4\sin^3\theta_3\sin^2\theta_2\sin\theta_1\ \nn\\
\gamma_3 &=&-\ii k_3 \Big(-\cot \theta_4 \sin \theta_3 \d \varphi \wedge \d \theta_1\wedge \d \theta_2\wedge \d \theta_4\wedge \d \theta_5
-\cos\theta_3  \d \varphi \wedge \d \theta_1\wedge \d \theta_2\wedge \d \theta_3\wedge \d \theta_5\Big)\nn\\
&&\hskip 2.truecm
\times \sin^5\theta_5 \sin^4\theta_4\sin^3\theta_3\sin^2\theta_2\sin\theta_1
\nn
\eeqa
and the  associated central charges
are  noted $k_1,k_2,k_3$.
The Lie brackets take the form (see \eqref{eq:KM-gen}
\beqa
\label{eq:KM-S6}
&\big[T_{a,n,m_1,m_2,m_3,\ell_1,\ell_2,\ell_3},T_{a',n',m'_1,m'_2,m'_3,\ell'_1,\ell'_2,\ell'_3}\big]=\nn\\
&\ii f_{a a'}{}^{a''}
c_{a,n,m_1,m_2,m_3,\ell_1,\ell_2,\ell_3,a',n',m'_1,m'_2,m'_3,\ell'_1,\ell'_2,\ell'_3}{}^{a'',n'',m''_1,m''_2,m''_3,\ell''_1,\ell''_2,\ell''_3}
T_{a'',n'',m''_1,m''_2,m''_3,\ell''_1,\ell''_2,\ell''_3} \nn\\
&+ (-1)^{m_1+n_2}(k_1 n'_1 + k_2 n_2' + k_3 n_3')
 g_{a a'} \delta_{n,n'}\delta_{\ell_1,\ell_1'} \delta_{\ell_2,\ell_2'}\delta_{\ell_3,\ell_3'}\delta_{m_1+m_1'} \delta_{m_2+m_2'}
\delta_{m_3+m_3'}  \ ,\nn \\
&\big[h_1,T_{a,n, m, n_1,n_2,\ell}\big]= n_1 T_{a,n, m, n_1,n_2,\ell} \ , \\
&\big[h_2,T_{a,n, m, n_1,n_2,\ell}\big]=n_2 T_{a,n, m, n_1,n_2,\ell} \ , \nn\\
&\big[h_3,T_{a,n, m, n_1,n_2,\ell}\big]= n_3 T_{a,n, m, n_1,n_2,\ell} \ , \nn
\eeqa

The $c_{IJ}{}^K$ coefficients can be obtained either from the  Clebsch-Gordan coefficients in the decomposition   
\beqa
{\cal D}_{0 ,n }\otimes {\cal D}_{0,n'}  = \bigoplus \limits_{0,N} D_{0, N} \ , \nn
\eeqa
associated to $G_2$, or using the decomposition
\beqa
{\cal D}_{n,0,0 }\otimes {\cal D}_{n',0,0}  = \bigoplus \limits_{N,0,0} D_{N,0,0} \ , \nn
\eeqa
corresponding to $SO(7)$.

A question that arises naturally in this context is whether for the non-compact real form $G_{2(2)}$ of the exceptional algebra $G_2$, a 
consistent construction can be obtained for  $G_{2(2)}/SL(3,\mathbb{R})$, which can be seen as a non-compact pseudo-Riemannian version
of the non-symmetric coset $G_2/SU(3)$. This case is of great physical relevance due to its relation with the super-Ehlers embedding of minimal supergravity 
without matter coupling in $4 + 1$ space-time dimensions (see \cite{ferrara}).


\section{Concluding remarks\label{Conclusion}}

We have considered a notion of generalised Kac-Moody algebras based on the set of smooth maps from an $n-$dimensional compact manifold  $\cal M$ (associated to a compact Lie group $G_c$) to a real or complex Lie group $G$, and studied the conditions that ensure that such generalisations admit  central extensions, that have been denoted $\hat{\mathfrak g}({\cal M})$.
From this point of view, it turns out that the harmonic analysis on the manifold ${\cal M}$ as well as the representation theory of $G_c$ constitute a crucial ingredient to properly express the commutators in the generalised Kac-Moody algebra.
\medskip

We also observed that the non-centrally extended algebras  ${\mathfrak g}({\cal M})$ can be obtained naturally from a $(4+n)$-dimensional Kaluza-Klein theory compactified on the compact manifold ${\cal M}$, from which it is easily deduced that to any unitary representation of $\mathfrak g$  there corresponds a uniquely determined unitary
representation of  ${\mathfrak g}({\cal M})$. The converse of this assertion also holds. This correspondence suggests to try an extrapolation of the condition obtained
for ${\cal M} = \mathbb T^r$  to the general case  $\hat{\mathfrak g}({\cal M})$, for the case of non-vanishing central charges. This may provide an alternative tool to inspect highest weight unitary representations.

\medskip
 In Section \ref{Sec2} we have seen that considering the set $\text{Diff}({\cal M})$  of vector fields on $\cal M$,
it is possible to define an algebra with a semidirect product structure $\text{Diff}({\cal M}) \ltimes {\mathfrak g}({\cal M})$,
in analogy with the commutator structure described by equation (\ref{eq:KM-vect}).   At this point, one may wonder whether  the centrally extended algebra $\hat {\mathfrak g}({\cal M})$
is compatible with $\text{Diff}({\cal M})$.  In this context, it turns out that the compatibility condition can be expressed in terms of the  two-cocycles associated to
the central extensions, leading to the constraint 
\beqa
\omega(L\cdot X,Y)+\omega(X,L\cdot Y)=0\ \ \forall X, Y \in  {\mathfrak g}({\cal M}), \forall L \in \text{Diff}({\cal M}) \ , \nn
\eeqa
where $L\cdot X$ denotes the natural action of $\text{Diff}({\cal M})$ on ${\mathfrak g}({\cal M})$.
Compatibility in the latter sense was discussed in \cite{Frap, RS, RS2}. On a different  footing, and in the context of bosonic membranes,
central extensions of $ \text{Diff}(\mathbb S^1 \times \mathbb S^1)$ and of $ \text{Diff}(\mathbb S^2)$  have been studied by several authors (see {\it e.g. } \cite{FI,adfi,bps} and references therein). A question that remains currently unanswered is whether the symmetric nature of the coset manifolds has any consequences for the structural properties of the generalised Kac--Moody algebras. Albeit it seems that the answer is in the negative, as can be suspected from the examples presented, a definitive answer requires a more detailed analysis, as well as a careful comparison with other examples, possibly in higher ranks. We hope to provide more evidence in this  respect in future work.  

\medskip 
Several additional possibilities emerge from the generic approach described in this paper, such as the problem whether this notion of generalised Kac-Moody algebra can be applied and leads to useful insights in the description of  extended objects, such as, for example, those arising in the framework of $M-$theory ($M_2-$  or  $M_5-$branes) \cite{bst}. The extension of these results to the non-compact case {\it via} the formalism provided by the Plancherel formula is certainly a problem worthy to be considered in detail, not only because of its geometrical significance, but also due to its current physical applications. In this context, it could be suggested that (super)membrane solutions of extended theories of (super)gravity in higher dimensions might be related to the various central extensions of the generalised Kac-Moody algebras introduced in the manuscript, like $M_2-$ and $M_5-$branes are central extensions of the $N = 1$, $D = 10 + 1$ $M$-theory superalgebra.  An eventual extension of the generalised Kac-Moody algebras to non-compact Lie groups or non-compact (and possibly pseudo-Riemannian) coset manifolds ({\it e.g.} non-Euclidean tori) could then be associated to exotic versions of the $M-$theory, such as the $M^\ast$-theory of the $M^\prime$-theory \cite{hull}.
 However, as commented in Section  \ref{sec:su2-u1}, the extension of this work to manifolds involving non-compact groups present subtleties that require additional techniques to surmount the difficulties posed by the non-compacity, the details of which have not yet been solved in fully satisfactory manner, but that  warrant further investigation.

As a final observation, also of physical interest,  we point out that the  motivation of the algebras ${\mathfrak g}({\cal M})$ and  
$\hat{\mathfrak g}({\cal M})$ in terms of current algebras  is an aspect that deserves to be analysed more in detail, considering for instance specific fields,  as it may lead to some concrete realisations of $\hat{\mathfrak g}({\cal M})$.
 Work in this direction is currently in progress.

\appendix

\section{Some identities}\label{ap:ID}
Let $\cal M$ be a an $n=p+q-$dimensional compact real manifold of volume $V$ with parametrisation
$y^A=(\varphi^i,u^r) = (\varphi^1,\cdots, \varphi^p, u^1,\cdots, u^q)$. Recall that there are two types of parameters.
Angles $\varphi^1,\cdots,\varphi^p$ such that functions on ${\cal M}$ are periodic in all $\varphi-$directions, as well
as parameters, $u^1,\cdots, u^p$ that do not correspond to angles and such that the functions on ${\cal M}$ are not
periodic in all the $u-$directions. (For instance, for the sphere $\mathbb S^2$, the two parameters are the angle $0 \le \varphi< 2 \pi$ and the
parameter $-1 \le u = \cos \theta \le 1$.)

 From the integration measure, we can write 
\beqa
\int_{{\cal M}} \d \mu({\cal M}) = \frac 1 V \int_{{\cal M}} \d^p \varphi\; d^q u =1 \ , \nn
\eeqa
and let ${\cal B} = \big\{\rho_I(\varphi,u)\ , I \in {\cal I}  \big\}$, where ${\cal I}$ is a countable set
(see Section \ref{sec:MLO}),
be a orthonormal Hilbert basis of $L^2({\cal M})$. Assume further  that all functions
are bounded. Since ${\cal B}$ is a complete orthonormal basis we have
\beqa
\int_{{\cal M}} \d \mu({\cal M}) \;\overline{\rho}^I(\varphi,u)\;\rho_J(\varphi,u) = \delta^I{}_J\nn
\eeqa
and 
\beqa
\label{eq:del}
\rho_I(\varphi,u)  \overline{\rho}^I(\varphi',u')=\delta^p(\varphi-\varphi')\delta^q(u-u')\ 
\eeqa
(the sum over repeated indices is implicit).
Since the functions are bounded and ${\cal B}$ is a complete Hilbert basis we have on the one hand
\beqa
\overline \rho^I(\varphi,u) = \eta^{IJ} \rho_J(\varphi,u) \ ,\nn\\
\rho_I(\varphi,u) = \eta_{IJ} \overline\rho^J(\varphi,u) \ , \nn
\eeqa
with
\beqa
\eta^{IJ} \eta_{JK} = \delta^I_K \ ,\nn
\eeqa
and on the other hand
\beqa
\label{eq:rr}
\rho_I(\varphi,u) \rho_J (\varphi,u)= c_{IJ}{}^K \rho_K(\varphi,u) \ , \\
\overline \rho^I(\varphi,u) \overline \rho^J (\varphi,u)= c^{IJ}{}_K \overline \rho^K(\varphi,u) \ , \nn
\eeqa
where
\beqa
\overline{c_{IJ}{}^K}= c^{IJ}{}_K = \eta^{IL} \eta^{JM} \eta_{KN} c_{LM}{}^N \ . \nn
\eeqa
We now assume that ${\cal M}$ is either $G_c$ of $G_c/H$, where $G_c$ is a compact Lie group and $H\subset G_c$, so that the coefficients  $c_{IJ}{}^K$ can be expressed by means of   Clebsch-Gordan coefficients.
By using the  standard Hilbert basis $\big\{ \big|I\big>\ , \ \ I \in {\cal I}\big\}$ corresponding to all unitary representation of $G_c$,
with the notations of Section \ref{sec:MLO},  and setting $\rho_I(\varphi,u)=\big<\varphi,u\big|I\big>$,
we can extend the usual techniques of quantum mechanics for the composition of spherical harmonics with $G_c=SU(2)$ to other groups $G_c$.
We thus obtain the relations
\beqa
\left\{\begin{array}{rcl}
\rho_I (\varphi,u)\rho_J (\varphi,u)&=& c_{IJ}{}^K \rho_K (\varphi,u)\ , \\
\rho_K(\varphi,u)&=& c^{IJ}{}_K \rho_I (\varphi,u)\rho_J(\varphi,u) \ ,
\end{array}
\right.\nn
\eeqa
and
\beqa
c_{IJ}{}^K c^{IJ}{}_L &=&\delta^K_L \ , \nn\\
c_{IJ}{}^K c^{LM}{}_K&=&\delta_I^L \delta_J^M \ . \nn 
\eeqa

\section{Missing label operators}\label{sec:ML}

\noindent As observed, it may be convenient to describe the representations of a semisimple Lie algebra $\frak{g}$ with respect to
some distinguished (semisimple) subalgebra $\frak{g}^{\prime}$ that may correspond to  an internal symmetry. The question that arises is whether in such a description the labels are sufficient to separate the degeneracies that may appear. This is known as the
 `internal labelling problem'  (see {\it e.g.}  \cite{Sah2,Sh3}). The subalgebra $\frak{g}^{\prime}$ provides
$\displaystyle \frac{1}{2}(\dim
\mathfrak{g}^{\prime}+\ell^{\prime}))$ labels, where it may happen that   $\frak{g}^{\prime}$ and  $\frak{g}$ have some Casimir operator in common.  Therefore, subtracting the number $\ell_{0}$ of such common functions, we still need 
\begin{equation}
n_0=\frac{1}{2}\left(
\dim\mathfrak{g}-\ell-\dim\mathfrak{g}^{\prime}-\ell^{\prime}\right)+\ell_{0} \nn
\end{equation}
operators  to separate the irreducible representations of $\mathfrak{g}^{\prime}$ that 
appear with multiplicity greater than one in the decomposition
of $\cal D$. Such operators must necessarily commute with the generators of  $\mathfrak{g}^{\prime}$, 
and are commonly called  `missing label operators'  or `subgroup scalars'. In order to 
prevent undesired interactions and to allow simultaneous diagonalisation, these operators are additionally required to commute with each other \cite{Sh2}. 
 Among the various approaches, differential operators constitute a convenient procedure to determine internal labelling operators \cite{Ra,Pe,Tro}: Given a Lie algebra $\mathfrak{g}$  with generators
$\left\{X_{1},\ldots,X_{n}\right\}$ and commutators \footnote{ Pay attention that there is no $\ii$ factor in the Lie brackets, which
is more convenient to identify the missing label operators.}
$\left[X_{i},X_{j}\right]=f_{ij}{}^{k}X_{k}$, the generators $X_{i}$ are
 realised as differential operators in the space $C^{\infty }\left( \mathfrak{g}^{\ast }\right) $
by:
\begin{equation}
\widehat{X}_{i}=-f_{ij}{}^{k}x_{k}\frac{\partial }{\partial x_{j}}\ ,\nn
\end{equation}
where $\left\{ x_{1},\ldots,x_{n}\right\}$ are the coordinates of
a covector in a dual basis of $\left\{X_{1},\ldots,X_{n}\right\}
$. The invariants of $\mathfrak{g}$ correspond to solutions of the system of
partial differential equations:
\begin{equation}
\widehat{X}_{i}F=0,\quad 1\leq i\leq n\ , \nn 
\end{equation}
with the number $\mathcal{N}(\mathfrak{g})$ of independent solutions given by the formula
\begin{equation} \mathcal{N}(\mathfrak{g}):=\dim \,\mathfrak{g}-
{\rm sup}_{x_{1},\ldots,x_{n}}{\rm rank}\left(
A(\mathfrak{g})\right)\ , \nn 
\end{equation}
where $A(\mathfrak{g})=\left(f_{ij}{}^{k}x_{k}\right)$ corresponds to the functional matrix associated with
the commutator table of $\mathfrak{g}$ over the given basis. For
polynomial solutions, the standard
symmetrisation map defined by 
\begin{equation}
\Lambda\left(x_{i_{1}}\ldots
x_{i_{p}}\right)=\frac{1}{p!}\sum_{\sigma\in
S_{p}}X_{\sigma(i_{1})}\ldots X_{\sigma(i_{p})} \nn 
\end{equation}
with $S_r$ the permutation group with $p$ elements, 
allows to recover the Casimir operators in their usual form
as elements belonging to the centre of the enveloping algebra $U(\mathfrak{g})$ \cite{Gee}. 

\medskip

\noindent If $\frak{g}^{\prime}\subset\frak{g}$ is an embedding of Lie algebras, it induces
branching rules of representations \cite{Pat2}.  In particular, the adjoint
representation of $\frak{g}$ decomposes as:
\begin{equation}
{\rm ad} (\frak{g})= {\rm ad} (\frak{g}^{\prime})\oplus
R\ ,\nn 
\end{equation}
where $R$ is a (completely reducible) representation of
$\frak{g}^{\prime}$ called the characteristic representation.\footnote{Complete reducibility is actually
ensured only if the subalgebra $\frak{g}^{\prime}$ is semisimple.} In order to compute the missing 
labels analytically, we can proceed as follows. Let $\left\{X_{1},\ldots,X_{m}\right\}$ be a basis of $\mathfrak{g}^{\prime}$ and 
extend it to a basis $\frak{B}=\left\{X_{1},\dots,X_{m},Y_1,\dots Y_{n-m}\right\}$ of $\frak{g}$. The brackets adopt the form:
\begin{equation*}
\left[ X_i,X_j\right]= f_{ij}{}^k X_k\ ,\quad \left[ X_i,Y_p\right]= g_{ip}{}^q Y_q\ ,\quad \left[ Y_p,Y_q\right]= E_{pq}{}^k X_k+F_{pq}{}^{r}Y_r\ ,
\end{equation*}
where $i,j,k\in\left\{1,\dots ,m\right\}$ and $p,q,r\in\left\{1,\dots ,n-m\right\}$. Now we consider those differential operators that are associated to generators of $\frak{g}^{\prime}$, {\it i.e.}, the system of partial
differential equations
\begin{equation}
\widehat{X}_{i}=-f_{ij}{}^{k}x_{k}\frac{\partial }{\partial x_{j}}-g_{ip}{}^{q}y_{q}\frac{\partial }{\partial y_{p}},\quad 1\leq i\leq
 m\ ,\label{Rep2}
\end{equation}
where $\left\{ x_{1},\ldots,x_{m},y_{1},\ldots,y_{n-m}\right\}$ are the coordinates in a dual basis of $\frak{B}$. We observe that solutions $F$ to the system (\ref{Rep2}) such that $\displaystyle \frac{\partial F}{\partial y_{p}}=0$ for all $1\leq p\leq n-m$ correspond to the Casimir invariants of the subalgebra, while a genuine missing label must explicitly depend on the variables $\left\{  y_{1},\ldots,y_{n-m}\right\}$. Now the system (\ref{Rep2}) has exactly $n-r^{\prime}$ independent solutions, where $r^{\prime}$ denotes the rank of the $m\times n$ polynomial coefficient matrix. From these solutions, $\ell+\ell^{\prime}-\ell_0$ correspond to the Casimir operators of either $\frak{g}$ or $\frak{g}^{\prime}$, so that the number of available labelling operators is given by $\chi=n-r^{\prime}-\ell-\ell^{\prime}+\ell_0$. It can be easily shown (see {\it e.g.} \cite{Pe}) that $m-r^{\prime}=\ell_0$, which implies that $\chi= 2n_0$, showing that there are $n_0$ more labels available than required. It should however be noted that among these $2n_0$ solutions, at most $n_0$ correspond to operators that commute with each other \cite{C97}.
 
\medskip 
\noindent Once a complete set of $\displaystyle \frac{\dim\frak{g}+\ell}{2}$ labelling operators has been found, they can be simultaneously diagonalised, from which an orthonormal basis of states for the representation $\cal D$ is obtained. A practical
recipe for the orthonormalisation can be  found {\it e.g.} in \cite{C97}.  
 
\bigskip \noindent \textbf{Acknowledgements.}  The authors thank P. Baseilhac, G. Bossard, E. Dudas, N. Mohammedi, M. Slupinski and specially P. Sorba for helpful discussions and suggestions on the manuscript. We are grateful to the anonymous reviewer for many helpful comments and stimulating suggestions that have greatly improved the presentation, as well as suggested prospective continuation of this work. RCS  acknowledges partial financial support by the research
grants MTM2016-79422-P (AEI/FEDER, EU) and PID2019-106802GB-I00/AEI/10.13039/501100011033 (AEI/ FEDER, UE).  MdeM is grateful to the Natural Sciences and Engineering Research Council (NSERC) of Canada for partial financial support (grant number RGPIN-2016-04309).

\bibliographystyle{utphys}
\bibliography{ref-rev}

\providecommand{\href}[2]{#2}\begingroup\raggedright\begin{thebibliography}{10}

\bibitem{dms}
P.~Di~Francesco, P.~Mathieu, and D.~Senechal,
  \href{http://dx.doi.org/10.1007/978-1-4612-2256-9}{{\em {Conformal Field
  Theory}}}.
\newblock Graduate Texts in Contemporary Physics. Springer: New York, 1997.

\bibitem{Kac}
V.~G. {Kac}, ``{Simple graded {L}ie algebras of finite growth},'' {\em Func.
  Anal. Appl.} {\bf 1} (1967)  82--83.

\bibitem{Kac2}
V.~G. {Kac}, {\em {Infinite Dimensional Lie Algebras. 3rd ed.}}
\newblock Cambridge University Press: Cambridge, MA, 1990.

\bibitem{Moo}
R.~V. Moody, ``Lie algebras associated with generalized {C}artan matrices,''
  \href{http://dx.doi.org/10.1090/S0002-9904-1967-11688-4}{{\em Bull. Amer.
  Math. Soc.} {\bf 73} (1967)  217--221}.
  \url{https://doi-org.scd-rproxy.u-strasbg.fr/10.1090/S0002-9904-1967-11688-4}.

\bibitem{Mdo}
I.~G. {Macdonald}, ``{Kac-Moody-algebras.}.'' {Lie {A}lgebras and {R}elated
  {T}opics, Proc. Semin., Windsor/Ont. 1984, CMS Conf. Proc. 5, 69-109
  (1986).}, 1986.

\bibitem{ps}
A.~{Pressley} and G.~{Segal}, {\em {Loop {G}roups.}}
\newblock Oxford University Press: Oxford, 1986.

\bibitem{go}
P.~Goddard and D.~I. Olive, ``{Kac-Moody and Virasoro Algebras in relation to
  Quantum Physics},'' \href{http://dx.doi.org/10.1142/S0217751X86000149}{{\em
  Int. J. Mod. Phys. A} {\bf 1} (1986)  303--404}.

\bibitem{bpz}
A.~Belavin, A.~Polyakov, and A.~Zamolodchikov, ``Infinite conformal symmetry in
  two-dimensional quantum field theory,''
  \href{http://dx.doi.org/https://doi.org/10.1016/0550-3213(84)90052-X}{{\em
  Nuclear Physics B} {\bf 241} (1984) no.~2, 333--380}.
  \url{https://www.sciencedirect.com/science/article/pii/055032138490052X}.

\bibitem{Fuks}
D.~Fuks, {\em {Cohomology of Infinite-Dimensional {L}ie {A}lgebras}}.
\newblock Springer: New York-Berlin, 1986.

\bibitem{KT}
R.~{H{\o}egh-Krohn} and B.~{Torresani}, ``{Classification and construction of
  quasisimple Lie algebras},'' {\em {J. Funct. Anal.}} {\bf 89} (1990) no.~1,
  106--136.

\bibitem{Frap}
L.~Frappat, E.~Ragoucy, P.~Sorba, F.~Thuillier, and H.~Hogaasen, ``{Generalized
  {Kac-Moody} algebras and the diffeomorphism group of a closed surface},''
  \href{http://dx.doi.org/10.1016/0550-3213(90)90663-X}{{\em Nucl. Phys. B}
  {\bf 334} (1990)  250--264}.

\bibitem{Borc}
R.~E. Borcherds, ``Central extensions of generalised {K}ac-{M}oody algebras,''
  \href{http://dx.doi.org/10.1016/0021-8693(91)90158-5}{{\em J. Algebra} {\bf
  140} (1991)  330--335}.

\bibitem{Griess}
R.~L. Griess, ``The friendly giant,''
  \href{http://dx.doi.org/10.1007/BF01389186}{{\em Inventiones Math.} {\bf 69}
  (1982)  1--102}.

\bibitem{Conw}
J.~H. Conway, ``A simple construction for the {F}ischer-{G}riess monster
  group,'' \href{http://dx.doi.org/10.1007/BF01388521}{{\em Inventiones Math.}
  {\bf 79} (1985)  513--540}.

\bibitem{Conw2}
J.~H. Conway and S.~P. Norton, ``Monstrous {M}oonshine,''
  \href{http://dx.doi.org/10.1112/blms/11.3.308}{{\em Bull. London Math. Soc.}
  {\bf 11} (1979)  308--339}.

\bibitem{Bor2}
R.~E. Borcherds, ``Monstrous moonshine and monstrous {L}ie superalgebras,''
  \href{http://dx.doi.org/10.1007/BF01232032}{{\em Inventiones Math.} {\bf 109}
  (1992)  405--444}.

\bibitem{Gann}
T.~Gannon, {\em {Moonshine Beyond the Monster}}.
\newblock Cambridge Univ. Press: Cambridge, MA, 2006.

\bibitem{ss}
A.~Salam and J.~Strathdee, ``{On Kaluza-Klein Theory},''
  \href{http://dx.doi.org/10.1016/0003-4916(82)90291-3}{{\em Annals Phys.} {\bf
  141} (1982)  316--352}.

\bibitem{bl}
D.~Bailin and A.~Love, ``{Kaluza-Klein Theories},''
  \href{http://dx.doi.org/10.1088/0034-4885/50/9/001}{{\em Rept. Prog. Phys.}
  {\bf 50} (1987)  1087--1170}.

\bibitem{dpn}
M.~J. Duff, B.~E.~W. Nilsson, and C.~N. Pope, ``{Kaluza-Klein Supergravity},''
  \href{http://dx.doi.org/10.1016/0370-1573(86)90163-8}{{\em Phys. Rept.} {\bf
  130} (1986)  1--142}.

\bibitem{ad}
S.~L. {Adler} and R.~F. {Dashen}, {\em {Current Algebras and Applications to
  Particle Physics}}.
\newblock Benjamin: New York, 1968.

\bibitem{gjt}
S.~Treiman, R.~Jackiw, and D.~J. Gross,
  \href{http://dx.doi.org/10.1515/9781400871506}{{\em {Lectures on Current
  Algebra and Its Applications}}}.
\newblock Princeton University Press: Princeton, NJ, 1972.

\bibitem{dd}
L.~Dolan and M.~Duff, ``{Kac-Moody symmetries of Kaluza-Klein theories},''
  \href{http://dx.doi.org/10.1103/PhysRevLett.52.14}{{\em Phys. Rev. Lett.}
  {\bf 52} (1984)  14--17}.

\bibitem{Scw}
J.~Schwinger, ``Field theory commutators,''
  \href{http://dx.doi.org/10.1103/PhysRevLett.3.296}{{\em Phys. Rev. Lett.}
  {\bf 3} (1959)  296--297}.
  \url{https://link.aps.org/doi/10.1103/PhysRevLett.3.296}.

\bibitem{HC}
Harish-Chandra, ``{Harmonic Analysis on Real Reductive Groups III, The
  Maas--Selberg relations and the Plancherel formula},''
  \href{http://dx.doi.org/10.2307/1971058}{{\em Ann. of Math.} {\bf 104} (1976)
   117--201}.

\bibitem{Bars}
I.~Bars, ``{Local charge algebras in quantum chiral models and gauge
  theories}.'' {I}n {V}ertex {O}perators in {M}athematics and {P}hysics, {E}d.
  {J}. {L}eponsky, {S}. {M}andelstam and {I}. {M}. {S}inger ({S}pringer,
  {B}erlin 1984), pp 373--391.

\bibitem{Har}
S.~M. Harrison, N.~M. Paquette, and V.~P., ``A {B}orcherds-{K}ac-{M}oody
  superalgebra with {C}onway symmetry,''
  \href{http://dx.doi.org/10.1007/s00220-019-03518-0}{{\em Comm. Math. Phys.}
  {\bf 370} (2019)  539--590}.

\bibitem{Azam}
S.~Azam, ``A new characterization of {K}ac-{M}oody-{M}alcev superalgebras,''
  \href{http://dx.doi.org/10.1142/S0219498817501444}{{\em J. Alg. Appl.} {\bf
  16} (2017)  1750144(15pp)}.

\bibitem{CFRS}
R.~Coquereaux, L.~Frappat, E.~Ragoucy, and P.~Sorba, ``{Extended
  super-Kac-Moody algebras and their super-derivation algebras},''
  \href{http://dx.doi.org/cmp/1104201313}{{\em Communications in Mathematical
  Physics} {\bf 133} (1990) no.~1, 1 -- 35}. \url{https://doi.org/}.

\bibitem{RS}
E.~Ragoucy and P.~Sorba, ``{Extended Kac-Moody algebras and applications},''
  \href{http://dx.doi.org/10.1142/S0217751X92001307}{{\em Int. J. Mod. Phys. A}
  {\bf 7} (1992)  2883--2972}.

\bibitem{PW}
F.~{Peter} and H.~{Weyl}, ``{Die Vollst\"andigkeit der primitiven Darstellungen
  einer geschlossenen kontinuierlichen Gruppe.},'' {\em {Math. Ann.}} {\bf 97}
  (1927)  737--755.

\bibitem{Schmid}
W.~Schmid, ``{Representations of semi-simple Lie groups}.'' {i}n
  {R}epresentation {T}heory of {L}ie {G}roups, {P}roceedings of the {SRC}/{LMS}
  {R}esearch {S}ymposium on {R}epresentations of {L}ie {G}roups, {O}xford, 28
  {J}une - 15 {J}uly 1977, {E}ds. {M}.{F}. {A}tiyah et al. ({C}ambridge
  {U}niversity {P}ress, {C}ambridge 1979), pp 185--235.

\bibitem{barut}
A.~O. Barut and R.~A. Raczka, {\em Theory of Group Representations and
  Applications, 2nd revised ed.}
\newblock Polish Scientific Publishers: Warszawa, 1980.

\bibitem{Ra}
G.~{Racah}, ``{Sulla caratterizzazione delle rappresentazioni irriducibili dei
  gruppi semisemplici di Lie},'' {\em {Atti Accad. Naz. Lincei, VIII. Ser.,
  Rend., Cl. Sci. Fis. Mat. Nat.}} {\bf 8} (1950)  108--112.

\bibitem{Co}
J.~Cornwell, {\em Group Theory in Physics, Volume 1}.
\newblock Academic Press: London, 1984.

\bibitem{Bdh}
L.~C. {Biedenharn}, ``{On the representations of the semisimple Lie groups. I:
  The explicit construction of invariants for the unimodular unitary group in
  \(n\) dimensions},'' {\em {J. Math. Phys.}} {\bf 4} (1963)  436--445.

\bibitem{Lou}
J.~D. {Louck}, {\em {Unitary {S}ymmetry and {Combinatorics}}}.
\newblock World Scientific: Hackensack, NJ, 2008.

\bibitem{bt}
R.~Bott and L.~W. Tu, \href{http://dx.doi.org/10.1007/978-1-4757-3951-0}{{\em
  {Differential Forms in Algebraic Topology}}}.
\newblock Springer: New York-Berlin, 1982.

\bibitem{gow}
M.~R. Gaberdiel, D.~I. Olive, and P.~C. West, ``{A Class of Lorentzian
  Kac-Moody algebras},''
  \href{http://dx.doi.org/10.1016/S0550-3213(02)00690-9}{{\em Nucl. Phys. B}
  {\bf 645} (2002)  403--437}, \href{http://arxiv.org/abs/hep-th/0205068}{{\tt
  arXiv:hep-th/0205068}}.

\bibitem{west}
P.~West, {\em {Introduction to Strings and Branes}}.
\newblock Cambridge University Press: Cambridge, 2012.

\bibitem{GT}
R.~Campoamor-Stursberg and M.~Rausch~de Traubenberg,
  \href{http://dx.doi.org/10.1142/11081}{{\em {Group Theory in Physics: A
  Practitioner's Guide}}}.
\newblock World Scientific: Singapore, 2019.

\bibitem{Edm}
A.~R. {Edmonds}, {\em {Angular {M}omentum in {Q}uantum {M}echanics. }}.
\newblock Princeton Univ. Press: Princeton NJ, 1996.

\bibitem{Van}
A.~P. {Jucys}, I.~B. {Levinson}, and V.~V. {Vanagas}, ``{Mathematical
  {A}pparatus of the {T}heory of {A}ngular {M}omentum.}.'' {Israel Program for
  Scientific Translations: Jerusalem}, 1962.

\bibitem{dewit1}
B.~de~Wit and A.~Van~Proeyen, ``{Broken sigma model isometries in very special
  geometry},'' \href{http://dx.doi.org/10.1016/0370-2693(92)91485-R}{{\em Phys.
  Lett. B} {\bf 293} (1992)  94--99}.

\bibitem{dewit2}
B.~de~Wit, F.~Vanderseypen, and A.~Van~Proeyen, ``{Symmetry structure of
  special geometries},''
  \href{http://dx.doi.org/10.1016/0550-3213(93)90413-J}{{\em Nucl. Phys. B}
  {\bf 400} (1993)  463--524}.

\bibitem{barg}
V.~Bargmann, ``Irreducible unitary representations of the {L}orentz group,''
  \href{http://dx.doi.org/10.2307/1969129}{{\em Ann. of Math. (2)} {\bf 48}
  (1947)  568--640}.
  \url{https://doi-org.scd-rproxy.u-strasbg.fr/10.2307/1969129}.

\bibitem{Campoamor-Stursberg:2014ffa}
R.~Campoamor-Stursberg and M.~Rausch~de Traubenberg, ``{Unitary representations
  of three dimensional Lie groups revisited: A short tutorial via harmonic
  functions},'' \href{http://dx.doi.org/10.1016/j.geomphys.2017.01.004}{{\em J.
  Geom. Phys.} {\bf 114} (2017)  534--553},
  \href{http://arxiv.org/abs/1404.4705}{{\tt arXiv:1404.4705 [math-ph]}}.

\bibitem{Kam}
J.~K. de~{F\'eriet}, {\em {Fonctions de la Physique Math\'ematique}}.
\newblock CNRS: Paris, 1957.

\bibitem{Ge}
J.~E. Avery and J.~S. Avery, \href{http://dx.doi.org/10.1142/10690}{{\em
  Hyperspherical Harmonics and their Physical Applications}}.
\newblock World Scientific: Singapore, 2018.
\newblock
  \href{http://arxiv.org/abs/https://www.worldscientific.com/doi/pdf/10.1142/10690}{{\tt
  https://www.worldscientific.com/doi/pdf/10.1142/10690}}.
\newblock \url{https://www.worldscientific.com/doi/abs/10.1142/10690}.

\bibitem{Pat}
J.~Patera and D.~Sankoff, {\em {Tables of Branching Rules for Representations
  of Simple Lie Algebras}}.
\newblock Presses de l'Universit\'e de Montr\'eal: Montr\'eal, 1973.

\bibitem{gms}
I.~M. Gel'fand, R.~A. Minlos, and Z.~Y. Shapiro, {\em Representations of the
  Rotation and Lorentz Groups and their Applications}.
\newblock Pergamon Press, Oxford, 1963.

\bibitem{otto}
U.~Ottoson, ``{A classification of the irreducible unitary representations of
  $SO_0(n,1)$},'' {\em Comm. Math. Phys.} {\bf 8} (1968)  228--244.

\bibitem{kostant}
B.~Kostant, ``{The principal three-dimensional subgroup and the Betti numbers
  of a complex simple Lie group},''
  \href{http://dx.doi.org/10.2307/2372999}{{\em Am. J. Math.} {\bf 81} (1959)
  973--1032}.

\bibitem{dynkin}
E.~Dynkin, ``{Maximal subgroups of the classical groups},'' {\em Amer. Math.
  Soc. Transl. Ser. 2} {\bf 6} (1957)  245--378.

\bibitem{vkac1}
V.~G. Kac, ``{Simple irreducible graded Lie algebras of finite growth},'' {\em
  Math. USSR-Izv.} {\bf 2} (1968)  1271--1311.

\bibitem{vkac2}
V.~G. Kac, ``{Automorphisms of finite order of semisimple Lie algebras},''
  \href{http://dx.doi.org/10.1007/BF01676631}{{\em Functional Anal. Appl.} {\bf
  3} (1969)  252--254}.

\bibitem{Vin}
E.~B. Vinberg, ``{The {W}eyl group of a graded Lie algebra },''
  \href{http://dx.doi.org/10.1070/IM1976v010n03ABEH001711}{{\em Izv. Akad. Nauk
  SSSR Ser. Mat.} {\bf 40} (1976)  488--526}.

\bibitem{rb}
M.~Beg and H.~Ruegg, ``A set of harmonic functions for the group {SU}$(3)$,''
  \href{http://dx.doi.org/10.1063/1.1704325}{{\em J. Math. Phys.} {\bf 6}
  (1965)  677--682}.

\bibitem{Wybo}
B.~G. Wybourne, ``Exceptional {L}ie groups in physics,'' {\em Lith. J. Phys.}
  {\bf 35} (1995)  123--132.

\bibitem{zz}
X.~Xu, \href{http://dx.doi.org/10.1007/978-981-10-6391-6}{{\em
  {R}epresentations of {L}ie {A}lgebras and {P}artial {D}ifferential
  {E}quations}}.
\newblock Springer: Singapore, 2017.

\bibitem{Sh2}
R.~T. {Sharp} and C.~S. {Lam}, ``{Internal-labeling problem},'' {\em {J. Math.
  Phys.}} {\bf 10} (1969)  2033--2038.

\bibitem{Sh3}
R.~T. {Sharp}, ``{Internal-labeling operators},'' {\em {J. Math. Phys.}} {\bf
  16} (1975)  2050--2053.

\bibitem{Pat2}
W.~G. {McKay} and J.~{Patera}, {\em {Tables of {D}imensions, {I}ndices, and
  {B}ranching {R}ules for {R}epresentations of {S}imple {L}ie {A}lgebras}},
  vol.~69.
\newblock CRC Press: Boca Raton, FL, 1981.

\bibitem{Gir}
Y.~Giroux, M.~Couture, and R.~T. Sharp, ``{Degenerate enveloping algebras of
  $SU(3)$, $SO(5)$, $G_2$ and $SU(4)$},''
  \href{http://dx.doi.org/10.1088/0305-4470/17/4/013}{{\em J. Phys. A: Math.
  Gen.} {\bf 17} (1984)  715}.

\bibitem{ferrara}
S.~Ferrara, A.~Marrani, and A.~Trigiante, ``{Super-Ehlers in Any Dimension},''
  \href{http://dx.doi.org/10.1007/JHEP11(2012)068}{{\em JHEP} {\bf 11} (2012)
  068}.

\bibitem{RS2}
E.~Ragoucy and P.~Sorba, ``{An Attempt to relate area preserving
  diffeomorphisms to Kac-Moody algebras},''
  \href{http://dx.doi.org/10.1007/BF00398331}{{\em Lett. Math. Phys.} {\bf 21}
  (1991)  329--342}.

\bibitem{FI}
E.~G. Floratos and J.~Iliopoulos, ``{A note on the classical symmetries of the
  closed bosonic membranes},''
  \href{http://dx.doi.org/10.1016/0370-2693(88)90220-1}{{\em Phys. Lett. B}
  {\bf 201} (1988)  237--240}.

\bibitem{adfi}
I.~Antoniadis, P.~Ditsas, E.~Floratos, and J.~Iliopoulos, ``{New realizations
  of the {V}irasoro algebra as membrane symmetries},''
  \href{http://dx.doi.org/10.1016/0550-3213(88)90612-8}{{\em Nucl. Phys. B}
  {\bf 300} (1988)  549--558}.

\bibitem{bps}
I.~Bars, C.~N. Pope, and E.~Sezgin, ``{Central extensions of area preserving
  membrane algebras},''
  \href{http://dx.doi.org/10.1016/0370-2693(88)90354-1}{{\em Phys. Lett. B}
  {\bf 210} (1988)  85--91}.

\bibitem{bst}
E.~Bergshoeff, E.~Sezgin, and P.~K. Townsend, ``{Properties of the
  eleven-dimensional super membrane theory},''
  \href{http://dx.doi.org/10.1016/0003-4916(88)90050-4}{{\em Annals Phys.} {\bf
  185} (1988)  330}.

\bibitem{hull}
C.~Hull, ``{Duality and the signature of space-time},''
  \href{http://dx.doi.org/10.1088/1126-6708/1998/11/017}{{\em JHEP} {\bf 11}
  (1998)  017}.

\bibitem{Sah2}
R.~T. {Sharp}, ``{Internal labelling: the classical groups},'' {\em {Proc.
  Camb. Philos. Soc.}} {\bf 68} (1970)  571--578.

\bibitem{Pe}
{Peccia, A. and Sharp,R. T.}, ``{Number of independent missing label
  operators},'' {\em {J. Math. Phys.}} {\bf 17} (1976)  1313--1314.

\bibitem{Tro}
E.~G. {Beltrametti} and A.~{Blasi}, ``{On the number of Casimir operators
  associated with any Lie group},'' {\em {Phys. Lett.}} {\bf 20} (1966)
  62--64.

\bibitem{Gee}
I.~M. {Gel'fand}, ``{Das Zentrum eines infinitesimalen Gruppenringes},'' {\em
  {Mat. Sb., Nov. Ser.}} {\bf 26} (1950)  103--112.

\bibitem{C97}
R.~{Campoamor-Stursberg}, ``{Internal labelling problem: an algorithmic
  procedure},'' {\em {J. Phys. A, Math. Theor.}} {\bf 44} (2011) no.~2, 18.
  Id/No 025204.

\end{thebibliography}\endgroup\newpage

\end{document}